\documentclass[11pt,a4paper]{article}

\usepackage[utf8]{inputenc}
\usepackage[T1]{fontenc}
\usepackage{lmodern}
\usepackage[margin=2.2cm]{geometry}
\usepackage{amsmath,amssymb,amsthm,mathtools}
\usepackage{bm}
\usepackage{graphicx}
\usepackage{xcolor}
\usepackage{booktabs}
\usepackage{longtable}
\usepackage[colorlinks=true,citecolor=blue,linkcolor=blue,urlcolor=blue]{hyperref}
\usepackage[numbers,sort&compress,square]{natbib}
\usepackage{setspace}
\usepackage{microtype}
\usepackage{authblk}
\usepackage{titlesec}

\newcommand{\Qsuper}{Q_{\mathrm{super}}}
\newcommand{\Qcl}{Q_{\mathrm{cl}}}
\newcommand{\gammarad}{\gamma_{\mathrm{rad}}}
\newcommand{\gammanr}{\gamma_{\mathrm{nr}}}

\newcommand{\Leff}{L_{\mathrm{eff}}}
\newcommand{\Vmode}{V_{\mathrm{mode}}}

\newcommand{\Fpur}{F_{\mathrm{P}}}

\newcommand{\im}{\mathrm{Im}\,}

\newcommand{\eg}{\textit{e.g.}}
\newcommand{\etal}{\textit{et~al.}}

\titleformat*{\section}{\large\bfseries}
\titleformat*{\subsection}{\normalsize\bfseries}
\titleformat*{\subsubsection}{\normalsize\itshape}
\theoremstyle{definition}
\newtheorem*{definition}{Definition}

\title{\bfseries Super-Resonance: Interference-Driven Suppression of
       Radiative Decay Across Wave Physics}

\author[1]{Igor V. Minin}
\author[1]{Oleg V. Minin}
\author[2]{Ilia L. Rasskazov}

\affil[1]{Novosibirsk Branch of Rzhanov Institute of Semiconductor Physics SBRAS "Technological Design Institute of Applied microelectronics", Lavrentev avenue, 2/1, Novosibirsk, 630090, Russia}

\affil[2]{Independent Researcher,
San Jose, CA 95124, USA}

\date{\today}

\begin{document}
\maketitle

\begin{abstract}
\noindent
The term \emph{super-resonance} has been coined independently in several
communities: by Tolstoy for collective modes of acoustic cavity arrays,
in photonics for extreme high-$Q$ states of dielectric resonators, in
magnonics for leaky-surface magnon-polaron bound states. It has since
come to be applied, often interchangeably, to phenomena that are
physically distinct. This analytical review is organised around the one usage that
admits a precise, field-agnostic definition with broad consensus behind
it: a resonant mode whose \emph{radiative} linewidth is driven toward
zero by destructive interference among its radiation channels,
$\gammarad/\gammarad^{\mathrm{cl}} \to 0$, while the bare single-channel
coupling to the environment remains generically large. The canonical
mathematical realisation is the Friedrich--Wintgen scenario of two modes
sharing a radiation continuum. The same non-Hermitian eigenvalue
structure recurs, with essentially identical mathematics, in Tolstoy's
super-resonant acoustic arrays, photonic bound states in the continuum
(BICs) and supercavity (quasi-BIC) modes, anapole states, dark modes of
coherent perfect absorbers, parity--time-symmetric devices near
exceptional points, magnonic dark modes, and topologically protected
edge modes whose radiation channels are closed by a bulk invariant. We
develop this operational definition, survey its realisations field by
field, and distil a compact set of transferable design rules and figures
of merit (quality factor, Purcell factor, cooperativity, effective
interaction length). We then explicitly disambiguate super-resonance
from two related but physically distinct families that share the
``super-'' prefix and are frequently conflated with it. The first is
\emph{coherent enhancement}, in which the coupling to the continuum is
collectively increased rather than suppressed: Dicke super-radiance,
superscattering, nuclear giant resonances, high-order Fano resonances of
mesoscale dielectric spheres, whispering-gallery modes. The second is
\emph{integer-commensurability locking}: mean-motion orbital resonances,
wave--particle resonances in plasmas, Floquet replicas in time-modulated
media. A third relative, \emph{amplification} --- a reflection
coefficient exceeding unity, as in black-hole super-radiance and its
laboratory analogues --- is likewise distinguished. Reading the
literature with this map in mind transfers design rules between
acoustics, photonics, and magnonics, identifies which figure of merit
each mechanism actually optimises, and removes a terminological
ambiguity that has begun to obscure genuinely common physics.
\end{abstract}

\vspace{1ex}

\tableofcontents

\section{Introduction}\label{sec:intro}

Few concepts in physics are as ubiquitous as resonance. Whenever a system
possesses a discrete eigenmode and that mode is excited at, or near, its
natural frequency, a disproportionate and often dramatic response ensues.
The phenomenon, named after the Latin \emph{resono} (``to re-sound''), was
known since antiquity --- exploited in the acoustics of the
fourth-century-BCE Greek theatre at Epidaurus \citep{rindel2025epidaurus}
and echoed in the legendary collapse of the walls of Jericho recounted in
the Book of Joshua \citep{kennedy2023jericho} --- yet it was first
described scientifically only by Galileo Galilei in 1602
\citep{frova2006galileo}; its kinematic completion in the work of
Helmholtz \citep{helmholtz1860resonator} and Rayleigh
\citep{rayleigh1894sound} provided the language in which acoustics, optics,
electrodynamics, and ultimately quantum mechanics would all be cast.
Resonance is so fundamental that even the modern interpretation of a
nuclear ``compound state'' rests on the Breit--Wigner formula
\citep{breit1936capture}, the simplest non-trivial example of a complex
eigenvalue of the scattering matrix. By the late twentieth century the
concept had become so total --- traversing acoustics, optics, classical and
quantum electrodynamics, nuclear and atomic physics, condensed matter,
celestial mechanics, and even cosmology --- that the Nobel committee, as
early as 1913, debated whether mechanical resonance alone merited the prize
in physics \citep{bleck2018mechanical,buchanan2019resonance}.

It is therefore remarkable that a more basic question has been asked
relatively recently and, until now, only in isolated communities: can a
resonance ever do \emph{better} than a textbook Lorentzian? It is generally
assumed that a physical resonance can significantly exceed the classical
Lorentzian line shape, depending on the metric of interest --- filter
sharpness, sensor sensitivity, or signal transmission
\citep{tang2024angle}. One specific
affirmative answer has been arrived at independently, with essentially
identical mathematics, by communities that ostensibly have nothing to do
with one another: when the radiative width of a mode is a coherent sum of
partial amplitudes, destructive interference among those amplitudes ---
the same interferometric (Friedrich--Wintgen) mechanism later recognised
as the origin of bound states in the continuum --- can
drive the \emph{radiative} linewidth toward zero even though the bare
coupling of the mode to its environment remains large. It is this
phenomenon --- and, we shall argue, only this phenomenon --- that deserves
the name \emph{super-resonance} in a strict sense. At the same time, the
prefix ``super-'' has been attached, in different communities, to at least
two further classes of resonant phenomena whose physics is materially
different, and part of the purpose of this review is to draw those
distinctions explicitly rather than to paper over them.

The interference phenomenon appeared first, under this name, in acoustics:
Tolstoy \citep{tolstoy1986superresonant,tolstoy1987properties} observed
that arrays of elastic spherical cavities can support collective scattering
modes whose radiation linewidth approaches zero in the ideal-array limit,
and coined the term \emph{super-resonance} on this occasion. The same
mathematical structure --- two or more modes sharing a radiation continuum,
with the dark combination decoupling from it --- had been formalised by
Friedrich and Wintgen \citep{friedrich1985interfering}, building on the
von~Neumann--Wigner construction of 1929, and it re-emerged independently
in photonics: bound states in the continuum (BICs) were identified
theoretically by Marinica \etal\ \citep{marinica2008bound} and demonstrated
by Hsu \etal\ \citep{hsu2013observation,hsu2016bound}; supercavity
(quasi-BIC) modes of subwavelength dielectric resonators by Rybin \etal\
\citep{rybin2017highq,klimov2019supercavity,rybin2024metaphotonics};
anapole (nonradiating) states of single nanoparticles by Miroshnichenko
\etal\ \citep{miroshnichenko2015nonradiating}; dark modes of coherent
perfect absorbers \citep{chong2010coherent,wan2011timereversed,
baranov2017coherent} and their super-resonant nested-cavity extension by
Malara \etal\ \citep{malara2016super}; parity--time-symmetric and
exceptional-point devices \citep{bender1998real,chen2017exceptional,
hodaei2017enhanced,miri2019exceptional}; and, most recently,
\emph{super-resonant} states in leaky surface magnon polaron spectra
\citep{sukhorukova2022superresonant} together with dark magnon modes of
multi-sphere cavity magnonics \citep{zhang2015magnon}. In every one of
these systems the same structural feature is present: the radiative width
of a mode is a coherent sum of partial amplitudes, and interference drives
it toward zero while the bare coupling remains large.

The same prefix has, however, also been attached to phenomena of a
different character. Wan, Huang, and Innanen \citep{wan2001pluto} named a
1:1 commensurability between Pluto's argument-of-perihelion circulation
and its 3:2 mean-motion libration the \emph{Pluto super-resonance} ---
a phase-locking phenomenon of conservative dynamics. Dicke
super-radiance \citep{dicke1954coherence,gross1982superradiance} and
superscattering \citep{ruan2010superscattering} denote collective
\emph{enhancement} of radiative coupling --- the rate is increased, not
suppressed. Black-hole super-radiance
\citep{press1972rotating,bekenstein1973black,brito2020superradiance}
denotes \emph{amplification}, a reflection coefficient exceeding unity.
The high-order Fano resonance of mesoscale dielectric spheres
\citep{wan2019highorder,minin2024discovery} concentrates the spectral
weight of the Mie expansion into a single high-multipole mode. In
nonlinear terahertz photonics, Babushkin \etal\
\citep{babushkin2023metallic} apply the same word to the merging of
closely spaced quantum-confinement resonances of a metallic
nanostructure into one composite peak whose position and width follow
the particle geometry --- again a consolidation of coupling, not a
suppression of radiative width (Sec.~\ref{sec:plasmonics}). Most
recently, the term \emph{hyperbolic super-resonance} has been introduced
for the extreme high-$Q$ polariton modes of hyperbolic metamaterials
proposed as mediators of long-range, ultrastrong qubit--qubit coupling
\citep{narimanov2024hyperbolic} --- once again an \emph{enhancement} of a
coupling rather than the suppression of a radiative width. These are
all striking resonant phenomena, and each is surveyed below; but they are
not the same physics as interference-driven radiative suppression, and we
will be explicit about the distinctions.

The interference phenomena listed two paragraphs above share a single
quantitative criterion: the \emph{radiative} linewidth of the mode,
$\gammarad$, is driven below the single-channel value
$\gammarad^{\mathrm{cl}}$ that the same mode would have in the absence of
interference, with the ratio vanishing in the ideal limit. The figures of
merit that follow --- the quality factor $Q$, the radiative bandwidth
$\Delta\omega_{\mathrm{rad}}$, the Purcell factor $\Fpur$ at fixed mode
volume, the cooperativity $C$, and the effective interaction length
$\Leff$ --- all diverge together in that limit. We refer to this class as
\emph{super-resonance}, and we shall argue that it admits a precise,
field-agnostic mathematical definition.

The literature on each of these sub-topics is, by itself, enormous, and
several outstanding reviews exist for individual fields. Hsu \etal\
\citep{hsu2016bound} and Azzam and Kildishev \citep{azzam2021photonic}
provide comprehensive accounts of bound states in the continuum; Limonov
\etal\ \citep{limonov2017fano} and Miroshnichenko \etal\
\citep{miroshnichenko2010fano} of Fano resonances; Rybin and Limonov
\citep{rybin2019resonance} and Rybin and Kivshar
\citep{rybin2024metaphotonics} of resonance effects in photonic crystals
and metamaterials; Frisk Kockum \etal\ \citep{frisk2019ultrastrong} and
Forn-D{\'\i}az \etal\ \citep{forn2019ultrastrong} of ultrastrong coupling;
El-Ganainy \etal\ \citep{elganainy2018non} and \"Ozdemir \etal\
\citep{ozdemir2019paritytime} of non-Hermitian physics; Brito, Cardoso, and
Pani \citep{brito2020superradiance} of black-hole super-radiance; and
Murray and Dermott \citep{murray1999solar} of solar-system dynamics. What
these field-specific accounts do not provide --- and what we attempt here
--- is two things: a cross-disciplinary treatment of the
interference-suppression mechanism itself, which is the same eigenvalue
problem whether it is solved in an acoustic array, a dielectric
metasurface, or a YIG sphere; and an explicit disambiguation of that
mechanism from the enhancement, locking, and amplification phenomena that
share its name.

Our central thesis is therefore deliberately narrower than a universal
umbrella: one mechanism --- destructive interference among radiation
channels --- admits a closed-form, field-agnostic mathematical expression
with broad consensus behind it, and the mapping from that expression back
to the individual fields reveals concrete design rules and transferable
FOMs. To make the argument we present the
mathematics \emph{first}, in Section~\ref{sec:definition}, and
then survey the individual fields against that template in
Section~\ref{sec:manifestations}. The motivation for this ordering is
pedagogical: only when the operational definition is internalised does the
common thread among, say, a Tolstoy acoustic array, a photonic quasi-BIC
metasurface, a coherent-perfect-absorber dark mode, and a magnonic dark
mode become impossible to miss --- and, equally, does it become clear why
a Laplace chain of Galilean moons or a Dicke super-radiant burst, for all
their interest, instantiate different physics.

The review is organised as follows.
Section~\ref{sec:definition} develops the operational definition of
super-resonance --- mechanism M1, interference-driven radiative
suppression --- exhibits its canonical Friedrich--Wintgen mathematics, and
then introduces, for contrast, the two relative families with which it is
most often conflated: coherent enhancement (family M2) and
integer-commensurability locking (family M3), together with the
amplification phenomena of rotating backgrounds.
Section~\ref{sec:manifestations} then surveys the realisations across the
disciplines that have, historically, hosted the term: acoustics and
elastodynamics (Sec.~\ref{sec:acoustics}); celestial mechanics and
gravitational physics, included for disambiguation
(Sec.~\ref{sec:celestial}); high-order Fano resonance in mesoscale
dielectric spheres (Sec.~\ref{sec:mie}); bound states in the continuum and
supercavity modes (Sec.~\ref{sec:bic}); plasmonics, whispering-gallery, and
photonic-crystal cavities (Sec.~\ref{sec:plasmonics}); non-Hermitian
photonics including coherent perfect absorption, parity--time symmetry, and
exceptional points (Sec.~\ref{sec:pt}); topological and Floquet
super-resonance (Sec.~\ref{sec:topological}); cavity QED, atomic
super-radiance, nuclear giant resonances, and magnetic resonance
(Sec.~\ref{sec:quantum}); magnonic super-resonance in spin-wave and
cavity-magnonic systems (Sec.~\ref{sec:magnonics}); plasma wave--particle
resonance and non-inductive current drive (Sec.~\ref{sec:plasmaSR}); and
emerging frontiers in time-modulated media, cavity optomechanics,
gravitational-wave readout, and ultralight dark-matter searches
(Sec.~\ref{sec:emerging}). Section~\ref{sec:applications}
surveys the application domains in which interference-suppressed radiative
decay has been demonstrated experimentally, with emerging directions
collected separately and labelled as such. Section~\ref{sec:design} distils
a compact set of design rules and identifies natural cross-field transfers,
and Section~\ref{sec:outlook} closes with the open problems whose
resolution will define super-resonance research over the coming decade.

Two caveats must be stated up front. First, the literature on resonance
itself is so vast that no review can hope to be exhaustive; our selection
emphasises, on one hand, the works most central to establishing the
conceptual unity, and on the other hand, the works closest to our own
research programme on mesoscale dielectric resonators
\citep{wan2019highorder,minin2024freezing,minin2024discovery,
minin2023cenosphere,minin2022janus,geints2024magnetic,yue2020teflon,
yue2019poynting,minin2022fano,minin2022extreme,minin2025mesotronics}.
Second, throughout the review we are careful to distinguish a genuine
super-resonance from a merely large or sharp ordinary resonance; the
delimitation is given in Sec.~\ref{sec:taxonomy} and revisited in
Sec.~\ref{sec:design}.

\section{Operational Definition, the Mechanism, and Its Relatives}\label{sec:definition}

\subsection{Non-Hermitian framework}\label{sec:nonherm}

Let a generic physical system be described by an effective non-Hermitian
Hamiltonian (or, equivalently, a dynamical operator) of the form
\begin{equation}
\widehat{H}_{\mathrm{eff}} = \mathbf{M} - \mathrm{i}\,\bm{\Gamma},
\qquad
\bm{\Gamma}=\bm{\Gamma}_{\mathrm{rad}}+\bm{\Gamma}_{\mathrm{nr}},
\label{eq:Heff}
\end{equation}
where $\mathbf{M}$ is the Hermitian part encoding the conservative dynamics
and $\bm{\Gamma}$ collects all loss channels, partitioned into radiative
coupling to a continuum ($\bm{\Gamma}_{\mathrm{rad}}$) and non-radiative
dissipation ($\bm{\Gamma}_{\mathrm{nr}}$). This single equation already
spans nearly all of the resonance literature: the matrix $\mathbf{M}$ may
be the Hermitian core of a tight-binding model, the elastodynamic
operator of a continuum medium, the Hamiltonian of a multi-level atom
coupled to a quantised field, the Riccati--Bessel propagator of Mie
scattering, or even the Teukolsky operator on a Kerr background. The
matrix $\bm{\Gamma}$ encodes the openness of the system: it is what
distinguishes a closed eigenvalue problem (with real eigenfrequencies and
infinite lifetimes) from the realistic situation in which energy can leak
into a continuum.

The complex eigenvalues
$\widetilde\omega_n=\omega_n - \mathrm{i}\gamma_n/2$ of \eqref{eq:Heff}
define the system's resonances, and the spectral response to a coherent
drive takes the generalised Lorentzian form
\begin{equation}
\chi(\omega)=\sum_n \frac{A_n}{\omega-\omega_n+\mathrm{i}\gamma_n/2},
\qquad
\gamma_n=\gamma_n^{\mathrm{rad}}+\gamma_n^{\mathrm{nr}},
\label{eq:lorentz}
\end{equation}
where $\chi(\omega)$ is the linear response (susceptibility) to a coherent
drive at frequency $\omega$, $A_n$ is the (generally complex) excitation
amplitude of the $n$-th mode, $\omega_n$ is its resonant frequency,
$\gamma_n=\gamma_n^{\mathrm{rad}}+\gamma_n^{\mathrm{nr}}$ its total linewidth
(the sum of radiative and non-radiative parts), and $Q_n=\omega_n/\gamma_n$
the corresponding quality factor \citep{breit1936capture,
fano1961interaction}. For each isolated mode there exists a
\emph{classical bound} $\Qcl$ on the quality factor: the value of $Q$
obtained when the mode is coupled to its environment through every
available channel without interference. This bound is the textbook
Lorentzian limit; it is what is being exceeded when a system becomes
super-resonant.

\subsection{Operational definition}\label{sec:def}

\begin{definition}[Super-resonance]
A resonant mode of an open physical system is said to be
\emph{super-resonant} when (i) its coupling to the radiation continuum is
generically large --- the mode would, in the absence of interference, decay
at the single-channel rate $\gammarad^{\mathrm{cl}}$ --- and (ii) destructive
interference among the partial amplitudes that compose the radiative width
drives that width below the single-channel value, with the ratio vanishing
in the ideal limit:
\begin{equation}
\boxed{\;\gammarad/\gammarad^{\mathrm{cl}}\to 0
\quad \Longrightarrow\quad
\Qsuper/\Qcl \;\xrightarrow{\;\text{ideal limit}\;}\;\infty.\;}
\label{eq:SRdef}
\end{equation}
Here $\Qsuper$ and $\Qcl$ denote the super-resonant and classical quality
factors of the mode and $\gammarad$, $\gammarad^{\mathrm{cl}}$ the
corresponding radiative linewidths. The derived FOMs --- the
inverse radiative bandwidth $\Delta\omega_{\mathrm{rad}}^{-1}$, the Purcell
factor $\Fpur$ at fixed mode volume $\Vmode$, the effective interaction
length $\Leff$, and the cooperativity $C$ --- diverge together with $Q$ in
the same limit.
\end{definition}

Equation~\eqref{eq:SRdef} is the central operational criterion of this
review. The FOMs it enumerates are not independent: the
Purcell factor, the cooperativity, the effective interaction length, and
the radiative bandwidth are all functions of $Q$ and $\Vmode$, related by
known identities that we shall invoke throughout (Section~\ref{sec:keyeqs}).
What unites them is that each one becomes \emph{arbitrarily large} when
the radiative width $\gammarad$ is driven to zero by the same coherent
mechanism. We deliberately exclude the mode volume itself from the list:
reducing $\Vmode$ is a geometric optimisation, not a radiative bound
exceeded by interference, and it does not diverge in the
$\gammarad\!\to\!0$ limit. Condition (i) is essential: it is what
distinguishes a super-resonance from an ordinary resonance that is narrow
merely because its bare coupling or its material loss is small
(Sec.~\ref{sec:taxonomy}). The constructive content --- the recipe for how
to reach $\Qsuper$ --- is the interference mechanism M1 of the next
subsection. The two relative families M2 and M3 surveyed thereafter do
\emph{not} satisfy this definition; they are included because the
literature labels them with the same prefix, and because the contrast with
them sharpens the definition itself.

It is worth flagging one assumption implicit in the bandwidth-related
figures of merit above. A higher quality factor is conventionally taken to
entail a narrower spectral bandwidth, the two being reciprocally tied
through the resonance linewidth \citep{fu2025subwavelength}. Tsakmakidis
\emph{et al.}\ \citep{tsakmakidis2017breaking} have argued theoretically,
however, that this reciprocal trade-off is strictly a property of
\emph{reciprocal} (Lorentz-symmetric) systems: in a non-reciprocal system
--- one whose response changes when source and detector are interchanged
--- the time--bandwidth product need not be bounded in the customary way,
and the more pronounced the asymmetry, the further the conventional limit
can in principle be exceeded, so that a high-$Q$ mode need not be
spectrally narrow. The claim has been contested: Mann, Sounas, and
Al\`u \citep{mann2019nonreciprocal} argued that a linear, time-invariant
cavity obeys the conventional time--bandwidth bound whether or not it is
reciprocal, locating any genuine violation in time-variance rather than
non-reciprocity as such. Whether a super-resonant mode can be combined
with broken reciprocity (or temporal modulation) to decouple $\Qsuper$
from the radiative bandwidth
$\Delta\omega_{\mathrm{rad}}$ is, to our knowledge, an open question.

\subsection{Mechanism M1 --- destructive interference of radiation channels}\label{sec:M1}

\begin{figure}[t]
\centering
\includegraphics[width=0.52\linewidth]{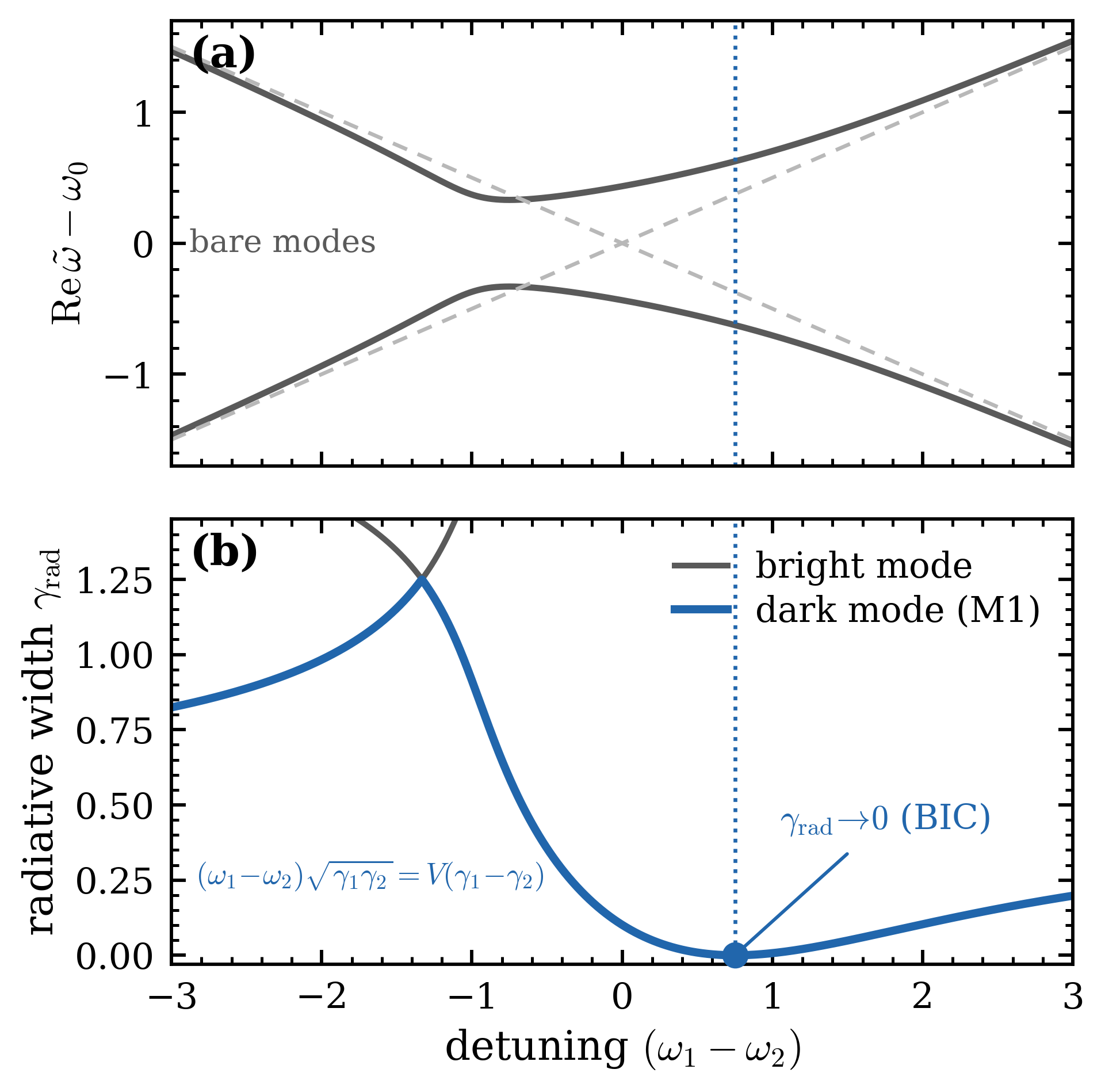}
\caption{Friedrich--Wintgen interference (mechanism M1), computed from
Eq.~\eqref{eq:fw_imag} for $\gamma_1=1$, $\gamma_2=0.25$, $V=0.5$
(dimensionless units). (a)~Real parts of the two hybrid eigenfrequencies
against the bare-mode detuning $\omega_1-\omega_2$: an avoided crossing
(bare modes dashed). (b)~The corresponding radiative widths
$\gammarad=-2\,\im\widetilde\omega$. At the Friedrich--Wintgen condition
$(\omega_1-\omega_2)\sqrt{\gamma_1\gamma_2}=V(\gamma_1-\gamma_2)$
(dotted vertical line) the width of the dark branch vanishes exactly ---
a bound state in the continuum realising the operational definition
\eqref{eq:SRdef} --- although both bare modes remain strongly coupled to
the continuum.}
\label{fig:fw}
\end{figure}

If the radiative loss rate $\gammarad$ is itself a sum of partial
amplitudes that can interfere coherently,
\(
\gammarad = |\sum_\alpha c_\alpha|^2,
\)
then a judicious choice of the coupling coefficients $c_\alpha$ can make
$\gammarad \to 0$ even while $\omega_n$ remains finite. This is the
mechanism behind bound states in the continuum
\citep{friedrich1985interfering,hsu2016bound,marinica2008bound}, supercavity
modes in subwavelength dielectric resonators \citep{rybin2017highq,
koshelev2018asymmetric,rybin2024metaphotonics}, anapole states
\citep{miroshnichenko2015nonradiating}, intracavity coherent
perfect absorption \citep{chong2010coherent,wan2011timereversed,
baranov2017coherent,malara2016super},
parity--time-symmetric and exceptional-point devices
\citep{bender1998real,chen2017exceptional,hodaei2017enhanced,
elganainy2018non,ozdemir2019paritytime,miri2019exceptional,yuan2021super},
super-resonant states of leaky surface magnon polarons
\citep{sukhorukova2022superresonant}, dark magnon modes
\citep{zhang2015magnon}, and topological photonic cavities
\citep{ozawa2019topological,lu2014topological,harari2018topological},
in which the symmetry-allowed set of radiation channels is emptied by a
bulk invariant rather than by parametric fine-tuning. (Black-hole
super-radiance, often grouped with these phenomena, is \emph{not} an
instance of M1: there the reflection coefficient exceeds unity and the
mode is amplified rather than darkened; see
Secs.~\ref{sec:taxonomy} and~\ref{sec:celestial}. Likewise, the
gain-compensated magnetic-dipole resonance of coated nanoparticles
\citep{liberal2014magnetic,gordon2007coated} achieves its narrow line by
\emph{active loss compensation}, which is a form of amplification, not of
passive channel interference.) Formally, the
mechanism corresponds to an avoided crossing in parameter space at which
$\im \widetilde\omega_n=0$: generically the vanishing of one complex
coupling amplitude, a codimension-two condition, which reciprocity (real
couplings) reduces to codimension one --- a single real condition, as in
the Friedrich--Wintgen model below. The canonical
mathematical realisation is the Friedrich--Wintgen scenario
\citep{friedrich1985interfering}, in which two interfering resonances
coupled to a single continuum produce the complex eigenvalues
\begin{equation}
\widetilde\omega_\pm =
\frac{\omega_1+\omega_2}{2}-\mathrm{i}\,\frac{\gamma_1+\gamma_2}{2}
\pm\frac{1}{2}\sqrt{\bigl[(\omega_1-\omega_2)-\mathrm{i}(\gamma_1-\gamma_2)\bigr]^2
+4\bigl(V-\mathrm{i}\sqrt{\gamma_1\gamma_2}\bigr)^{2}},
\label{eq:fw_imag}
\end{equation}
where $\gamma_1,\gamma_2$ are the radiative decay rates (half-widths;
the bare complex eigenvalues are $\omega_j-\mathrm{i}\gamma_j$, so the
full linewidth of bare mode $j$ is $2\gamma_j$) and
$\omega_1,\omega_2$ the bare resonant frequencies of the two interfering
modes, $V$ is their direct (near-field) inter-mode coupling, and
$\sqrt{\gamma_1\gamma_2}$ is the coupling mediated by the shared radiation
continuum. At the Friedrich--Wintgen condition
\(
(\omega_1-\omega_2)\sqrt{\gamma_1\gamma_2} = V(\gamma_1-\gamma_2),
\)
the imaginary part of one eigenvalue vanishes exactly: the corresponding
eigenmode is a genuine bound state in the continuum, and its $Q$ diverges. The
robustness of this construction --- it requires only \emph{two} coupled
modes and a single shared continuum --- is what makes M1 the most widely
implemented super-resonance mechanism in modern photonics.
Figure~\ref{fig:fw} traces \eqref{eq:fw_imag} across the avoided
crossing: exactly at the Friedrich--Wintgen condition the radiative
width of the dark branch touches zero, while the bright branch absorbs
the entire coupling of the pair to the continuum.

\subsection{Relative family M2 --- coherent enhancement and spectral-weight
concentration (not super-resonance in the strict sense)}\label{sec:M2}

The first relative family --- which we label M2 for continuity with the
survey of Section~\ref{sec:manifestations}, but which does \emph{not}
satisfy the operational definition \eqref{eq:SRdef} --- collects phenomena
in which coherence \emph{increases} the coupling of one collective mode to
the continuum, or concentrates the spectral weight of a multimode expansion
into a single dominant term. Where M1 suppresses a radiative rate, M2
enhances one; the two have opposite signs of the relevant rate change, and
conflating them under a single criterion would render the concept
meaningless. The total response of a multimode resonator can be
reorganised so that the
amplitude of a single high-multipole mode comes to dominate the entire
spectral weight. In a Mie-resonant dielectric microsphere this happens
at discrete values of the size parameter $q=2\pi a/\lambda$ and of the
refractive-index contrast $m=n_{\mathrm{sphere}}/n_{\mathrm{medium}}$: the
internal scattering coefficients $c_n$ and $d_n$ exhibit sharp peaks in a
single order $n^\ast$, whose partial amplitude stands orders of magnitude
above every other
\citep{wan2019highorder,minin2023cenosphere,minin2024discovery,
yue2019poynting,minin2024freezing}. The internal-field intensities then
follow
\begin{equation}
\frac{|\mathbf{E}|^2_{\max}}{|\mathbf{E}_0|^2}
\;\propto\;\bigl|c_{n^\ast}\bigr|^2 \cdot Q_{n^\ast},
\qquad
\frac{|\mathbf{H}|^2_{\max}}{|\mathbf{H}_0|^2}
\;\propto\;\bigl|d_{n^\ast}\bigr|^2 \cdot Q_{n^\ast},
\label{eq:mie_intensity}
\end{equation}
and reach $10^6$--$10^7$ for visible-frequency water droplets and low-absorbing
borosilicate spheres of size $a\!\approx\!5$--$10\,\mu$m
\citep{wan2019highorder,minin2022borosilicate,minin2024discovery,
minin2023cenosphere}. Whispering-gallery modes of toroidal, microdisk,
and silica microsphere resonators \citep{vahala2003optical,
armani2003ultrahigh,spillane2002ultralow,matsko2006optical} are themselves
M2 modes operating at very high azimuthal order; Dicke super-radiance
\citep{dicke1954coherence,gross1982superradiance,scheibner2007superradiance,
angerer2018superradiant,cong2016dicke,jordao2025single}, in which $N$
in-phase emitters radiate at peak rate $\propto N^2$ rather than $N$,
belongs to the same class --- the in-phase combination of single-emitter
amplitudes maximally couples to the collective mode while the orthogonal
combinations decouple from the radiation continuum. The nuclear giant
dipole resonance \citep{harakeh2001giant,bortignon1998giant,
ishkhanov2000giant,savran2013experimental}, in which the coherent
superposition of single-particle dipole transitions exhausts the
energy-weighted sum rule, is the same phenomenon at the femtometre scale.

\begin{figure}[t]
\centering
\includegraphics[width=0.58\linewidth]{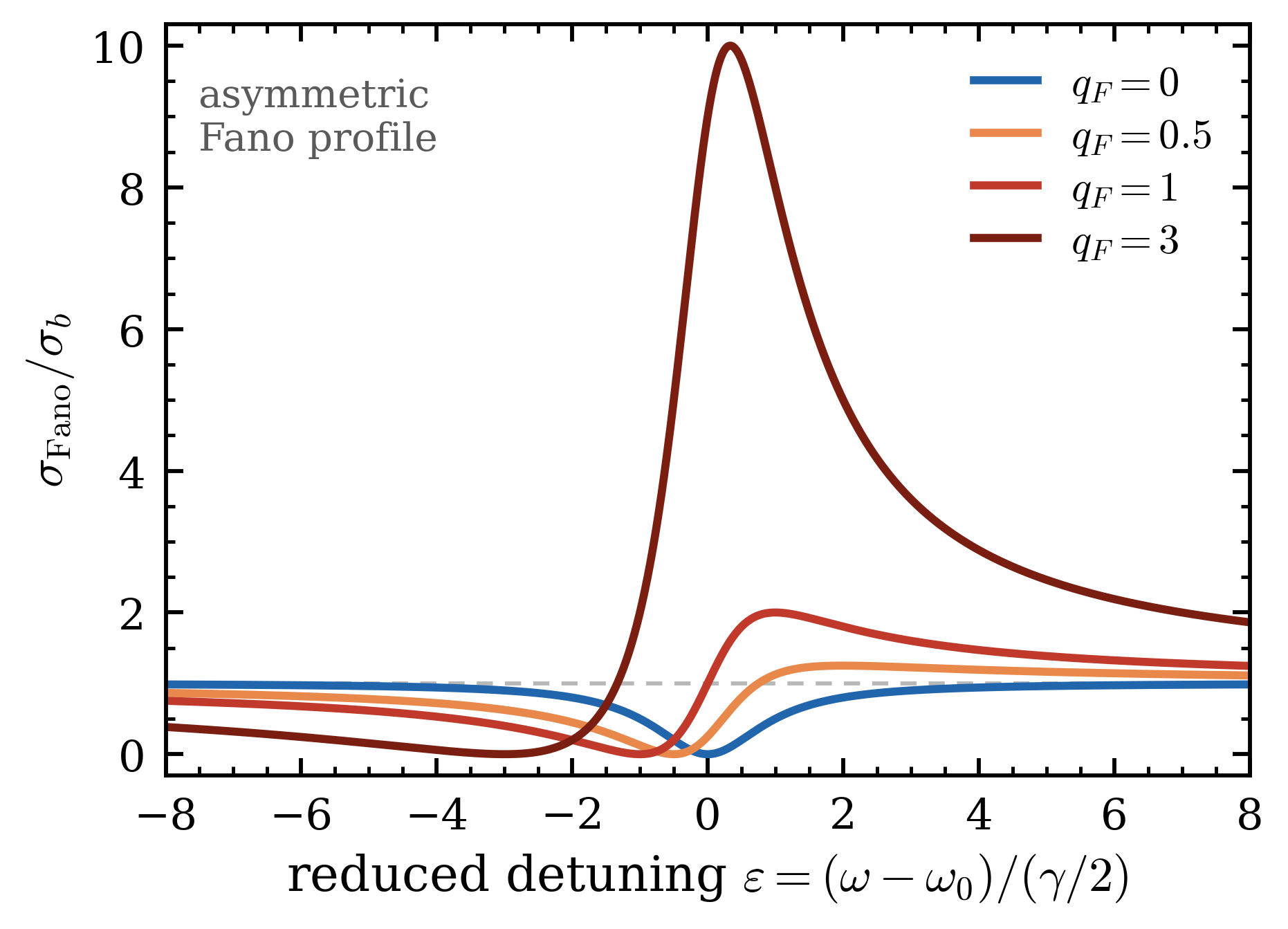}
\caption{The Fano line-shape family of Eq.~\eqref{eq:fano} against the
reduced detuning $\varepsilon$. At $q_F=0$ the interference of the
narrow mode with the background carves a symmetric dip (transparency
window); $q_F\simeq 1$ yields the maximally asymmetric profile whose
steep spectral slope underlies Fano-resonant refractometry
(Sec.~\ref{sec:sensing}); for $q_F\gg1$ the resonant channel dominates
and the profile approaches a Lorentzian of peak height $1+q_F^2$. The
dashed line marks the bare background level.}
\label{fig:fano}
\end{figure}

Mathematically, M2 is the dual of M1: rather than driving
one diagonal element of $\bm{\Gamma}$ to zero, it concentrates the
\emph{coupling} so that a single mode exhausts the spectral weight while
the other modes remain spectators. The duality is instructive but should
not be mistaken for identity. In Dicke super-radiance (see Sec.~\ref{sec:dicke} below) the collective decay
rate is \emph{enhanced} to $\propto N^2$ --- a fast, broad burst, the
antithesis of a vanishing radiative width --- even though the $N-1$
orthogonal (subradiant) combinations do satisfy the M1 criterion and are,
in our language, genuine super-resonances. Superscattering and
super-directivity likewise exceed single-channel bounds by \emph{opening
and phasing additional} radiation channels, not by closing one. The
high-$Q$ character of whispering-gallery and high-order Mie modes, for its
part, originates in the exponentially weak radiative coupling of a
high-order multipole --- a small bare coupling, which the exclusion
criteria of Sec.~\ref{sec:taxonomy} place outside the strict definition.
What makes the M2 family scientifically close to M1 is the shared
multipole-expansion machinery and the frequent coexistence of bright
(enhanced) and dark (suppressed) collective modes in one and the same
structure. The canonical line shape, when a narrow mode interferes with a
smooth background continuum, is the Fano line
shape \citep{fano1961interaction,limonov2017fano,miroshnichenko2010fano},
\begin{equation}
\sigma_{\mathrm{Fano}}(\omega)=\sigma_b
\,\frac{(q_F+\varepsilon)^2}{1+\varepsilon^2},
\qquad
\varepsilon=\frac{\omega-\omega_0}{\gamma/2},
\label{eq:fano}
\end{equation}
where $q_F$ is the Fano asymmetry parameter. Figure~\ref{fig:fano}
shows the family \eqref{eq:fano}: the profile interpolates from a
symmetric transparency dip at $q_F=0$ (the EIT-like window of
Sec.~\ref{sec:plasmonics}) through the maximally asymmetric shape near
$q_F\simeq 1$, whose steep slope refractometric sensors exploit
(Sec.~\ref{sec:sensing}), toward an increasingly Lorentzian peak as
$q_F$ grows. The high-order Fano
resonance of mesoscale dielectric spheres
\citep{wan2019highorder,minin2024discovery,minin2022fano,limonov2017fano}
is the analytic signature of M2 in the optical regime.

\subsection{Relative family M3 --- integer-commensurability locking
(not super-resonance in the strict sense)}\label{sec:M3}

\begin{figure}[t]
\centering
\includegraphics[width=0.52\linewidth]{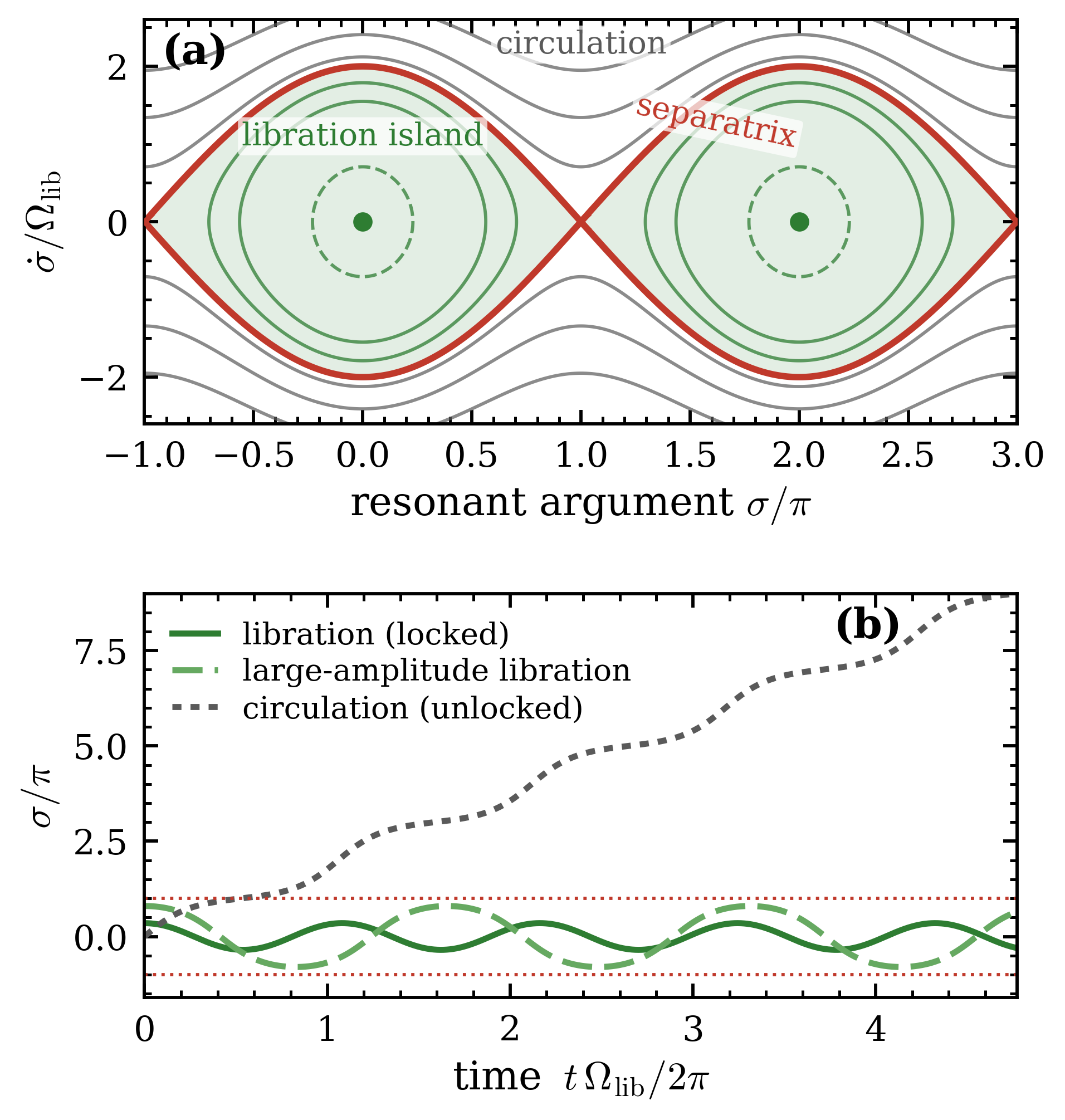}
\caption{Integer-commensurability locking (family M3) in the universal
one-resonance (pendulum) approximation, in which the resonant argument
$\sigma$ of Eq.~\eqref{eq:mmr} obeys
$\ddot\sigma=-\Omega_{\mathrm{lib}}^2\sin\sigma$. (a)~Phase portrait:
libration islands (shaded) are bounded by the separatrix (thick line)
that passes through the hyperbolic fixed points (crosses); outside the
separatrix the argument circulates. (b)~Numerically integrated
$\sigma(t)$: small- and large-amplitude libration remains bounded
(locked), whereas an initial condition just outside the separatrix
($\dot\sigma_0=2.02\,\Omega_{\mathrm{lib}}$) circulates without bound.
No radiative width enters anywhere: the longevity of the locked state is
a property of conservative phase-space topology, which is why M3 falls
outside the strict criterion \eqref{eq:SRdef}.}
\label{fig:locking}
\end{figure}

The second relative family --- labelled M3, again for continuity, and
again falling outside the operational definition \eqref{eq:SRdef} ---
collects phenomena of \emph{phase locking}. Here there is typically no
radiation continuum at all, and the longevity of the locked state derives
from the topology of a resonant island in classical phase space, not from
interference among decay amplitudes.
When two or more characteristic frequencies of a non-linearly coupled
system satisfy an exact integer relation $n\omega_1 = m\omega_2$, the
resulting phase locking traps the system into a long-lived collective
state whose effective lifetime greatly exceeds that of either uncoupled
subsystem. This is the origin of the Laplace resonance in the Galilean
satellites \citep{barnes2011laplace,paita2018element,goldreich1965explanation,
murray1999solar}, the 1:1 super-resonance of Pluto's argument of perihelion
\citep{wan2001pluto}, sub- and super-harmonic resonances of cracked beams
\citep{bovsunovskii2006damping}, multi-photon resonances in driven
quantum systems \citep{gat2013resonance}, magnetic resonance phenomena
\citep{rabi1938new,purcell1946resonance,kubo1969generalized}, lower-hybrid
plasma current drive \citep{stix1992waves,fisch1987theory,karney1979stochastic},
and the Floquet replicas that appear when an optical medium is modulated
periodically in time \citep{galiffi2022photonics,lyubarov2022amplified,
rudner2020band}. The canonical mathematical form is the slow-time
adiabatic invariant: for two coupled oscillators of natural frequencies
$\omega_1, \omega_2$ the mean-motion super-resonance condition reads
\begin{equation}
n\,\omega_1 - m\,\omega_2 \approx 0,
\qquad
|n\omega_1 - m\omega_2|\;\ll\;\Omega_{\mathrm{lib}},
\label{eq:mmr}
\end{equation}
where $\Omega_{\mathrm{lib}}$ is the libration frequency around the
resonant island in phase space. When satisfied, the libration amplitude
is bounded for cosmologically long times --- loosely, the
celestial-mechanics analogue of an infinite-$Q$ resonance. The
universal phase-space geometry behind \eqref{eq:mmr} is drawn in
Fig.~\ref{fig:locking}: trajectories inside the separatrix librate about
the exact commensurability and stay locked, while those outside
circulate freely.

 The analogy
with M1 is real but limited: in both cases the system finds a long-lived
state that exists \emph{only as a consequence of the coupling}, but with
M3 the state is a phase-locked island of conservative dynamics rather than
a destructively interfering pair of decay amplitudes, the relevant
``lifetime'' is set by the slow chaotic diffusion out of the island (or by
weak dissipation), and no radiative bound of the form
\eqref{eq:SRdef} is being exceeded. A wave--particle resonance in a
plasma, for example, is sharp because the collisional decorrelation rate
is small --- precisely the kind of small-bare-loss sharpness that the
exclusion criteria of Sec.~\ref{sec:taxonomy} rule out. We survey M3
phenomena below both because the term \emph{super-resonance} was
independently coined for one of them \citep{wan2001pluto} and because the
phase-locking mathematics transfers usefully to Floquet-engineered systems
(Sec.~\ref{sec:topological}); but we do not claim they obey the
definition.

\subsection{What super-resonance is \emph{not}}\label{sec:taxonomy}

To prevent the concept from becoming all-encompassing and therefore
meaningless, we adopt the following exclusion criteria.

\textit{First}, an ordinary high-$Q$ resonance --- a mechanical pendulum
in vacuum, a typical optical microcavity --- is \emph{not} super-resonant,
because the single-channel coupling itself is small. Super-resonance
requires that the channel coupling is generically large but is brought to
zero by interference. This exclusion also covers whispering-gallery and
high-order Mie modes considered in isolation: their high $Q$ derives from
the exponentially weak radiative coupling of a high-order multipole, not
from interference among large partial amplitudes.

\textit{Second}, a resonance whose $Q$ has been raised solely by
reducing material losses is \emph{not} super-resonant; lowering
$\gammanr$ does not modify $\gammarad/\gammarad^{\mathrm{cl}}$. The
operational criterion \eqref{eq:SRdef} is explicit about this: only the
\emph{radiative} bound is being exceeded. The same applies to a
wave--particle or orbital resonance whose sharpness reflects a small
decorrelation or diffusion rate.

\textit{Third}, a system that exceeds a classical bound on a
FOM by \emph{adding and phasing} radiation channels --- a superscattering
nanoparticle that exceeds the single-channel scattering cross-section
limit \citep{ruan2010superscattering,ruan2011design,qian2019experimental},
or a super-directional antenna that approaches or exceeds practical
Chu--Harrington-type limits \citep{krasnok2015superdirective} --- is
\emph{not} super-resonant under the present definition, however
remarkable: the radiative coupling is being coherently \emph{increased},
which is the M2 enhancement family. The colloquial association of the
prefix ``super-'' with any bound-beating resonance is exactly the
ambiguity this review aims to remove.

\begin{figure}[t]
\centering
\includegraphics[width=0.78\linewidth]{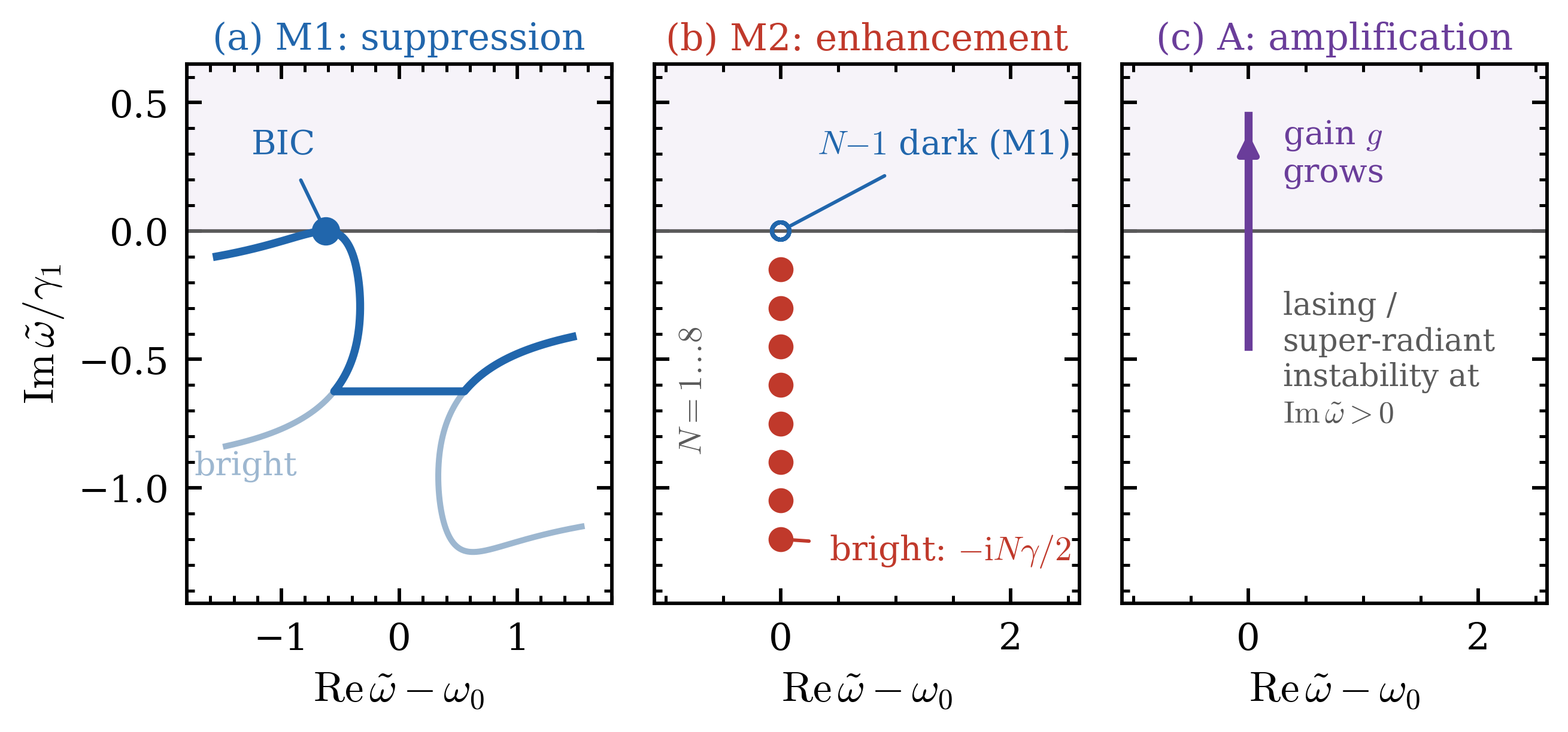}
\caption{Three motions of the complex eigenvalue
$\widetilde\omega$, each computed from the minimal model of its class;
the shaded strip is the gain half-plane $\im\widetilde\omega>0$.
(a)~Suppression (M1): the locus traced by the two hybrid eigenvalues of
the Friedrich--Wintgen Hamiltonian of Fig.~\ref{fig:fw} as the detuning
is swept (thick curve: the darker of the two eigenvalues at each
detuning). The dark branch is driven onto the real axis --- the BIC ---
while both bare modes stay strongly coupled to the continuum.
(b)~Enhancement (M2): eigenvalues of the rank-one Dicke radiative
matrix $\Gamma_{jk}=\gamma$ for $N=1\ldots8$ point emitters. The bright
eigenvalue marches \emph{deeper}, $-\mathrm{i}N\gamma/2$, while the
$N-1$ dark combinations sit on the real axis --- the M1 flip side of
the same algebra (Sec.~\ref{sec:quantum}). (c)~Amplification (A): a
single mode with gain, $\widetilde\omega=\omega_0-\mathrm{i}(\gamma-g)/2$,
crosses into the gain half-plane as $g$ grows --- the black-hole-bomb
and laser instability scenario of Sec.~\ref{sec:celestial}.}
\label{fig:taxonomy}
\end{figure}

\textit{Fourth}, a mode whose amplitude grows because the system supplies
energy --- black-hole super-radiance and its analogue-gravity
realisations \citep{press1972rotating,bekenstein1973black,
brito2020superradiance,torres2017rotational}, gain-compensated
resonators \citep{liberal2014magnetic}, and parametric amplification in
time-modulated media \citep{galiffi2022photonics,lyubarov2022amplified}
--- is an \emph{amplification} phenomenon: the relevant eigenvalue moves
into the gain half-plane, $\im\widetilde\omega > 0$, rather than onto the
real axis. Amplification, enhancement (M2), and suppression (M1) are
three different signs of the same imaginary part, and only the third is
super-resonance in the sense of \eqref{eq:SRdef}. Figure~\ref{fig:taxonomy}
draws the three motions side by side, each computed from the minimal
model of its class.

\paragraph{Superoscillation versus super-resonance.}
Superoscillation is a distinct but related wave phenomenon, in which a
globally band-limited signal oscillates locally much faster than its
highest constituent frequency. When combined with high-$Q$ resonances in
photonics, superoscillatory excitation enables deeply sub-wavelength field
localisation, but it is fundamentally limited by how strongly the
resonator absorbs these rapid local oscillations. In the mesotronics
context the connection between superoscillation and the high-order Fano
resonance has been made explicit \citep{wan2019highorder,minin2022fano}:
the high-order Fano resonance can be read as the resonant counterpart of a
superoscillatory field confined within a mesoscale particle.

Table~\ref{tab:taxonomy} summarises the mechanism-by-field
mappings that will guide the survey of Section~\ref{sec:manifestations}.

\begin{table}[ht!]
\centering
\caption{Classification of resonant phenomena often grouped under the
``super-'' prefix, by field. M1: radiation-channel interference ---
super-resonance in the strict sense of Eq.~\eqref{eq:SRdef}. M2: coherent
enhancement / spectral-weight concentration (relative family, not strict
super-resonance). M3: integer-commensurability locking (relative family,
not strict super-resonance). A: amplification, $\im\widetilde\omega>0$
(relative family, not strict super-resonance). casc.: multiplicative
coupled-cavity cascade (an engineering composition,
Sec.~\ref{sec:design}). SR: super-resonance; PhC: photonic
crystal.}\label{tab:taxonomy}
\small
\begin{tabular}{@{}p{2.8cm}p{1.0cm}p{3.0cm}p{6.5cm}@{}}
\toprule
\textbf{Field} & \textbf{Mech.} & \textbf{Phenomenon} & \textbf{Representative refs.} \\
\midrule
Acoustics      & M1/M2 & Tolstoy super-resonance     & \cite{tolstoy1986superresonant,tolstoy1987properties,minin2019extreme,michavila2019super,norris1986scattering} \\
Elastodynamics & M1/M3 & elastic flow-control SR     & \cite{harris2026super,bovsunovskii2006damping,lee2025surface,hussein2014dynamics} \\
Celestial      & M3 & Laplace / Pluto / Neptune   & \cite{williams1971neptune,wan2001pluto,paita2018element,barnes2011laplace,goldreich1965explanation,laskar1989numerical,smirnov2020double} \\
Gravitational  & A & black-hole super-radiance   & \cite{press1972rotating,bekenstein1973black,brito2020superradiance,kobes2006superresonance,basak2003superresonance,torres2017rotational,mcdonald2024resonant,thorne1994blackholes,anacleto2013noncommutative} \\
Photonics (Mie)& M2 & high-order Fano in spheres  & \cite{wan2019highorder,minin2024discovery,minin2023cenosphere,minin2022borosilicate,minin2024freezing,yue2020teflon,yue2019poynting,minin2022fano,minin2022extreme} \\
Photonics (BIC)& M1 & supercavity / quasi-BIC     & \cite{rybin2017highq,koshelev2018asymmetric,hsu2013observation,hsu2016bound,bogdanov2019bound,marinica2008bound,koshelev2020subwavelength,azzam2021photonic,rybin2024metaphotonics,sadrieva2017transition} \\
Anapole / superscatt.& M1/M2 & nonradiating sources & \cite{miroshnichenko2015nonradiating,baryshnikova2019optical,savinov2018optical,ruan2010superscattering,ruan2011design,qian2019experimental,lukosiunas2023extremely} \\
Plasmonics     & M2 & plasmonic Fano; composite THz res. & \cite{lukyanchuk2010fano,lassiter2010fano,miroshnichenko2010fano,limonov2017fano,stockman2011nanoplasmonics,khurgin2015deep,babushkin2023metallic} \\
WGM / PhC cavity & M2/M1 & ultra-high-$Q$ WGM; nanocavity & \cite{vahala2003optical,armani2003ultrahigh,matsko2006optical,akahane2003ultrahigh,sekoguchi2014photonic,geints2024magnetic,minin2022janus,spillane2002ultralow} \\
Coherent abs.  & M1 & CPA / super-resonant CPA    & \cite{chong2010coherent,wan2011timereversed,baranov2017coherent,malara2016super} \\
PT / EP        & M1/A & PT super-resonance / EP     & \cite{yuan2021super,elganainy2018non,miri2019exceptional,ozdemir2019paritytime,chen2017exceptional,hodaei2017enhanced,bender1998real} \\
Cavity QED     & M2 & ultrastrong coupling      & \cite{frisk2019ultrastrong,forn2019ultrastrong,thompson1992observation,reithmaier2004strong,walther2006cavity,englund2007controlling,jaynes1963comparison,larson2024jaynes,lecamp2007very} \\
Atomic / mol.  & M2 & Dicke super-radiance        & \cite{dicke1954coherence,gross1982superradiance,scheibner2007superradiance,angerer2018superradiant,jordao2025single,cong2016dicke,meiser2009prospects} \\
Nuclear        & M2 & giant dipole / pygmy        & \cite{harakeh2001giant,bortignon1998giant,ishkhanov2000giant,savran2013experimental,breit1936capture} \\
Magnonics      & M1/M2 & magnon polaron / dark mode / polariton & \cite{sukhorukova2022superresonant,huebl2013high,lachancequirion2019hybrid,tabuchi2014hybridizing,zhang2014strongly,zhang2015magnon,soykal2010strong,lachance2017resolving,lachance2020entanglement,rameshti2022cavity,yuan2022quantum,chumak2015magnon,kittel1948theory} \\
Plasma         & M3 & Landau / lower-hybrid       & \cite{stix1992waves,fisch1987theory,karney1979stochastic,landau1946vibrations,oneil1965collisionless,bonoli2014review,prater2004heating,brambilla1998kinetic,fisch1992interaction,wesson2011tokamaks} \\
Topological    & M1 & topological cavity / laser  & \cite{ozawa2019topological,lu2014topological,harari2018topological} \\
Time-varying   & M3/A & photonic time crystals   & \cite{galiffi2022photonics,lyubarov2022amplified,rudner2020band} \\
Gravity-wave   & casc. & coupled-cavity readout            & \cite{aasi2015advanced,abbott2016observation,drever1983laser} \\
\bottomrule
\end{tabular}
\end{table}

\subsection{Key quantitative relations}\label{sec:keyeqs}

For the rest of the review the following identities will be invoked
repeatedly. None of them is new; they are collected here because each one
converts the suppressed radiative width of an M1 super-resonance into a
measurable figure of merit, and because the same identities, applied
loosely, are often the source of the M1/M2/M3 conflation this review aims
to undo.

\paragraph{Purcell enhancement.}
The Purcell factor for spontaneous emission into a single mode of an
optical cavity is
\begin{equation}
\Fpur
=\frac{3}{4\pi^2}\left(\frac{\lambda}{n}\right)^{\!3}\frac{Q}{\Vmode},
\label{eq:purcell}
\end{equation}
\citep{purcell1946spontaneous,vahala2003optical}. An M1 super-resonance
increases $Q$ at essentially fixed $\Vmode$, so $\Fpur$ grows in direct
proportion to the radiative suppression; subwavelength dielectric
supercavities are attractive precisely because the quasi-BIC $Q$
enhancement arrives in an already small mode volume
\citep{rybin2017highq,koshelev2018asymmetric}. We emphasise that the $Q$
and $\Vmode$ factors in \eqref{eq:purcell} have different physical
origins --- interference versus geometry --- and only the former is a
super-resonance effect. A concrete measure of what the interference
buys, and of what it costs, is provided by the dielectric nanocavities
of Hoang, Chu, Garc\'{\i}a-Vidal, and Png \citep{hoang2022nanocavity}.
Designed from exact multipole solutions for collective Mie resonances of
``photonic molecules'', their cavity supports a Feshbach-type BIC mode
--- the construction descends from Feshbach's unified theory of nuclear
reactions \citep{feshbach1958unified}, of which the Friedrich--Wintgen
scenario \eqref{eq:fw_imag} is the two-level distillation --- in which
the overlap of two resonances suppresses the radiative loss channels by
destructive interference. The design is predicted to enhance the
emission rate of an embedded dipolar emitter by orders of magnitude
while channelling the emission into an in-plane bidirectional beam of
$\sim\!10^\circ$ divergence. The cost appears in the tolerance budget:
at near-infrared frequencies the geometry must be fabricated to
$\pm5$\,nm, a demanding target for mass production, and only the
mid-infrared version of the design (an L3 photonic-crystal defect
cavity re-optimised in the Feshbach framework) relaxes the requirement
to $\pm50$\,nm. We return to this trade-off as Rule~5 of
Sec.~\ref{sec:design}.

\paragraph{Strong coupling and cooperativity.}
For an emitter of decay rate $\gamma_e$ coupled to a cavity of linewidth
$\kappa$ with coupling $g$, the cooperativity is $C=4g^2/(\kappa\gamma_e)$
and strong coupling requires $g\gg\{\kappa,\gamma_e\}$
\citep{thompson1992observation,reithmaier2004strong,walther2006cavity,
jaynes1963comparison,larson2024jaynes}. The cooperativity is the figure
of merit through which an M1 super-resonance benefits cavity QED: driving
$\kappa_{\mathrm{rad}}$ down by interference raises $C$ in direct
proportion. Ultrastrong coupling
($g/\omega_0\gtrsim 0.1$), by contrast, is a coupling-\emph{enhancement}
regime --- the rotating-wave approximation breaks down and the
Jaynes--Cummings ladder is replaced by a richer dressed-state spectrum
\citep{frisk2019ultrastrong,forn2019ultrastrong} --- and belongs to the
M2 family, not to super-resonance proper. The same identities apply
unchanged in solid-state platforms: an InAs quantum dot tuned to
resonance with a GaAs micropillar cavity mode \citep{munsch2009continuous}
exhibits the same Purcell-enhanced emission as an atomic emitter.

\paragraph{Effective interaction length.}
For hybrid cavities such as a Fabry--P\'erot resonator nested inside a
ring resonator, the effective interaction length is the \emph{product}
(not the sum) of the two finesses,
\begin{equation}
\Leff \simeq \mathcal{F}_{\mathrm{ring}}\,\mathcal{F}_{\mathrm{FP}}\,\ell,
\label{eq:Leff}
\end{equation}
where $\mathcal{F}_{\mathrm{ring}}$ and $\mathcal{F}_{\mathrm{FP}}$ are the
finesses of the ring and Fabry--P\'erot resonators, respectively, and
$\ell$ is the physical (single-pass) cavity length
\citep{malara2016super}; the same multiplicative geometry has recently
been demonstrated on a silicon-photonics platform by Selim and Anwar
\citep{selim2023enhanced} using nested ring resonators, where the
effective length and quality factor of the composite cavity exceed
those of either constituent. Two caveats keep this identity honest.
First, the multiplicative composition holds for specific nested-cavity
geometries operated at mutual resonance \citep{malara2016super,
selim2023enhanced}; it is a design outcome of coupled-cavity
engineering, not a universal law of cascaded resonances, and in general a
coupled-cavity system must be analysed through its full transfer
function. Second, the analogy frequently drawn with gravitational-wave
interferometers should be stated carefully: power- and signal-recycling
mirrors reshape the circulating power and the signal response of the
Michelson--Fabry--P\'erot system rather than multiplying arm finesse into
a single scalar \citep{aasi2015advanced,drever1983laser}, and we use the
language of coupled-cavity transfer functions, not of finesse products,
when we return to that application in Sec.~\ref{sec:gravitational}.

\paragraph{Sub-wavelength imaging.}
Resonant field localisation can encode sub-wavelength spatial information
in narrow, well-separated spectral or temporal features, transferring the
inverse problem from direct spatial resolution to spectrometric or
time-resolved detection. A graded-index
photonic-crystal lens perforated with an optimised single defect focuses
to a spot as small as $\sim\!\lambda/75$ \citep{li2016ultrasharp};
a resonant metalens based on a dense array of coupled subwavelength
resonators was shown, in a numerical feasibility study, to deliver
$\lambda/30$-scale subwavelength imaging from far-field time-reversed
signals at microwave frequencies \citep{li2014resonant}; and resonant
multiples of seismic waves enable far-field imaging of reflector
boundaries beyond the classical resolution limit \citep{guo2016seismic}.
We stress that no universal scaling of spatial resolution with modal $Q$
follows from these examples: each scheme rests on its own
system-specific inverse problem, and higher $Q$ buys finer spectral
discrimination --- not, in general, proportionally finer spatial
resolution. The role of M1 super-resonance in this context is to supply
the narrow, long-lived modes on which such encoding schemes depend
(Sec.~\ref{sec:imaging}).

\paragraph{Black-hole super-radiance (for contrast).}
For a bosonic field of frequency $\omega$ and azimuthal number $m$
incident on a rotating background with horizon angular velocity
$\Omega_H$, amplification (super-radiance) occurs whenever
\begin{equation}
0 < \omega < m\,\Omega_H,
\label{eq:superradianceBH}
\end{equation}
\citep{press1972rotating,bekenstein1973black,brito2020superradiance,
basak2003superresonance,torres2017rotational,thorne1994blackholes}.
We record this condition here precisely because it is \emph{not} an
instance of \eqref{eq:SRdef}: the reflection coefficient exceeds unity
and the relevant eigenvalue acquires a positive imaginary part when the
mode is confined --- by a mirror or by a massive-field potential ---
yielding the exponentially growing ``black-hole bomb'' instability. It
is the canonical member of the amplification family of
Sec.~\ref{sec:taxonomy}.

\paragraph{Three canonical forms.}
Each of the manifestations to be surveyed in
Section~\ref{sec:manifestations} is an instance of one of the following
three canonical forms --- of which only the first is super-resonance in
the strict sense of \eqref{eq:SRdef}.

\textit{Canonical form I (M1, radiation suppression --- super-resonance
proper).}
Two coupled resonances coupled to a single continuum give
\(\im\widetilde\omega_- \!\to\!0\) on the Friedrich--Wintgen manifold
\eqref{eq:fw_imag}. The corresponding super-resonant mode has
$\gammarad / \gammarad^{\mathrm{cl}}\!\to\!0$.

\textit{Canonical form II (M2, modal concentration --- enhancement
family).}
A high-order partial amplitude $|c_{n^\ast}|$ peaks sharply within a
multipole expansion, exhausting essentially all of the spectral weight
\eqref{eq:mie_intensity}, or a collective in-phase combination of $N$
emitters couples to the continuum at an enhanced rate. The coupling is
increased, not suppressed.

\textit{Canonical form III (M3, commensurability --- locking family).}
An integer relation $n\omega_1=m\omega_2$ traps the system into a
phase-locked state \eqref{eq:mmr}, with libration frequency
$\Omega_{\mathrm{lib}}$ --- the frequency of small oscillations of the
resonant argument about the stable fixed point --- parametrising a slow
drift around the resonant island. No radiative bound is involved.

The three forms share a common non-Hermitian language --- the eigenvalue
problem \eqref{eq:Heff} --- and can be evaluated by the same numerical
machinery; but they describe different motions of the complex eigenvalue,
and we will keep the distinction explicit throughout. Only form~I
realises the operational definition; forms~II and~III are surveyed for
contrast and for the (real, but limited) transfer of mathematical
technique between them.

\section{Manifestations Across the Spectrum of Physics}\label{sec:manifestations}

The operational definition of Section~\ref{sec:definition} is now exercised
across the fields in which the term super-resonance --- or one of its
``super-'' relatives --- has appeared. The
arrangement is broadly chronological, but within each subsection we
identify which class (M1 super-resonance proper, M2 enhancement, M3
locking, or amplification) the phenomenon instantiates, and we flag
explicitly the cases in which a field's customary usage of the prefix
differs from the strict definition. Although the depth of
exposition necessarily varies --- some fields, such as Mie-resonant
photonics and BICs, currently set the experimental pace, while others,
such as celestial mechanics, have already passed through their golden age
and serve here primarily as historical anchors or as contrasting
relatives --- the reader will notice how regularly the M1 interference
equations recur.

\subsection{Acoustics and elastodynamics}\label{sec:acoustics}

\subsubsection{Tolstoy's super-resonant arrays}

\begin{figure}[t]
\centering
\includegraphics[width=0.52\linewidth]{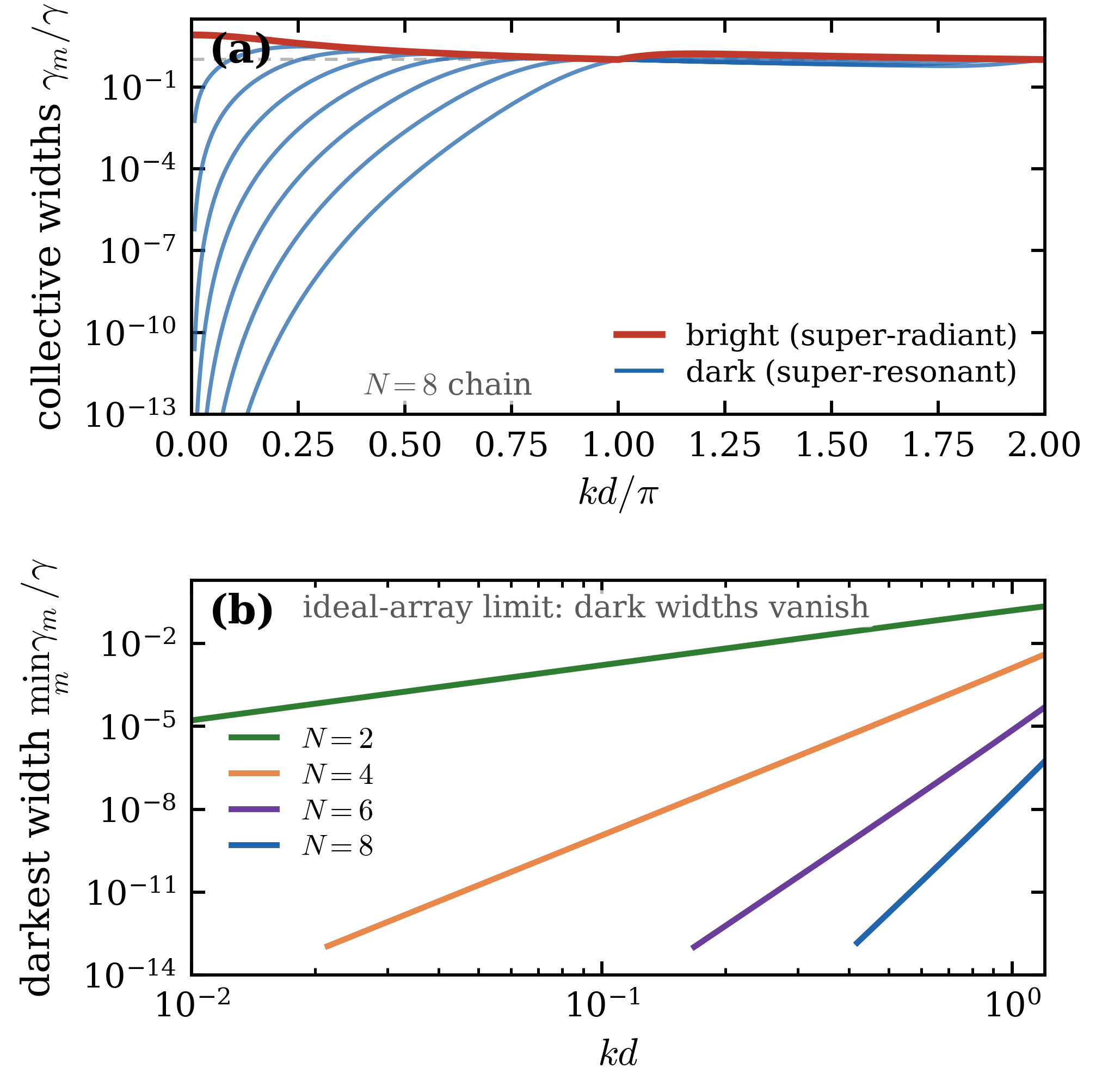}
\caption{Collective radiative widths of $N$ identical monopole
resonators sharing the three-dimensional scalar radiation continuum ---
the minimal model of a Tolstoy super-resonant array
\citep{tolstoy1986superresonant,tolstoy1987properties}. The widths are
the eigenvalues of the radiative coupling matrix
$\Gamma_{jk}=\gamma\,j_0(k r_{jk})$ for a linear chain of pitch $d$.
(a)~All widths versus $kd$ for $N=8$: in the sub-wavelength limit one
bright mode carries the full coupling $N\gamma$ while $N-1$ modes
darken without bound; near $kd=\pi$ the nearest-neighbour coupling
$j_0(\pi)=0$ regroups all widths near the single-cavity value $\gamma$
(dashed line). (b)~The darkest width for $N=2$--$8$: numerically the
darkening follows $(kd)^{2(N-1)}$ (fitted exponents $2.0$, $6.0$,
$10.2$, $14.7$), so even a short sub-wavelength chain realises
$\gammarad/\gammarad^{\mathrm{cl}}$ many orders of magnitude below
unity. The same rank-structure underlies the Dicke dark states of
Fig.~\ref{fig:taxonomy}(b).}
\label{fig:tolstoy}
\end{figure}

A generation before the term \emph{super-resonance} entered the
literature, related collective-resonance physics was already being probed
experimentally: in the
late 1950s and early 1960s, pioneer measurements of the response of
single bubbles and bubble clouds in water to weak and strong impulsive
loads were carried out in the Soviet Union by V.~F.~Minin; a recent
brief historical account is given in Ref.~\citep{minin2024history}. Read in
hindsight, some of the non-linear acoustic phenomena uncovered there can
be interpreted as precursors of collective cavity resonances, although
the connection is retrospective and rests on a secondary source rather
than on the original measurements. The term itself enters the literature
with Tolstoy's 1986--87 papers
\citep{tolstoy1986superresonant,tolstoy1987properties},
in which arrays of elastic spherical cavities --- each of which sustains
its own low-frequency monopole and dipole resonances --- are shown to
interact collectively in a way that drives the radiation linewidth of
certain array eigenmodes to zero, and the corresponding scattering
cross-section to extreme values. The essential physics is already
visible in the minimal two-cavity system: when two air bubbles in water
pulsate close to one another, their acoustic interaction splits the
single-bubble breathing resonance at the Minnaert frequency
\citep{minnaert1933musical} into two distinct peaks, and the behaviour
of the pair differs sharply according to whether it oscillates in phase
--- the symmetric combination, whose radiation damping is collectively
enhanced --- or out of phase, the antisymmetric combination whose
radiative width is suppressed: the $N{=}2$ germ of the bright/dark
dichotomy that Tolstoy's arrays generalise to $N$ cavities. The
mechanism, recognised in modern
language as a discrete realisation of the Friedrich--Wintgen BIC
\eqref{eq:fw_imag}, has the inter-cavity coupling $V$ playing the
role of the off-diagonal matrix element: in an array of $N$ identical
cavities the symmetric and antisymmetric combinations of single-cavity
eigenmodes acquire markedly different radiation widths, and the dark
(antisymmetric) state has $\gammarad/\gammarad^{\mathrm{cl}} \to 0$ in
the ideal-array limit. It is therefore a canonical acoustic M1
super-resonance, formulated --- in hindsight --- a quarter of a century
before the experimental demonstration of photonic BICs (the
Friedrich--Wintgen theory itself dates from the same period
\citep{friedrich1985interfering}). The construction is quantified in
Fig.~\ref{fig:tolstoy} for the simplest faithful model, $N$ identical
monopole resonators sharing the scalar radiation continuum: as the array
pitch shrinks below the wavelength, one collective mode absorbs the
entire coupling ($\gammarad\to N\gamma$, the super-radiant partner)
while the remaining $N-1$ modes darken as steep powers of $kd$ ---
Tolstoy's super-resonance in its distilled form, and simultaneously the
discrete ancestor of the subradiant states of atomic arrays
(Sec.~\ref{sec:quantum}).

\subsubsection{Single-particle acoustic super-resonance}

Norris \citep{norris1986scattering} provided the rigorous low-frequency
expansion for elastic spherical inclusions, in which the secular determinant
degenerates exactly at the super-resonance condition. More recently, Minin
and Minin demonstrated super-resonance in a \emph{single} mesoscale
dielectric sphere immersed in water \citep{minin2019extreme,michavila2019super}
--- the first ``Tolstoy resonance in one element''. The geometry is the
acoustic analogue of the Mie supercavity of Sec.~\ref{sec:mie}, with the
elastic wavefield playing the role of the electromagnetic mode and the
sphere acting as a self-interfering resonator that supports both M1 (dark
modes of the surface elasticity) and M2 (high-order monopole and dipole
concentration) signatures simultaneously.

\subsubsection{Single-bubble cavitation and sonoluminescence}

A distinct acoustic phenomenon in which collective dynamics concentrate
energy into a single resonator is single-bubble sonoluminescence
\citep{margulis2000sonoluminescence}. In a standing acoustic field the
primary Bjerknes force gathers a cloud of in-phase pulsating cavitation
bubbles together until a single, phase- and intensity-stable bubble
remains, pulsing coherently for as long as several hours and emitting a
flash of light at each collapse; the emission is acutely sensitive to
acoustic pressure, dissolved-gas content, and the host liquid. We note it
for completeness, because the literature occasionally files such cavitation
effects alongside acoustic ``super-resonance,'' but it lies \emph{outside}
the strict definition \eqref{eq:SRdef}: the longevity and the extreme
energy concentration at collapse are governed by nonlinear cavitation
dynamics and resonant energy \emph{absorption} within one resonator --- an
M2-type concentration --- not by destructive interference among radiation
channels. It is included here as a related contrast, not as an instance of
M1.

\subsubsection{Bandwidth-breaking elastic super-resonance}

The most striking recent result is the elastic super-resonance demonstrated
by Harris, Kianfar, Roca, Yago, Brehm, and Hussein
\citep{harris2026super}. Their building block is a coiled, locally
resonant elastic metamaterial architected so that several internal
energy-transport pathways converge at a single structural point; the
spatial convergence of these pathways sustains modal dominance over a much
wider band than a single resonator can. Concretely, the authors showed
that the antiphase modal response is preserved across a frequency range
more than five times wider than that of an equivalent uncoiled resonator
of the same footprint --- a direct breaking of the textbook resonance
time--bandwidth limit. The defining condition can be written
\begin{equation}
\phi_{\mathrm{super}}(\omega) \approx -\pi,\quad
\forall\,\omega\in[\omega_{\mathrm{low}},\omega_{\mathrm{high}}],
\qquad
(\omega_{\mathrm{high}}-\omega_{\mathrm{low}}) \gg \Delta\omega_{\mathrm{conv}},
\label{eq:phi_super}
\end{equation}
where $\Delta\omega_{\mathrm{conv}}$ is the conventional Lorentzian
bandwidth of a single elastic resonator and the phase
$\phi_{\mathrm{super}}$ is measured between the driving force and the
acoustic-radiation response. The application demonstrated in
\cite{harris2026super} is broadband laminar-flow control: a super-resonant
phononic subsurface interfacing with a channel flow passively suppresses
four distinct unstable flow perturbations simultaneously over the extended
band, where a conventional resonance can address only one. In our
taxonomy this construction is M1-\emph{adjacent}: the interference of
converging energy-transport pathways is the same ingredient that powers
radiative suppression, but the figure of merit being engineered is the
\emph{phase} of the response over a broad band rather than a radiative
linewidth, so the strict criterion \eqref{eq:SRdef} applies only in a
generalised sense. It is nonetheless the closest elastodynamic
counterpart of the broadband quasi-BIC engineering
pursued in photonics (Sec.~\ref{sec:bic}).

\subsubsection{Cracked-beam sub- and super-harmonics}

Cracked-beam vibrations \citep{bovsunovskii2006damping} provide a
complementary realisation based on M3: the nonlinear stiffness modulation
introduced by the closing crack generates sub- and super-harmonic
resonances whose frequencies are integer combinations of the drive and the
linearised eigenfrequency. These integer-commensurate satellites are
themselves narrow and large, and they are sensitive probes of structural
integrity. The FOM here is not $Q$ but rather the
narrow-band amplitude of the satellite peak, which obeys an inequality of
the form \eqref{eq:mmr}. Surface-acoustic-wave direction sensing in a
single micropillar \citep{lee2025surface} exploits modal interference of
the M1 type within a single resonator; the construction is, again, the
elastic analogue of a one-dimensional photonic supercavity.

\subsubsection{Acoustic metamaterials and the acoustic diffraction limit}

Local resonances were introduced in acoustic metamaterials by Liu \etal\
\citep{liu2000locally}, and the subsequent flowering of acoustic
metamaterials and metasurfaces \citep{cummer2016controlling,ma2016acoustic,
hussein2014dynamics} has produced numerous architectures in which
(i)~radiation channels interfere to yield extreme transmission, (ii)~modal
hybridisation concentrates spectral weight, or (iii)~periodic modulation
unlocks Floquet ladders of effectively commensurate modes --- each of
these falls under one of the three mechanisms M1--M3. The acoustic
diffraction limit has been discussed in this context by Maznev and Wright
\citep{maznev2015upholding}, who establish a complementary
``super-resolution'' bound that is broken precisely by configurations
exhibiting M1 interference or M2 modal concentration. Acoustic
super-resonance is therefore mature in concept and ripe for
super-resolution imaging at sub-MHz frequencies, an application we
discuss in Sec.~\ref{sec:acousticTHz}.

Resonant amplification of mechanical waves also operates at the largest
geophysical scales. Shallow-water (tsunami) waves are strongly amplified
when an underwater landslide of slowly varying mass moves in resonance
with the wave it generates \citep{didenkulova2011tsunami}, and
trans-oceanic tsunamis resonate with the natural modes of bays and
continental shelves to produce locally extreme run-up
\citep{wang2021tsunami} --- a hydrodynamic counterpart of the
commensurability (M3) and cavity-mode (M2) mechanisms catalogued above.

\subsection{Celestial mechanics and gravitational physics}\label{sec:celestial}

\subsubsection{Mean-motion and Laplace resonances}

In the solar system the orbital frequencies of distinct bodies satisfy a
remarkable number of small-integer commensurabilities. The Galilean
satellites Io, Europa, and Ganymede are locked in a three-body Laplace
resonance,
\[
n_{\mathrm{Io}} - 3 n_{\mathrm{Eu}} + 2 n_{\mathrm{Ga}}=0,
\]
\citep{barnes2011laplace,paita2018element}. Pluto and Neptune share a 3:2
mean-motion resonance \citep{williams1971neptune}, and Pluto in
particular exhibits a remarkable 1:1 commensurability between its
argument-of-perihelion circulation and the libration in the 3:2 resonance
--- the so-called \emph{Pluto super-resonance} of Wan, Huang, and Innanen
\citep{wan2001pluto}. Goldreich \citep{goldreich1965explanation} provided
the canonical explanation of why such commensurabilities are
over-represented: the slow secular dissipation of tidal energy drives
planetary systems inexorably towards integer ratios, which then act as
dynamical attractors. The libration around the resonant island plays the
role of an infinite-$Q$ ``ringing'' mode of the solar system on
cosmological time-scales --- mechanism M3 in its purest astronomical form
\citep{murray1999solar,laskar1989numerical}.

Mechanical analogues of the same phenomenon were noted as early as 1774 by
Daniel Bernoulli in his study of the double mathematical pendulum
\citep{smirnov2020double}: phase locking between the two arms confines the
energy to a small island of phase space when the natural frequencies are
commensurate. The mathematical structure of M3 is therefore older than
the celestial application by nearly a century, which underscores the
field-agnostic character of the mechanism.

\subsubsection{Black-hole super-radiance}

A rotating black hole amplifies an incident bosonic wave whenever the
super-radiance condition \eqref{eq:superradianceBH} holds
\citep{press1972rotating,bekenstein1973black,brito2020superradiance,
thorne1994blackholes}. The amplification draws on the rotational energy
of the hole, and in the taxonomy of Sec.~\ref{sec:taxonomy} it belongs
squarely to the \emph{amplification} family, not to super-resonance
proper: the reflection coefficient exceeds unity, and the relevant
eigenvalue moves into the gain half-plane rather than onto the real axis.
Confinement of the super-radiant mode --- by a mirror or by a
massive-field potential --- turns the amplification into an exponentially
growing instability, the ``black-hole bomb'' \citep{brito2020superradiance},
with profound astrophysical consequences for ultralight bosonic dark
matter. We include the phenomenon in this review because the literature
on its acoustic analogues has explicitly adopted the term
``super-resonance'' \citep{basak2003superresonance,kobes2006superresonance},
making it a prime example of the terminological collision this review
seeks to resolve.

The acoustic analogue developed by Basak and Majumdar
\citep{basak2003superresonance}, Kobes \citep{kobes2006superresonance},
and Anacleto, Brito, and Passos \citep{anacleto2013noncommutative}
established the universality of the effect within the domain of effective
acoustic geometries (``analogue gravity''). In the laboratory, Torres
\etal\ \citep{torres2017rotational} provided the first experimental
observation of rotational super-radiant scattering in a draining water
vortex --- rotational amplification instantiated in a kitchen sink. Tang,
Shi, Zhang \etal\ \citep{tang2024unusual} extended these considerations
to optical scattering by a rotating mesoscale sphere, linking
mechanical rotation to the photonic-hook phenomenology and showing
that rotation modifies the modal-coherence condition of
Sec.~\ref{sec:mie} (a non-reciprocal rotational effect, though no
reflection gain --- the defining signature of the amplification family
--- has been demonstrated in this setting).

\subsubsection{Gravitational-wave resonances and gravity--light coupling}

McDonald and Ellis \citep{mcdonald2024resonant} demonstrated that
gravitational waves passing through the magnetosphere of a neutron star
can be resonantly converted to electromagnetic radiation when the
gravitational-wave frequency matches a magnetospheric resonance. This
mechanism --- of the M2 kind --- was invoked to explain the puzzling
correlations between gamma-ray bursts and gravitational events in
compact-binary coalescences and provides one of the cleanest astrophysical
tests of the photon--graviton coupling.

The direct detection of gravitational waves by LIGO and Virgo
\citep{aasi2015advanced,abbott2016observation} rests on resonant optical
storage: the Pound--Drever--Hall-locked Fabry--P\'erot arm cavities
\citep{drever1983laser} achieve a finesse $\mathcal{F}\!\sim\!450$ and an
effective storage time that pushes the optical readout sensitivity toward
the standard quantum limit. Power- and signal-recycling cavities reshape
the circulating power and the signal transfer function of the coupled
system, enhancing the strain sensitivity by orders of magnitude --- a
coupled-cavity cascade in the engineering sense of
Sec.~\ref{sec:keyeqs}, though not an interference-suppressed radiative
width. We return to this application in Sec.~\ref{sec:gravitational}.

\subsection{High-order Fano resonance in mesoscale dielectric spheres
(M2 family)}\label{sec:mie}

\subsubsection{High-order Fano resonance of dielectric spheres}

\begin{figure}[t]
\centering
\includegraphics[width=0.52\linewidth]{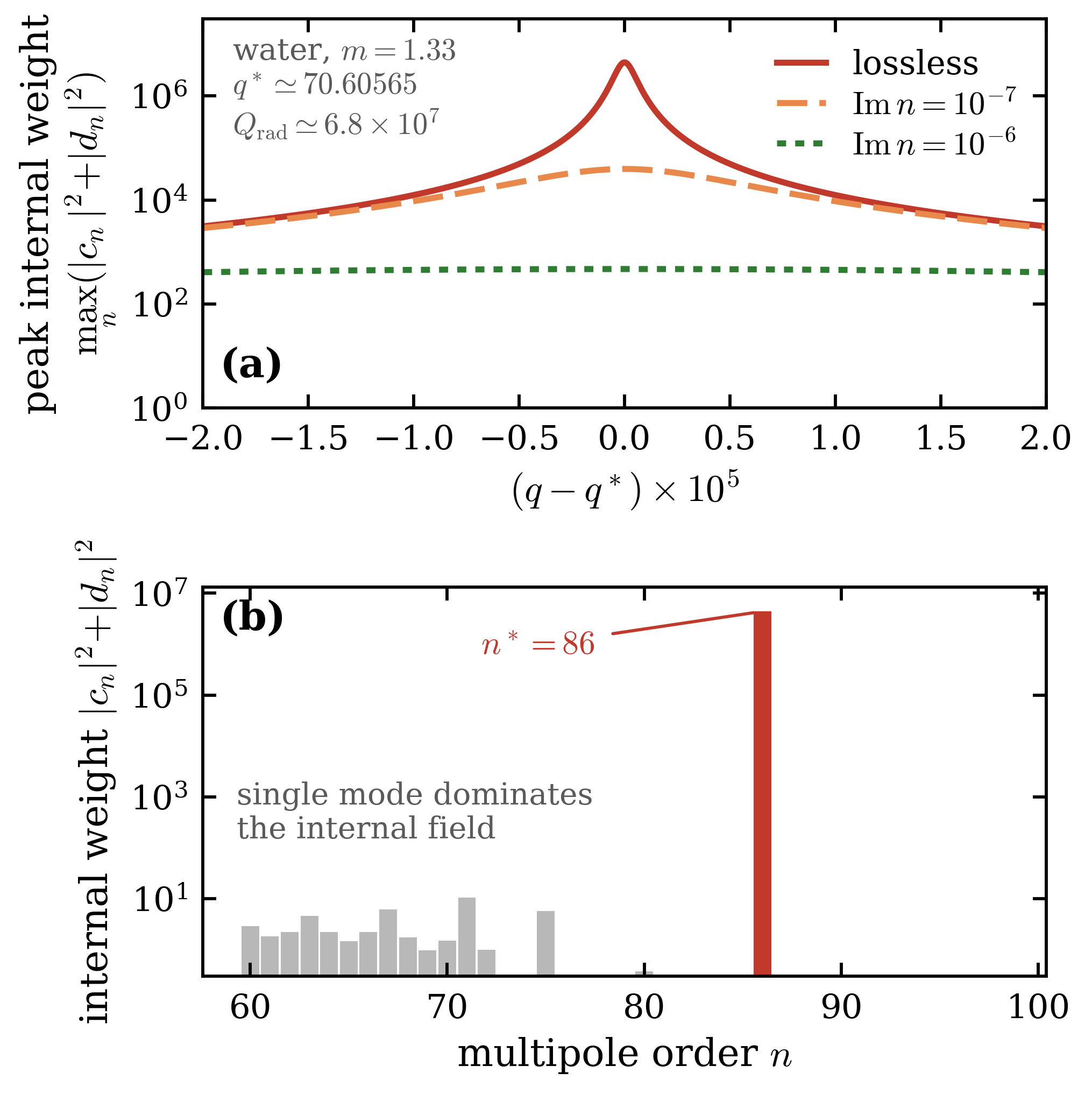}
\caption{High-order Fano resonance of a mesoscale water sphere
($m=1.33$; family M2), computed from the internal Lorenz--Mie
coefficients associated with \eqref{eq:Mie_ab}. (a)~Peak internal
spectral weight $\max_n(|c_n|^2{+}|d_n|^2)$ across the supercavity pole
at $q^\ast=70.60565$ --- the $q\!\approx\!70.6$ resonance discussed in
the text ($a\!\approx\!6\,\mu$m at $\lambda=534$\,nm); the numerical
linewidth gives a radiative $Q\simeq7\times10^{7}$. Absorption
representative of high-purity borosilicate glass,
$\mathrm{Im}\,n=10^{-7}$, suppresses the
peak weight by a factor $\sim\!10^{2}$ and broadens the line;
$\mathrm{Im}\,n=10^{-6}$ suppresses it by nearly four orders --- the
loss ceiling of Secs.~\ref{sec:mie} and~\ref{sec:outlook}.
(b)~Internal weight per multipole order at the pole: the single
magnetic-type order $n^\ast=86$ exhausts the internal spectral weight
(canonical form II of Sec.~\ref{sec:keyeqs}).}
\label{fig:mie}
\end{figure}

The single most direct laboratory realisation of the M2
spectral-weight-concentration family is the
high-order Fano resonance of mesoscale dielectric spheres
\citep{wan2019highorder,minin2024discovery,minin2023cenosphere,
minin2022borosilicate,minin2024freezing,minin2022fano,yue2019poynting}.
The Lorenz--Mie scattering amplitudes
\citep{mie1908beitrage,bohren1983absorption},
\begin{subequations}
\label{eq:Mie_ab}
\begin{align}
a_n &= \frac{m\psi_n(mq)\psi_n'(q) - \psi_n(q)\psi_n'(mq)}
            {m\psi_n(mq)\xi_n'(q) - \xi_n(q)\psi_n'(mq)},\\
b_n &= \frac{\psi_n(mq)\psi_n'(q) - m\psi_n(q)\psi_n'(mq)}
            {\psi_n(mq)\xi_n'(q) - m\xi_n(q)\psi_n'(mq)},
\end{align}
\end{subequations}
with $q\!=\!2\pi a/\lambda$ the size parameter of Sec.~\ref{sec:M2}
(denoted $x$ in Ref.~\citep{bohren1983absorption}),
$m\!=\!n_{\mathrm{sphere}}/n_{\mathrm{medium}}$,
and $\psi_n,\xi_n$ the Riccati--Bessel functions, exhibit isolated narrow
zeros of the denominator at the supercavity points $(q,m)$
\citep{wan2019highorder,klimov2019supercavity}. At such a point one of
$|a_n|$ or $|b_n|$ approaches unity at very high order $n^\ast$ while the
other partial amplitudes remain $\mathcal{O}(10^{-2})$ --- the canonical
form II of Section~\ref{sec:keyeqs}. The internal field computed at such
a point \citep{wan2019highorder,minin2024discovery,
yue2019poynting,minin2024freezing} is dominated by a single high-multipole
mode and forms hot spots in the vicinity of the polar points of the
sphere where the intensity reaches
\begin{equation}
|E|^2/|E_0|^2 \sim 10^{6},\qquad
|H|^2/|H_0|^2 \sim 10^{7},
\label{eq:Eintensity}
\end{equation}
for a water microdroplet of size parameter $q\!=\!70.6$ ($a\!\approx\!6\,\mu$m
at $\lambda\!=\!534$\,nm). Two qualifications must accompany these
numbers. First, they are obtained from Lorenz--Mie computations for an
\emph{idealised lossless} sphere: the corresponding modal $Q$ is a
property of the lossless pole --- Ref.~\citep{minin2024discovery} quotes
a value of order $6\times 10^{8}$, while our own recomputation at the
refined pole position yields a radiative $Q\simeq7\times10^{7}$
(Fig.~\ref{fig:mie}) --- and the finite absorption of real water
($\mathrm{Im}\,n\!\sim\!10^{-8}$--$10^{-7}$ in the visible) or
borosilicate reduces the attainable peak intensity by orders of magnitude
(see the loss discussion below and Sec.~\ref{sec:outlook}). Second,
direct experimental measurement of the internal hot-spot intensity at
optical frequencies remains an open challenge; the experimental evidence
to date is indirect (scattering spectra, microwave-frequency analogues,
and freezing-droplet dynamics). With these caveats the high-order Fano
resonance remains the sharpest known single-particle realisation of
spectral-weight concentration. Both of its defining features --- the
dominance of a single multipole order and the acute sensitivity to
absorption --- are computed explicitly in Fig.~\ref{fig:mie} for the
canonical water-droplet pole: a magnetic-type order $n^\ast=86$ towers
more than five orders of magnitude above every other partial amplitude,
and material loss representative of high-purity borosilicate glass,
$\mathrm{Im}\,n\!\sim\!10^{-7}$, already erases two orders of magnitude
of the peak internal weight.

A salient feature emphasised by Minin and Minin
\citep{minin2024discovery,minin2023cenosphere} is that the effect is
\emph{not} restricted to high-index materials: the high-order Fano
resonance has been
identified in low-index spheres ($n\!\approx\!1.3$, water droplets), in
intermediate-index spheres ($n\!\approx\!1.5$, borosilicate
\citep{minin2022borosilicate}; cenosphere shells
\citep{minin2023cenosphere}; Teflon \citep{yue2020teflon}), and in
high-index spheres ($n\!>\!2$, Si). The condition for the resonance is
a discrete condition on the pair $(q,m)$ and not a continuous parameter
regime. Klimov \citep{klimov2019supercavity} pointed out, importantly,
that the analogous supercavity condition is \emph{not} supported by
infinite cylinders --- the dimensionality matters --- but is present in
spherical and ellipsoidal geometries. The result invites a finite-element-style
classification of which geometries support which mechanism, a programme
that remains largely open.

A direct consequence of the Lorenz--Mie eigenvalue structure
\eqref{eq:Mie_ab} is that the size parameter at which a multipole of
given order resonates scales inversely with the particle's refractive
index, $q^\ast\!\sim\!n^\ast/n$ (the inner turning point of the
centrifugal barrier, up to an Airy-zone correction). Driving $n$ to extreme values, $n\!\sim\!10$,
therefore pulls the resonance of a strongly subwavelength particle
($q\!\lesssim\!1$) down to the dipole order, $n^\ast\!=\!1$ --- a
regime in which the internal magnetic-field intensity nevertheless
remains anomalously large. The effect was first demonstrated
experimentally in the microwave range by Luk'yanchuk \etal\
\citep{lukyanchuk2021colossal}, who reported \emph{colossal} magnetic
fields inside a high-index dielectric sphere, and was subsequently
corroborated by the independent confirmation experiments of Minin and
Minin \citep{minin2026confirmation}. The high-$n$ limit therefore
provides a complementary experimental route to extreme internal-field
concentration, distinct
from the high-$n^\ast$ mesoscale pathway but governed by the same
canonical form II.

The same Fano logic extends from single spheres to engineered
metasurfaces. Coupling an ultra-narrow high-$Q$ quadrupolar magnetic mode
to the two broad electric-dipole and electric-quadrupole modes of a
silicon metasurface realises a multi-mode
``super-Fano'' resonance \citep{superfano2023} --- a construction that
in fact combines M2 spectral-weight concentration with genuine M1
interference between the narrow and broad channels. In an amorphous-silicon
metasurface of elliptical nanoresonators, H\"ahnel \etal\ showed that this
multi-mode interference concentrates the spectral weight into the narrow
collective mode and thereby boosts third-harmonic generation by a factor of
$\sim\!900$ relative to an unpatterned silicon film of the same thickness,
reaching an absolute conversion efficiency of $\sim\!2.8\times10^{-7}$ at a
pump intensity of $1.2\,\mathrm{GW\,cm^{-2}}$ --- a metasurface-scale
demonstration of the M2 modal-concentration mechanism applied to nonlinear
conversion (Sec.~\ref{sec:extreme}).

\subsubsection{Time-domain dynamics and Janus geometries}

The time-domain dynamics of the high-order Fano resonance was investigated
for the first time in a freezing water microdroplet by Minin \etal\
\citep{minin2024freezing}: the core--shell structure that develops
spontaneously upon freezing modulates the size parameter of the underlying
sphere and therefore sweeps the Fano resonance through its asymmetric
line shape, producing a transient super-resonance that can be observed
non-invasively. Janus geometries, in which a segment of the sphere is
removed \citep{minin2022janus,minin2017diffractive}, support
whispering-gallery sub-modes of super-resonant character and generate
sub-wavelength hot spots on the truncated flat surface, with hundred-fold
enhancement of the magnetic-field intensity. The magnetic
whispering-gallery super-resonance \citep{geints2024magnetic} extends the
same logic to the magnetic-field component, which is in many ways the
cleaner channel because plasmonic enhancements are typically electric.

\subsubsection{Sensitivity, loss, and refractometric response}

Three further results sharpen the framework. \textit{First}, the
precision with which the size parameter $q$ must be tuned matters
acutely: Chy\l ek, Pendleton, and Pinnick \citep{chylek1985internal}
established already in 1985 that the peak internal field varies by
several orders of magnitude over fractional size-parameter variations of
$10^{-6}$. \textit{Second}, finite material loss has a counter-intuitive
consequence: introducing dissipation $k\!\sim\!10^{-5}$ suppresses the
maximum intensity by two orders of magnitude, but the localisation of the
surviving response is actually enhanced \citep{yue2020teflon,
yue2019poynting,wan2019highorder}; the Poynting-vector vortex flow
inside the particle provides the physical explanation
\citep{yue2019poynting}. Moreover, for sufficiently small but non-zero
loss the situation is even more striking: there exists a narrow regime
in which both the peak internal field \emph{and} its spatial localisation
are simultaneously enhanced relative to the lossless limit
\citep{minin2024discovery,minin2024freezing}, an effect that runs
counter to the standard intuition that dissipation can only degrade a
high-$Q$ super-resonance. \textit{Third}, the FOM is
extremely sensitive to environmental refractive-index variations
\citep{minin2022extreme,geints2024magnetic}, which opens the way to
super-resonance refractometric sensing of unprecedented sensitivity ---
an application we develop quantitatively in Sec.~\ref{sec:sensing}.

\subsubsection{Gain-compensated ``magnetic-dipole super-resonance''
(amplification family)}

A construction that the photonics literature explicitly calls
``super-resonance'', but that our taxonomy places in the
\emph{amplification} family, is the
magnetic-dipole resonance of gain-loaded coated nanoparticles introduced
by Liberal, Ederra, Gonzalo, and Ziolkowski \citep{liberal2014magnetic,
gordon2007coated}. A silicon core of radius $75\,$nm is encased in a
shell of radius $120\,$nm doped with gain medium; the Mie coefficient
$b_1$ for the magnetic-dipole channel is driven to exceed its
single-channel passive limit because the gain actively compensates the
radiation loss --- the eigenvalue is pushed toward (and potentially past)
the real axis by energy input, not by passive interference. The
mechanical
force exerted on a probe by such a particle exceeds by orders of magnitude
that exerted by a passive silicon nanoparticle of the same size,
illustrating that extreme resonant states are not merely spectroscopic
curiosities but produce concrete mechanical consequences --- a point we
develop in Sec.~\ref{sec:trapping}.

\subsection{Bound states in the continuum and supercavity modes}\label{sec:bic}

\subsubsection{Origin: from von Neumann--Wigner to Friedrich--Wintgen}

Bound states in the continuum are the cleanest known realisation
of mechanism M1. First proposed mathematically by von~Neumann and
Wigner in 1929, and re-derived in interferometric form by Friedrich and
Wintgen \citep{friedrich1985interfering}, BICs in photonics were
identified theoretically by Marinica \etal\ \citep{marinica2008bound} and
demonstrated experimentally by Hsu \etal\ \citep{hsu2013observation}; the
field has been comprehensively reviewed in
\citep{hsu2016bound,azzam2021photonic,rybin2024metaphotonics,
rybin2019resonance}. The radiative width vanishes exactly at the BIC
point and grows quadratically with the symmetry-breaking parameter as the
system departs from the ideal condition; the resulting \emph{quasi-BIC}
or supercavity mode is what the photonic-resonator community uses as a
practical super-resonance. A particularly clean realisation of the
Friedrich--Wintgen scenario is the avoided resonance crossing (ARC)
between two proximity resonances of a microcavity, which has been
proposed by Gandhi and Ghosh \citep{gandhi2020ultrahigh} as an
ultra-high-$Q$ platform for refractive-index sensing.

\subsubsection{Quasi-BIC in subwavelength dielectric resonators}

\begin{figure}[t]
\centering
\includegraphics[width=0.58\linewidth]{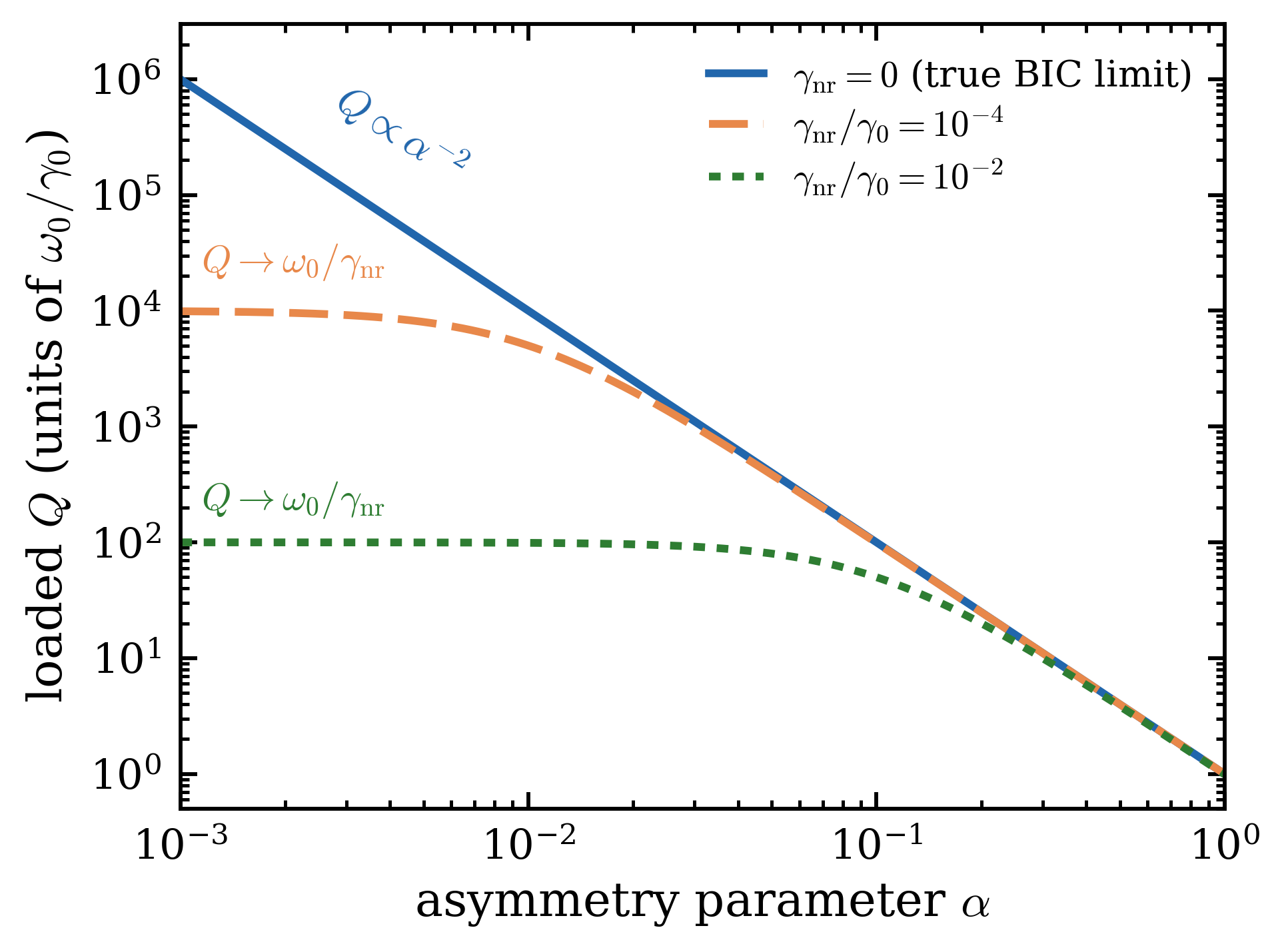}
\caption{From BIC to quasi-BIC (mechanism M1) in the minimal
coupled-mode model. Breaking the protecting symmetry by a dimensionless
asymmetry $\alpha$ opens the radiative channel at
$\gammarad=\gamma_0\alpha^2$, so the loaded quality factor obeys
$Q/Q_0=1/(\alpha^2+\gammanr/\gamma_0)$ with $Q_0=\omega_0/\gamma_0$:
the inverse-square law demonstrated for asymmetric quasi-BIC
metasurfaces \citep{koshelev2018asymmetric}. Non-radiative loss
$\gammanr$ caps the attainable $Q$ at $\omega_0/\gammanr$ (plateaus),
reproducing the transition from a true BIC to a high-$Q$ leaky
resonance in real structures \citep{sadrieva2017transition}. Only the
lossless $\alpha\to0$ limit realises \eqref{eq:SRdef} exactly.}
\label{fig:qbic}
\end{figure}

In subwavelength dielectric resonators, Rybin, Koshelev, Sadrieva,
Samusev, Bogdanov, Limonov, and Kivshar \citep{rybin2017highq}
demonstrated that a single isolated finite cylinder supports supercavity
modes at the avoided crossing of its Mie-type and Fabry--P\'erot-type
modes: for permittivity $\varepsilon=80$ the computed $Q$ reaches
$6.5\times10^{4}$ at subwavelength dimensions, the maximum $Q$ growing
as a power of $\varepsilon$ (so that silicon, $\varepsilon\approx13$,
yields $Q\approx200$ --- still an order of magnitude beyond the ordinary
Mie resonances of the same resonator), with subwavelength operation
available for $\varepsilon\gtrsim3.5$. Koshelev, Lepeshov, Liu, Bogdanov, and Kivshar
\citep{koshelev2018asymmetric} generalised the construction to
metasurfaces composed of asymmetric meta-atoms, showing that the
quasi-BIC $Q$ scales as the inverse square of the asymmetry parameter and
that the non-linear conversion efficiency benefits enormously
\citep{koshelev2020subwavelength}. Figure~\ref{fig:qbic} reproduces this
scaling in the minimal coupled-mode model --- symmetry breaking opens
the radiative channel at the rate $\gammarad=\gamma_0\alpha^2$, so the
loaded $Q$ climbs as $\alpha^{-2}$ until non-radiative loss caps it ---
which is also the cleanest way to see how the true BIC degrades into a
finite-$Q$ leaky resonance in real samples.

 Bogdanov \etal\
\citep{bogdanov2019bound} emphasised that the strong-coupling regime is
the natural habitat of BIC super-resonance, while Sadrieva \etal\
\citep{sadrieva2017transition} clarified the role of substrate and
surface roughness in transitioning from a true BIC to a high-$Q$ leaky
resonance. Lukosiunas \etal\ \citep{lukosiunas2023extremely} showed that
critical narrowing of resonance peaks occurs as their eigenfrequencies
approach the edge of the continuum; combining two coupled modes near
such an edge then yields the so-called superscattering phenomenon
\citep{ruan2010superscattering,ruan2011design,qian2019experimental,
rybin2024metaphotonics,shalin2023photonic}.

\subsubsection{Anapoles, superscattering, and the duality of M1 and M2}

The anapole state of a single dielectric nanoparticle
\citep{miroshnichenko2015nonradiating,baryshnikova2019optical,
savinov2018optical} is an M1-type local super-resonance: the destructive
interference of electric-dipole and toroidal-dipole moments causes the
far-field radiation to vanish while the near-field intensity is finite
and large. Superscattering --- the converse phenomenon in which several
multipole moments are tuned to resonate constructively and the total
scattering cross-section exceeds the single-channel bound --- belongs to
the M2 enhancement family \citep{ruan2010superscattering,ruan2011design,
qian2019experimental,krasnok2015superdirective}, and was experimentally
confirmed by Qian \etal\ \citep{qian2019experimental}. The two phenomena
are dual: anapole suppresses the radiation channel that superscattering
maximally exploits, and both rely on the same multipole-expansion
machinery of Mie theory \eqref{eq:Mie_ab}.

\subsubsection{Mie-tronics and the broader metaphotonics landscape}

The renaissance of all-dielectric photonics
\citep{kuznetsov2016optically,baranov2017alldielectric,kivshar2018meta,
barreda2019applications} has identified high-index, low-loss
semiconductors as the natural platform for super-resonant devices.
Decker \etal\ \citep{decker2015highefficiency} demonstrated
high-efficiency dielectric Huygens metasurfaces; Vaskin \etal\
\citep{vaskin2018directional} exploited Mie-resonant arrays for spectral
and directional emission shaping; and Koshelev and Kivshar
\citep{koshelev2019meta} surveyed the resulting dielectric resonant
metaphotonics. The unifying message is that mechanisms
M1 and M2 are operative across nearly the entire metaphotonics
literature, and the boundary between an ``ordinary'' high-$Q$ resonance
and a super-resonance is precisely the condition \eqref{eq:SRdef}.

\subsection{Plasmonics, whispering-gallery, and photonic-crystal cavities}\label{sec:plasmonics}

\subsubsection{Plasmonic Fano resonance}

In plasmonics, the M2 mechanism manifests as the plasmonic Fano resonance
of Luk'yanchuk \etal\ \citep{lukyanchuk2010fano,limonov2017fano} and
Lassiter \etal\ \citep{lassiter2010fano}. A dark plasmonic mode embedded
in a broad bright continuum produces an asymmetric line shape with steep
spectral slopes; the near-field enhancement on the slope can exceed the
maximum-on-resonance of either constituent mode, with the slope
steepness $d\sigma/d\omega$ scaling as $Q$ in much the same way as for
the high-order Fano resonance of dielectric spheres
\eqref{eq:fano}. This is what makes plasmonic super-resonances especially
attractive for refractometric sensors, which exploit the slope rather
than the peak \citep{stockman2011nanoplasmonics}. A representative
realisation couples a plasmonic stub to a circular
metal--insulator--metal cavity: interference between the ultra-narrow
discrete resonance and the broad cavity continuum yields an ultra-sharp
Fano line shape with a refractometric FOM of order $10^{4}$
\citep{yun2016fano}. We notice that the
dominant practical limitation in plasmonic super-resonance is loss in the
metal \citep{khurgin2015deep}, a fact which has driven much of the recent
shift to all-dielectric (Sec.~\ref{sec:bic}) and hybrid metal--dielectric
architectures \citep{barreda2019applications}.

A closely related manifestation is the analogue of electromagnetically
induced transparency (EIT): when a narrow dark mode is coupled to a broad
bright mode, the interference that produces a Fano peak can instead open
an ultra-narrow transparency window in an otherwise opaque background.
Such EIT-like super-resonances have been realised in all-dielectric
silicon terahertz metasurfaces \citep{wang2021eit} and in cascaded
micro-ring resonators \citep{zhou2025eit}; their steep dispersion
underlies slow-light and high-sensitivity refractometric applications.

\subsubsection{Composite confinement-induced super-resonance in
metallic nanostructures}

Metallic nanostructures host one further, independent coinage of the
term. Well below the plasmon frequency, the optical response of a
nanoparticle is governed not by the collective plasmon but by
transitions between the quantum-confined states of the conduction
electrons; the particle acts as an \emph{electronic billiard}, and the
statistics of its level spacings depend on whether the billiard is
integrable or chaotic. Babushkin \etal\ \citep{babushkin2023metallic}
predicted that for typical (integrable or nearly integrable) geometries
the closely spaced confinement-induced resonances join into a single
composite resonance --- which they name a super-resonance --- whose
position and width are controlled by the geometry of the nanostructure,
and which carries a giant low-frequency nonlinearity. Their simulations
indicate that this nonlinearity permits efficient down-conversion of an
optical pump to terahertz and mid-infrared frequencies in
sub-micrometre devices based on nanoparticle composites. In the
taxonomy of Sec.~\ref{sec:taxonomy} the phenomenon is M2: many partial
amplitudes merge into one dominant peak, a spectral-weight
consolidation with the coupling enhanced rather than a radiative width
interfered away. Its interest for the present review is twofold --- as
a rare bridge between mesoscopic electron dynamics and resonant
photonics, and as one more community reusing the word for physics
distinct from the M1 definition.

\subsubsection{Hyperbolic super-resonance}

The most recent independent coinage belongs to hyperbolic media:
Narimanov and Demler \citep{narimanov2024hyperbolic} introduced the term
\emph{hyperbolic super-resonance} for the extreme high-$Q$ polariton
modes of hyperbolic metamaterials, proposed as mediators of long-range,
ultrastrong qubit--qubit coupling. In the taxonomy of
Sec.~\ref{sec:taxonomy} the construction belongs with the enhancement
family: the figure of merit being driven to extremes is the
\emph{coupling} between emitters mediated by the high-density polariton
spectrum, not an interference-suppressed radiative width. We record it
here both for completeness of the terminological map and because the
proposal connects the resonance physics of this section to the
ultrastrong-coupling cavity QED of Sec.~\ref{sec:quantum}.

\subsubsection{Whispering-gallery microcavities}

\begin{figure}[t]
\centering
\includegraphics[width=0.58\linewidth]{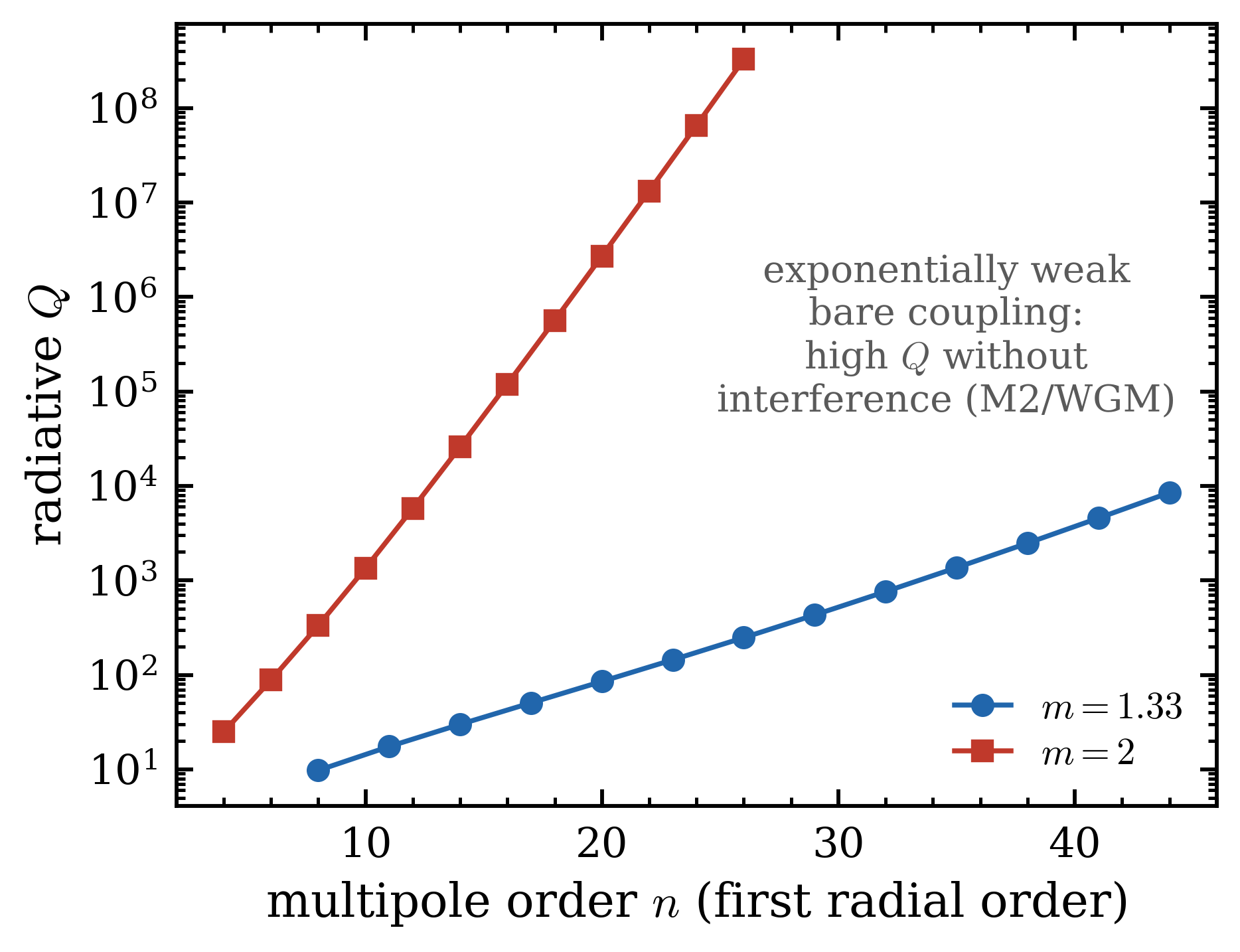}
\caption{Radiative $Q$ of whispering-gallery-type Mie resonances
(first radial order, TE polarisation) versus multipole order $n$, for
relative refractive indices $m=1.33$ (water/silica-like) and $m=2$.
Each point is computed by locating the pole of the $b_n$ Mie
denominator and measuring its numerical linewidth. The exponential
growth --- one decade of $Q$ per $\sim\!3$ multipole orders at $m=2$,
one per $\sim\!12$ at $m=1.33$ --- is the curvature
suppression of the bare radiative coupling that powers ultra-high-$Q$
microspheres and toroids \citep{vahala2003optical,armani2003ultrahigh}.
Because the bare coupling itself is small, these modes fall under the
first exclusion of Sec.~\ref{sec:taxonomy} rather than under the
super-resonance criterion \eqref{eq:SRdef}; contrast the quasi-BIC
route of Fig.~\ref{fig:qbic}, where $Q$ is raised by interference at
fixed (wavelength-scale) size.}
\label{fig:wgm}
\end{figure}

Whispering-gallery modes (WGMs) of toroidal, micropillar, microdisk, and
microsphere resonators \citep{vahala2003optical,matsko2006optical,
armani2003ultrahigh,spillane2002ultralow} have provided some of the
highest experimentally measured $Q$-factors in any optical resonator:
$Q$ approaching $10^{10}$ in silica microspheres
\citep{vahala2003optical} and in excess of $10^8$ in
chip-scale silica toroids \citep{armani2003ultrahigh}. These WGMs are M2-type
high-order modes: the field is concentrated near the surface in a
single high-azimuthal-order mode whose radiation losses are exponentially
suppressed by curvature. Figure~\ref{fig:wgm} quantifies this route to
high $Q$ directly from Lorenz--Mie theory: the radiative $Q$ of the
first-radial-order resonances grows exponentially with the multipole
order $n$, at a rate that itself grows steeply with index contrast ---
roughly one decade of $Q$ per three multipole orders at $m=2$, one per
twelve at $m=1.33$. This is precisely the ``small bare coupling'' mechanism that the
first exclusion criterion of Sec.~\ref{sec:taxonomy} distinguishes from
super-resonance: no interference is involved, and the price of the high
$Q$ is a large resonator circumference (high $n$) rather than a
fine-tuned condition.

The connection to the Mie super-resonance of Sec.~\ref{sec:mie} is direct:
the supercavity modes of Refs.~\citep{wan2019highorder,minin2024discovery,
minin2023cenosphere} are themselves WGMs of extremely high order
$n^\ast\!\sim\!40$--$100$. The Janus-particle WGM construction
\citep{minin2022janus,minin2017diffractive} exploits the modal selection
imposed by truncating a portion of the sphere to amplify specific WGMs
and concentrate the field on the cut. A similar logic holds for the magnetic-field component in the magnetic
whispering-gallery super-resonance \citep{geints2024magnetic}. Vollmer and Arnold \citep{vollmer2008whispering}
established WGM biosensing as the de-facto application of super-resonant
microcavities, demonstrating single-molecule detection down to the
attomole level; we develop this application in Sec.~\ref{sec:sensing}.

\subsubsection{Photonic-crystal nanocavities}

Photonic-crystal slabs and three-dimensional photonic crystals
\citep{joannopoulos2008photonic} support nanocavities with localised modes
that exhibit some of the strongest spatial confinements ever reported.
Akahane, Asano, Song, and Noda \citep{akahane2003ultrahigh} introduced
the gentle-confinement design that achieved $Q\!\sim\!4.5\times 10^4$;
subsequent refinements pushed the experimental $Q$ to
$Q\!\sim\!9\times 10^6$ in silicon at telecom wavelength
\citep{sekoguchi2014photonic}, with mode volumes of order $(\lambda/n)^3$.
These cavities are M1-type super-resonances: the photonic band gap
forbids the propagation channels through which the bare mode would
otherwise leak. The combination of high $Q$ with sub-wavelength $V$
yields Purcell factors of order $10^5$ \eqref{eq:purcell} and motivates
much of cavity QED on chip
\citep{lecamp2007very,englund2007controlling,reithmaier2004strong}.

\subsubsection{Kerker effects and superdirectional scattering}

The Kerker conditions \citep{kerker1983electromagnetic}, generalised to
dielectric meta-atoms with overlapping electric and magnetic Mie
resonances, produce superdirectional scattering. Krasnok \etal\
\citep{krasnok2015superdirective} demonstrated an enhanced Purcell
factor in all-dielectric antenna structures built on this principle. The
underlying mechanism is again the constructive coherent combination of
multiple Mie modes (M2), making the scattering pattern resemble that of
a much larger antenna. Alaee \etal\ \citep{alaee2020kerker} extended
the same construction to atomic antennas, showing that lattices of
electric-dipole emitters can exhibit Kerker-type directional scattering,
superscattering, and scattering dark states --- a discrete-emitter
realisation of the M1/M2 duality of Sec.~\ref{sec:bic}.

\subsection{Non-Hermitian super-resonance: CPA, PT-symmetry, and exceptional points}\label{sec:pt}

\subsubsection{Coherent perfect absorption}

Coherent perfect absorption (CPA) is the time-reversed counterpart of
lasing \citep{chong2010coherent}: two counter-propagating coherent beams
fed into a lossy resonator are absorbed with unit efficiency when their
relative phase realises the time-reverse of a laser eigenmode. The
phenomenon was demonstrated by Wan, Chong, Ge, Noh, Stone, and Cao
\citep{wan2011timereversed}; its applications to interferometric control
of absorption and to memory devices are reviewed in
Ref.~\cite{baranov2017coherent}.

\subsubsection{Super-resonant coherent perfect absorption}

Malara \etal\ \citep{malara2016super} took the construction a decisive
step further by nesting a Fabry--P\'erot inside a ring resonator, so that
the interfering fields are themselves super-resonant modes of the
coupled-resonator system. Their hybrid structure produces split
eigenmodes of mixed bright/dark character; the dark
``super-resonant'' mode exhibits an effective interaction length
\eqref{eq:Leff} given by the \emph{product} of the two finesses, vastly
exceeding what either resonator achieves alone. This is the most direct
experimental realisation of multiplicative M1 super-resonance on a single
chip, and the construction has since served as a template for the
multi-cavity gravitational-wave readout architectures
(Sec.~\ref{sec:gravitational}).

\subsubsection{Parity--time symmetry and exceptional points}

The PT-symmetry programme initiated by Bender and Boettcher
\citep{bender1998real} has become a central organising principle of
non-Hermitian physics \citep{elganainy2018non,ozdemir2019paritytime,
miri2019exceptional}. In the unbroken-symmetry regime, two coupled
gain/loss resonators have real eigenvalues despite the non-Hermitian
Hamiltonian; at the \emph{exceptional point} the two eigenvalues and
their corresponding eigenvectors coalesce, and small parameter changes
produce a square-root (rather than linear) response in the eigenvalues,
yielding an enhanced spectral sensitivity
\citep{chen2017exceptional,hodaei2017enhanced}. An important caveat
accompanies the sensing claim: the eigenvalue \emph{splitting} is
enhanced, but so, in general, is the noise. Langbein
\citep{langbein2018no} argued that the measurement precision of an
EP sensor need not exceed that of a conventional (diabolic-point)
sensor once linewidths and signal-to-noise are accounted for, and Lau
and Clerk \citep{lau2018fundamental} showed within a full quantum-noise
treatment that enhanced signal power requires gain rather than EP
proximity as such, identifying broken reciprocity --- not the EP ---
as the resource that can beat conventional bounds. The
splitting-versus-precision distinction is now standard in the EP-sensing
literature, and we adopt it when the application is revisited in
Sec.~\ref{sec:sensing}.

Yuan \etal\ \citep{yuan2021super} demonstrated explicitly an ``optical
super-resonance'' in a customised PT-symmetric coupled-meta-atom system:
at the exceptional point the transmission curve is broad, yet the
localised-field quality factor is very high --- precisely the
super-resonance signature. PT-symmetric photonic super-resonance has an
M1 core: the radiation loss of one of the two coupled modes is
converted into a gain channel, and the interference of these two
contributions makes the net loss vanish, restoring \eqref{eq:SRdef}.
A definitional subtlety should be flagged, consistent with the fourth
exclusion of Sec.~\ref{sec:taxonomy} and with the M1/A entry of
Table~\ref{tab:taxonomy}: because the balance is maintained by
\emph{active} gain, the construction carries an amplification-family
ingredient, and only the inter-mode interference --- not the gain itself
--- is M1 content in the strict sense.

\subsection{Topological and Floquet super-resonance}\label{sec:topological}

Topological and Floquet engineering supply a different route to
super-resonance, in which the radiation channels are closed not by a
fine-tuned amplitude cancellation but by a global property of the band
structure --- a topological invariant or a Floquet replica. The
resulting super-resonant modes inherit, for free, the robustness to
disorder that ordinary M1 super-resonances achieve only by precision
fabrication. The conceptual overlap with both M1 (interference-driven
radiation suppression) and M3 (integer-locked replicas) is intentional:
topological and Floquet super-resonances sit at the intersection of the
two mechanisms, and it is here that the unifying framework of
Section~\ref{sec:definition} gains some of its most striking predictive
power.

\subsubsection{Photonic topological insulators}\label{sec:topophotonics}

The transposition of topological-band physics from condensed matter to
photonics was opened by Haldane and Raghu \citep{haldane2008possible},
who showed that a magneto-optical photonic crystal with broken
time-reversal symmetry supports one-way edge states analogous to the
chiral channels of the integer quantum Hall effect. The proposal was
realised experimentally a year later by Wang, Chong, Joannopoulos, and
Solja\v{c}i\'c \citep{wang2009observation} in a gyromagnetic
photonic-crystal slab at microwave frequencies: a back-scattering-immune
edge mode propagating around a defective boundary was observed
directly, with the radiation amplitude into the bulk continuum
suppressed by the same topological argument that protects the quantum
Hall current. In the language of this review, the edge mode is an M1
super-resonance whose vanishing radiation rate is enforced by a
\emph{global} property of the band structure, not by a local
fine-tuning of inter-mode coupling --- the radiation channel is closed
because the symmetry-allowed set of channels is empty, not because two
amplitudes destructively interfere.

The construction was rapidly generalised to non-magnetic time-reversal-
invariant systems. Khanikaev \etal\ \citep{khanikaev2013photonic}
demonstrated photonic topological insulators in which polarisation
degrees of freedom play the role of pseudo-spin, and Hafezi \etal\
\citep{hafezi2013imaging} imaged topologically protected edge states
directly in a silicon-on-insulator ring-resonator array. These platforms
operate at telecom wavelengths in CMOS-compatible materials, which
makes the topological super-resonance an industrial as well as a
fundamental result. Comprehensive reviews of the rapidly growing field
are provided by Lu \etal\ \citep{lu2014topological} and Ozawa \etal\
\citep{ozawa2019topological}. In every case the figure of merit is the
back-scattering length, which can exceed the disorder-limited transport
length of an ordinary waveguide by orders of magnitude; it plays the
role that $Q$ plays for the Friedrich--Wintgen BIC of
Section~\ref{sec:bic}, with only the microscopic mechanism differing. The same protection can be conferred
on the Fano super-resonance itself: a \emph{topological} Fano resonance,
whose ultra-sharp asymmetric line shape is guaranteed by design and
rendered immune to geometrical imperfections, was demonstrated in an
acoustic system by Zangeneh-Nejad and Fleury
\citep{zangenehnejad2019topological}.

\subsubsection{Topological lasers and non-Hermitian topology}\label{sec:topolaser}

When the topological edge mode is pumped beyond threshold, a
\emph{topological laser} is obtained --- the first instance of a
super-resonance laser whose lasing mode is protected against
disorder by a bulk invariant rather than by a fabrication tolerance.
The theoretical proposal of Harari, Bandres, Lumer \etal\
\citep{harari2018topological} was demonstrated experimentally by
Bandres, Wittek, Harari \etal\ \citep{bandres2018topological} in an
InP-based microring lattice with a winding gain landscape;
single-mode lasing was achieved over the entire edge of the lattice,
and the output remained single-mode even when defects, missing rings,
or fabrication imperfections were deliberately introduced. The
super-resonance FOM here is the side-mode suppression
ratio in the presence of disorder, which exceeds by orders of
magnitude that of an ordinary distributed-feedback laser of comparable
finesse.

The combination of gain and loss that drives a topological laser
inevitably introduces non-Hermiticity, and a fertile body of work has
arisen at the intersection of topology and non-Hermitian physics
\citep{bergholtz2021exceptional}. Exceptional points
(Sec.~\ref{sec:pt}) become topological objects of their own
\citep{bergholtz2021exceptional,miri2019exceptional}, and the
bulk-boundary correspondence acquires anomalous-skin-effect
modifications that are absent in Hermitian topology. The relevant
super-resonance FOM becomes the winding number of the
spectrum around an exceptional point, which controls the
square-root-enhanced response of the EP sensors of
Refs.~\citep{chen2017exceptional,hodaei2017enhanced} and is itself a
field-agnostic invariant of the joint topological--non-Hermitian
Hamiltonian \eqref{eq:Heff}. A quantum extension was delivered by
Mittal \etal\ \citep{mittal2018topological}, who demonstrated that a
topological photonic lattice can serve as a robust source of
indistinguishable photon pairs --- super-resonance applied to quantum
light.

\subsubsection{Floquet band engineering}\label{sec:floquet}

A driven quantum system whose Hamiltonian is periodic in time,
$H(t+T)=H(t)$, is described by quasi-energies $\varepsilon$ defined
modulo the drive frequency $\Omega=2\pi/T$, in close analogy to crystal
momentum in a spatial lattice \citep{rudner2020band}. The static
band structure is therefore replaced by a Floquet band structure
whose eigenmodes are temporally periodic linear combinations of the
$\{n\}$-photon dressed states of the undriven system. When the drive
amplitude and frequency are tuned to an integer multi-photon resonance,
\(
n_1\omega_1 = n_2\omega_2 = \ldots
\)
between several characteristic frequencies of the unperturbed system,
the corresponding Floquet replicas are forced to coincide, and a new
gap opens at the avoided crossing. The condition is precisely the M3
integer-commensurability locking of Eq.~\eqref{eq:mmr}, transported
from celestial mechanics to a driven quantum system. Lindner, Refael,
and Galitski \citep{lindner2011floquet} showed that the resulting
Floquet bands can carry non-zero Chern numbers even when the
undriven system is topologically trivial --- the Floquet drive is a
constructive route to topological order. Rudner and Lindner
\citep{rudner2020band} provide the canonical review.

\subsubsection{Floquet topological insulators and lasers}\label{sec:floquettopo}

The photonic realisation of the Lindner--Refael--Galitski proposal was
delivered by Rechtsman \etal\ \citep{rechtsman2013photonic} in a
helical-waveguide array: the periodic modulation along the propagation
direction acts as a Floquet drive, and the resulting effective
two-dimensional band structure carries topological invariants that
protect optical edge states from disorder. The photonic
Floquet topological insulator is an M1/M3 hybrid
super-resonance: the radiation channels of the edge state are closed
by the bulk-edge correspondence (M1, topological), but the
\emph{existence} of those bands depends on the integer commensurability
of the Floquet drive with the unperturbed band structure (M3). The
combination explains why the experimental implementation has proved
robust to substantial disorder in the waveguide array --- deliberately
introduced defects included --- far
beyond what an ordinary M1 BIC tolerates.

Photonic time crystals
\citep{galiffi2022photonics,lyubarov2022amplified,lustig2018topological}
extend this construction to optical media whose permittivity is
modulated in time rather than along a propagation direction.
Lyubarov \etal\ \citep{lyubarov2022amplified} predicted amplified
emission inside a photonic time crystal: the Floquet replicas of vacuum
fluctuations interfere constructively at the band edge of the time
crystal, and a coherent gain emerges without an inversion medium. We
stress that this prediction awaits experimental realisation at optical
frequencies, where achieving sufficiently deep and fast permittivity
modulation remains the central obstacle; demonstrated time-modulation
physics is so far confined to microwave and RF metamaterial platforms. The
amplification threshold, $g/\omega_0$, is a hybrid M1/M3 figure of
merit: it falls when the time-modulation amplitude is increased (M3
strengthening the commensurability) and when the spectral overlap of
the replicas with the dispersion-relation gap is improved (M1
strengthening the interference). Lustig, Plotnik \etal\
\citep{lustig2018topological} closed the loop by showing that photonic
time crystals can themselves carry Floquet topological invariants --- a
hybrid (topological)+(time-crystal)+(super-resonance) construction
whose FOMs multiply along three orthogonal axes.

\subsubsection{Connection to M1 and M3, and disorder-robust applications}\label{sec:topom1m3}

We close this subsection by making explicit the cross-mechanism content
that has been implicit throughout. \textit{First}, every topological
super-resonance is, at its mathematical core, an M1 phenomenon: the
radiation channels of the edge or topological-bound state are
suppressed because the symmetry-allowed coupling matrix
$\bm{\Gamma}_{\mathrm{rad}}$ has a zero block dictated by the bulk
invariant. Where ordinary M1 super-resonance (e.g.\ the
Friedrich--Wintgen BIC, Eq.~\eqref{eq:fw_imag}) achieves the zero by
fine-tuning a single off-diagonal element, the topological route
achieves it by a global selection rule that survives any continuous
deformation of the parameters within a topological phase.
\textit{Second}, every Floquet super-resonance is at its core an M3
phenomenon: the Floquet replica band structure exists only because the
drive frequency stands in an integer commensurability with the
underlying time scale of the system \eqref{eq:mmr}, and the same
mathematical structure that produces the Pluto super-resonance
\citep{wan2001pluto} produces the Floquet topological band of
Refs.~\citep{rechtsman2013photonic,rudner2020band,lustig2018topological}.
\textit{Third}, the cross-product M1$\times$M3 --- a Floquet
topological super-resonance --- inherits the robustness of both: it
is protected against generic disorder (topological) and against drift
of the operating point (Floquet locking), and the relevant tolerance
budget (cf.\ Rule 5 of Sec.~\ref{sec:design}) is loosened by orders of
magnitude compared with that of either mechanism in isolation.

The practical consequence is that topological and Floquet
super-resonances should be the platform of choice whenever disorder is
expected to dominate the tolerance budget: ultra-broadband optical
delay lines for quantum information \citep{hafezi2013imaging,
mittal2018topological}, mode-stable on-chip lasers
\citep{harari2018topological,bandres2018topological}, photonic
quantum networks (cf.\ Sec.~\ref{sec:quantumnet}), THz emitters
operating across temperature drifts of $\pm 50\,$K, and proposed
gravitational-wave readout architectures in which the signal-recycling
cavity is replaced by a Floquet-engineered topological waveguide. A
speculative but well-defined target is the on-chip cascade of a
mesoscale resonator (Sec.~\ref{sec:mie}) with an M1 quasi-BIC
metasurface (Sec.~\ref{sec:bic}) and a Floquet
topological waveguide; whether the multiplicative composition of
\eqref{eq:Leff} survives across such heterogeneous elements is an open
question, but the appeal is that the tolerance to disorder would be set
by the topological gap rather than by lithographic precision.

\subsection{Cavity QED, atomic, and nuclear super-resonance}\label{sec:quantum}

\subsubsection{Cavity QED and ultrastrong coupling}

The Jaynes--Cummings model \citep{jaynes1963comparison,larson2024jaynes}
describes the canonical strong-coupling regime, $g\!\gg\!\kappa,\gamma_e$,
in which a single atom and a single cavity photon are reversibly
exchanged. When the coupling strength becomes comparable to the bare
mode frequency, $g/\omega_0 \gtrsim 0.1$, the rotating-wave approximation
fails and the system enters the ultrastrong-coupling regime
\citep{frisk2019ultrastrong,forn2019ultrastrong}. The ground state of
the joint system is no longer the vacuum and the spectrum acquires
non-trivial dressed-state structure, with cooperativities
$C\!\gtrsim\!10^4$ readily achieved in superconducting circuits.
Ultrastrong coupling belongs to the M2 enhancement family: the
coupling itself is the FOM being increased, and at large $g/\omega_0$
both the radiation lifetime and the modal density of states deviate
qualitatively from their Markovian Jaynes--Cummings expectations. It is
surveyed here for contrast with the genuine M1 constructions of cavity
QED --- the suppression of $\kappa_{\mathrm{rad}}$ by photonic-crystal
band gaps and quasi-BIC cavity designs.

Single-quantum-dot strong coupling was demonstrated by Reithmaier \etal\
\citep{reithmaier2004strong} and Englund \etal\
\citep{englund2007controlling}, and cavity QED with single neutral atoms
by Thompson, Rempe, and Kimble \citep{thompson1992observation}. The
broader landscape is reviewed in Walther \etal\
\citep{walther2006cavity}; small-mode-volume photonic-crystal cavities
\citep{lecamp2007very} provide the platform on which the cooperativity
record will be advanced. Across these platforms the figure of merit is
the cooperativity, $C=4g^2/(\kappa\gamma_e)$; an M1 super-resonance
enters as a means of suppressing $\kappa_{\mathrm{rad}}$, while the
collective enhancement of $g$ belongs to the M2 family.

\subsubsection{Dicke super-radiance and single-photon super-radiance}\label{sec:dicke}

Dicke \citep{dicke1954coherence} showed in 1954 that $N$ in-phase
emitters in a volume smaller than a wavelength cubed radiate at a peak
rate $\propto N^2$ instead of $N$; the modern theory was systematised by
Gross and Haroche \citep{gross1982superradiance}. Super-radiance is the
purest example of the M2 enhancement family: the radiative coupling of
the symmetric collective state is coherently \emph{increased}, the exact
opposite of the suppression that defines super-resonance. The flip side
of the same algebra, however, is genuinely M1: the $N-1$ orthogonal
\emph{subradiant} combinations carry vanishing net dipole moment and
decouple from the radiation continuum --- collective dark states whose
radiative width is interference-suppressed in precisely the sense of
\eqref{eq:SRdef}. Bright and dark collective states are thus the M2 and
M1 faces of a single Hamiltonian, which is why the two families are so
persistently conflated; Fig.~\ref{fig:taxonomy}(b) displays the two
faces in the complex-eigenvalue plane, and Fig.~\ref{fig:tolstoy} shows
how the ideal point-emitter dichotomy softens, but survives, at finite
emitter spacing. Solid-state
realisations include super-radiance from quantum dots
\citep{scheibner2007superradiance,cong2016dicke,jordao2025single} and
from colour centres in diamond \citep{angerer2018superradiant}. Meiser
\etal\ \citep{meiser2009prospects} proposed that the collective physics
of the super-radiant line can be exploited to realise lasers
with millihertz linewidths --- a proposal, not yet a demonstrated
device, that we revisit in Sec.~\ref{sec:clocks}.

Single-photon super-radiance \citep{jordao2025single} is a particularly
striking instantiation: even a single photon can be coherently shared
across $N$ emitters such that the collective spontaneous-emission rate
into a specific mode scales as $N$ while the rates into all other modes
remain unchanged. This is M2 operating at the level of a single quantum
excitation, and is the natural building block of an emitter-array
photonic network (Sec.~\ref{sec:quantumnet}).

\subsubsection{Nuclear giant resonances}

Giant resonances in atomic nuclei
\citep{harakeh2001giant,bortignon1998giant,ishkhanov2000giant} are
collective oscillations of neutrons against protons (giant dipole
resonance, GDR), of monopole or quadrupole shape (GMR, GQR), and of
pygmy character at lower excitation energies
\citep{savran2013experimental}. The GDR is a textbook member of the M2
enhancement family:
single-particle dipole transitions, when coherently superposed by the
nuclear interaction, concentrate essentially the entire dipole strength
into a single broad peak whose centroid is approximately
$79 A^{-1/3}$\,MeV in the simplest one-term estimate (modern empirical
fits use a two-term mass dependence \citep{harakeh2001giant}). The strength
of the GDR exhausts the energy-weighted sum rule
(Thomas--Reiche--Kuhn) almost completely, in close analogy to the way
the high-order Mie partial amplitude exhausts the total scattering
cross-section in Eq.~\eqref{eq:mie_intensity} --- a striking cross-field
parallel between optics at micrometre scales and nuclear physics at
femtometre scales. The pygmy dipole resonance
\citep{savran2013experimental} is a precursor to the GDR with comparable
structural interpretation. The same Breit--Wigner machinery
\citep{breit1936capture} that underwrites the nuclear analysis is the
direct ancestor of the generalised Lorentzian \eqref{eq:lorentz} used
throughout this review.

\subsubsection{Magnetic resonance and multi-photon resonances}

Multi-photon resonance phenomena in qubits and many-photon wave packets
\citep{gat2013resonance} provide a clean M3 example: when the
many-photon centre frequency satisfies $n\omega = \omega_{\mathrm{qubit}}$
with integer $n$, the effective $n$-photon Rabi coupling generates a
sharp commensurability peak in the absorption cross-section, displaced from the
fundamental and narrowing as the drive is tuned onto the exact integer
commensurability --- the qubit analogue of the cracked-beam
super-harmonics of Sec.~\ref{sec:acoustics} and of the Floquet replicas
of Sec.~\ref{sec:floquet}.

Magnetic resonance --- the magnetic-moment reorientation first detected in
a molecular beam by Rabi \etal\ \citep{rabi1938new} and observed in
condensed matter by Purcell \etal\ \citep{purcell1946resonance} --- is the
historical antecedent of all coherent-resonance spectroscopies, with the
bare condition $\omega = \gamma B_0$ relating the drive frequency to the
gyromagnetic ratio $\gamma$ and the static field $B_0$, and it remains the
workhorse of quantum-information experiments. We stress that a nuclear line
that is merely \emph{narrow} because spin--lattice relaxation is slow is
not super-resonant by our criterion \eqref{eq:SRdef}: a long $T_2$ lowers
$\gammanr$ but does not touch the radiative bound (cf.\ the second
exclusion of Sec.~\ref{sec:taxonomy}). The genuinely super-resonant
phenomenon of magnetic resonance is \emph{motional narrowing}
\citep{kubo1969generalized}: rapid stochastic modulation of the local
field averages out inhomogeneous broadening so that the observed linewidth
falls \emph{below} the static-disorder bound, $\Delta\omega_{\mathrm{obs}}
\sim \langle\delta\omega^2\rangle\,\tau_c \ll \sqrt{\langle\delta\omega^2
\rangle}$ for short correlation time $\tau_c$. We present motional
narrowing as an instructive \emph{analogy} to M1 rather than an instance
of it: the suppressed channel is inhomogeneous (dephasing) broadening,
not radiative decay, so the strict criterion \eqref{eq:SRdef} does not
apply; but the lesson --- that a broadening channel can be averaged or
interfered away while the underlying coupling stays large --- is the
spin-physics ancestor of the interference-based linewidth suppression
that recurs throughout this review.

\subsection{Magnonic super-resonance: spin waves, cavity magnonics, and dark modes}\label{sec:magnonics}

Magnonics --- the physics and engineering of spin waves in ferromagnetic
or ferrimagnetic media --- has matured over the last decade into a
laboratory platform that hosts essentially every super-resonance
mechanism catalogued in this review. Magnons are bosonic quanta of
collective spin precession that propagate with low loss in
single-crystal yttrium iron garnet (YIG) at microwave frequencies, and
they couple linearly to both microwave photons (Zeeman coupling),
acoustic phonons (magnetoelastic coupling), and optical photons
(Faraday and inverse-Faraday effects). This four-way coupling makes
magnonic systems uniquely positioned to mediate cross-domain quantum
information \citep{lachancequirion2019hybrid,rameshti2022cavity,
yuan2022quantum}, and the same coupling underwrites a series of
super-resonance constructions --- BIC-type dark modes, ultrastrong
cavity-magnon polaritons, super-radiant magnonic ensembles, and
single-magnon-resolved quantum experiments --- each of which is, in
our framework, an instance of M1, M2, or their composition.

\subsubsection{Spin waves and magnetostatic modes in ferromagnets}\label{sec:spinwaves}

The starting point is Kittel's 1948 derivation of the uniform-precession
mode of a ferromagnet \citep{kittel1948theory}, whose frequency,
$\omega_K = \gamma\sqrt{(H_0+(N_y-N_x)M)(H_0+(N_z-N_x)M)}$ for an
ellipsoidal sample of demagnetising tensor $\{N_i\}$ and saturation
magnetisation $M$, statically magnetised along the $x$ axis by the
field $H_0$, defines the ferromagnetic resonance (FMR) line that
underlies all subsequent magnonics. The natural extension to spatially
varying spin precessions yields the magnon dispersion
$\omega(\mathbf k)$, which interpolates smoothly between
magnetostatic surface and volume modes (Damon--Eshbach type) at long
wavelengths and exchange-dominated propagation
$\omega \propto A k^2$ at short wavelengths. In single-crystal YIG the
exchange stiffness and dipolar coupling are well-characterised across
nine orders of magnitude in wavevector
\citep{chumak2015magnon,yuan2022quantum}, and the Gilbert damping
parameter $\alpha\!\sim\!10^{-5}$ is the lowest of any known
ferromagnet, conferring magnon $Q$-factors that approach $10^4$ at
gigahertz frequencies.

A distinctive feature of magnonic systems is the coexistence of
several propagating bosonic species --- magnons, photons, and phonons
--- whose interaction generates hybrid quasi-particles. The
magnetoelastic coupling between magnons and phonons of matched
wavevector produces the \emph{magnon polaron} of Sukhorukova \etal\
\citep{sukhorukova2022superresonant}; the Zeeman coupling to a cavity
mode produces the \emph{cavity-magnon polariton} of Huebl, Tabuchi, and
Zhang \citep{huebl2013high,tabuchi2014hybridizing,zhang2014strongly};
and the simultaneous presence of both produces the
magnon--photon--phonon trilinear polariton that is the canonical
hybrid quantum platform of cavity magnonics
\citep{lachancequirion2019hybrid,rameshti2022cavity}. Each of these
hybrid quasi-particles inherits a non-Hermitian eigenvalue structure
identical to \eqref{eq:Heff}, and each can be tuned to satisfy the
super-resonance criterion \eqref{eq:SRdef}.

\subsubsection{Cavity magnonics and the strong-coupling regime}\label{sec:cavmagnon}

The theoretical proposal of Soykal and Flatt\'e
\citep{soykal2010strong} established that a YIG sphere placed at an
anti-node of a microwave cavity field is the magnonic equivalent of a
single emitter in an optical cavity: the coherent coupling
$g/2\pi\!\sim\!\sqrt{N}\,g_0\!\sim\!100\,$MHz scales as the square root
of the number of unpaired spins, $N\!\sim\!10^{18}$ in a millimetre-scale
sphere, while the magnon and cavity decay rates remain in the
megahertz range. The strong-coupling regime, $g\!\gg\!\{\kappa,\gamma\}$,
is therefore reached effortlessly at room temperature, and was first
demonstrated experimentally by Huebl \etal\
\citep{huebl2013high}, Tabuchi \etal\ \citep{tabuchi2014hybridizing},
and Zhang \etal\ \citep{zhang2014strongly} in 2013--2014.
Cooperativities $C\!=\!4g^2/(\kappa\gamma_m)\!\gtrsim\!10^5$ have since
become routine at millikelvin temperatures
\citep{lachancequirion2019hybrid}, with $C\!\gtrsim\!10^4$ persisting
at room temperature in the most carefully prepared samples.

The cavity-magnon polariton combines an M2 ingredient
--- the coherent $\sqrt N$ scaling
of the collective spin-photon coupling, identical to the Dicke
construction \citep{dicke1954coherence} of Sec.~\ref{sec:quantum} ---
with a genuine
M1 ingredient: the radiation-channel suppression of the polariton
linewidth at the avoided crossing, the Friedrich--Wintgen
condition of Eq.~\eqref{eq:fw_imag} applied to a coupled microwave
photon and a uniform-precession magnon. The FOM is the
cooperativity. A natural --- and so far purely speculative --- next
target would nest a cavity-magnon polariton inside a second cavity (a
``magnonic Malara configuration'', cf.\ Sec.~\ref{sec:pt}) in the hope
of reproducing the multiplicative composition of Eq.~\eqref{eq:Leff};
whether the composition survives in the magnonic setting is an open
design question, not an established result. The comprehensive review of cavity magnonics by Rameshti
\etal\ \citep{rameshti2022cavity} surveys the state of the art and
the open problems.

\subsubsection{Magnon dark modes and BIC-type magnonic super-resonance}\label{sec:darkmagnon}

\begin{figure}[t]
\centering
\includegraphics[width=0.9\linewidth]{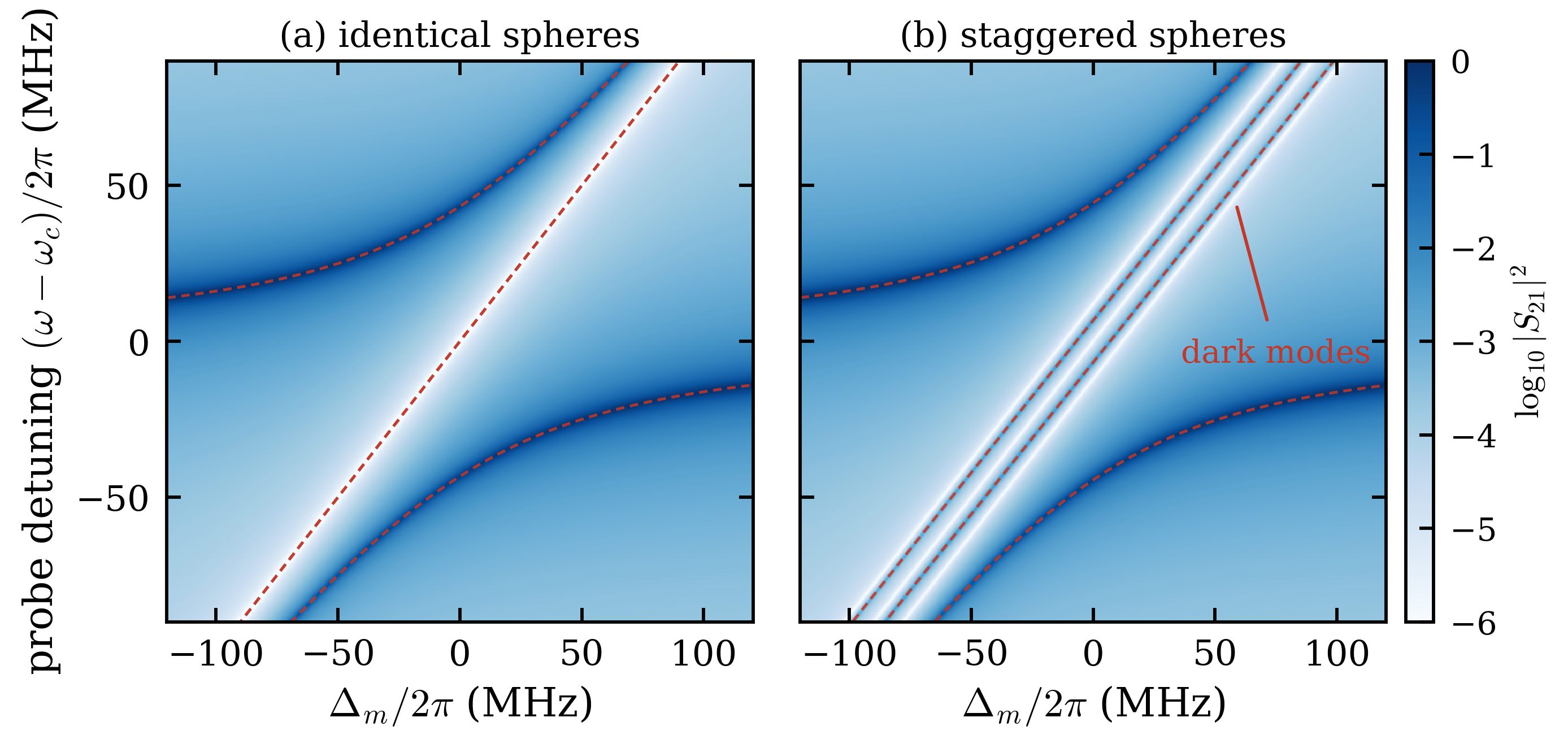}
\caption{Bright and dark collective modes of cavity magnonics, computed
from the input--output transmission of a single-mode cavity coupled to
$N=3$ magnon modes,
$S_{21}\propto[\,\mathrm{i}(\omega_c-\omega)+\kappa/2+
\sum_j g^2/(\mathrm{i}(\omega_{m,j}-\omega)+\gamma_m/2)\,]^{-1}$, with
parameters representative of YIG-sphere experiments
($g/2\pi=25$\,MHz, $\kappa/2\pi=2.5$\,MHz, $\gamma_m/2\pi=1$\,MHz;
dashed lines: real eigenfrequencies of the corresponding
$(N{+}1)\times(N{+}1)$ Hamiltonian \eqref{eq:Heff}).
(a)~Identical spheres: only the symmetric (bright) combination couples,
producing a cavity-magnon polariton anticrossing of collective
splitting $2\sqrt{N}\,g$ --- the M2 ingredient. (b)~Sphere frequencies
staggered by $\pm12$\,MHz: the $N-1=2$ dark combinations acquire a weak
residual coupling and appear as ultra-narrow lines inside the polariton
gap --- the M1 dark modes demonstrated with YIG-sphere arrays
\citep{zhang2015magnon}.}
\label{fig:magnon}
\end{figure}

A defining feature of multi-mode cavity-magnonics is the existence of
\emph{dark} collective modes that decouple from the radiation continuum
by a destructive-interference selection rule. Zhang, Zou, Zhu \etal\
\citep{zhang2015magnon} demonstrated that an array of YIG spheres
coupled to a common cavity mode supports $N-1$ dark modes for $N$
spheres: the symmetric combination couples maximally to the cavity
(the bright magnon polariton), while the $N-1$ orthogonal
combinations carry zero net dipole moment with respect to the cavity
field and therefore have radiative linewidths suppressed by orders of
magnitude. The construction is a direct discrete analogue of the
Friedrich--Wintgen BIC \eqref{eq:fw_imag} and the Tolstoy acoustic
array \citep{tolstoy1986superresonant}; it is an M1
super-resonance whose dark eigenmode constitutes a \emph{magnonic
gradient memory}, capable of storing a coherent microwave excitation
for times limited only by the bulk magnon dephasing rather than by
the cavity escape rate. Figure~\ref{fig:magnon} reproduces the
bright/dark anatomy from the input--output theory of the coupled
system: with identical spheres only the symmetric combination is
visible, as an anticrossing of collective splitting $2\sqrt{N}g$; a
slight stagger of the sphere frequencies makes the $N-1$ dark modes
weakly visible as ultra-narrow lines threading the polariton gap ---
the experimental signature by which they are identified and addressed.

The Sukhorukova \etal\ leaky-surface-magnon-polaron BIC
\citep{sukhorukova2022superresonant} is a continuous-spectrum
counterpart of the Zhang--Tang dark mode: hybridisation between
inhomogeneous-exchange and magnetoelastic interactions generates a
discrete state embedded in the continuum of bulk magnetoelastic
radiation, with a radiative width that vanishes in the
dissipation-free limit. It is, to our knowledge, the first work to use
the term \emph{super-resonant} for a magnonic state in the sense of
our operational definition, and it suggests that a fully magnonic
analogue of the BIC metasurface programme of Sec.~\ref{sec:bic} is
within experimental reach. The FOM is the magnon-polariton
linewidth, which sets the upper bound on the coherence time of any
quantum information protocol implemented in the dark mode.

\subsubsection{Quantum magnonics and single-magnon detection}\label{sec:quantummagnon}

The same magnetic-dipole coupling that produces the
classical cavity-magnon polariton can, when combined with a
superconducting qubit dispersively coupled to the cavity, project
the YIG sphere into a single-magnon Fock state. Lachance-Quirion
\etal\ \citep{lachance2017resolving} resolved individual quanta of
collective spin excitation in a millimetre-sized YIG sphere through
the dispersive shift they induce on a coupled superconducting qubit
--- the magnonic analogue of the dispersive photon-number readout of
circuit QED. A follow-up experiment by the same group
\citep{lachance2020entanglement} demonstrated entanglement-based
single-shot detection of a single magnon with detection efficiency
above $0.7$, completing the cavity-magnonics counterpart of the
single-photon experiments of cavity QED \citep{walther2006cavity}.

These experiments confirm the M2 character of
cavity magnonics in the single-quantum regime: the collective spin
excitation behaves as a single bosonic mode whose super-resonant
coupling to the cavity is dominated by the symmetric combination of
$10^{18}$ spin contributions, with the orthogonal combinations
manifest only as a discrete spectrum of dark modes
\citep{zhang2015magnon}. The Yuan \etal\ Physics Reports review
\citep{yuan2022quantum} elaborates on the quantum-information
applications, which include magnon-mediated entanglement of
otherwise-disjoint quantum systems, microwave-to-optical quantum
transduction, and remote-state preparation between superconducting
qubits.

\subsubsection{Magnonics as a super-resonance platform: applications}\label{sec:magnonicsapp}

Magnonics combines several practically attractive features into a
single experimental platform: long magnon coherence (microseconds in
YIG at millikelvin temperatures), simultaneous coupling to photons in
three different ranges (microwave, terahertz, optical), and the
ability to operate at temperatures from millikelvin to room
temperature. The super-resonance applications that follow from this
combination are best organised around the four-way magnon coupling.

\textit{Microwave--optical transduction.}\enspace The simultaneous
coupling of a YIG magnon to a microwave cavity (Zeeman) and to an
optical mode of a co-located whispering-gallery resonator (Faraday)
realises a coherent transducer between superconducting-qubit
microwave photons and telecom-band optical photons --- the missing
link between local quantum processors and a fibre-optic quantum
internet. Reported conversion
efficiencies remain at the percent level
\citep{lachancequirion2019hybrid,yuan2022quantum}. One can speculate
that combining a long-lived M1 dark magnon mode \citep{zhang2015magnon}
with a strongly coupled cavity-magnon polariton
\citep{tabuchi2014hybridizing,zhang2014strongly} would raise the
efficiency substantially; we flag this as an unquantified conjecture
rather than a prediction --- no analysis to date establishes either the
attainable efficiency or a specific threshold.

\textit{Magnonic logic and computing.}\enspace Spin waves can be used
as the carrier of information in a magnonic logic circuit
\citep{chumak2015magnon}, with the advantage of nanoscale wavelengths
and dissipationless interference. A magnonic BIC, in the spirit of
\citep{sukhorukova2022superresonant}, would provide a long-lived
register element whose FOM is the lifetime--energy product
of the dark mode --- an M1 super-resonance figure of merit identical
to that of a photonic quasi-BIC \citep{koshelev2018asymmetric}.

\textit{Magnon BEC and macroscopic quantum phenomena.}\enspace
Bose--Einstein condensation of magnons has been demonstrated in YIG
films at room temperature under parametric pumping
\citep{demokritov2006bose}; the resulting magnon condensate is a
macroscopically occupied single eigenmode. The condensation transition
is a thermodynamic phenomenon distinct from Dicke super-radiance, though
both produce macroscopic single-mode coherence; the relevant figure of
merit is the second-order coherence
$g^{(2)}(0)$, which approaches unity in well-prepared samples and
constitutes a magnonic counterpart of the photon laser.

\textit{Quantum sensing.}\enspace The ability to resolve single
magnons \citep{lachance2017resolving,lachance2020entanglement} converts
the magnonic platform into a quantum sensor for magnetic fields,
mechanical vibrations, and spin-wave-coupled phenomena including dark
matter searches. The sensitivity is set by the noise on the
magnon-number measurement; long-lived M1 dark modes lengthen the
interrogation time and thereby improve the standard-quantum-limit
sensitivity.

We notice that the cross-pollination of cavity magnonics with the
photonic-BIC and topological-photonics programmes of
Secs.~\ref{sec:bic}, \ref{sec:topological} has only just begun, and
that the most striking near-term opportunity is the construction of a
\emph{magnonic Floquet topological insulator}: a parametrically driven
YIG film in which the Floquet replicas of the Damon--Eshbach surface
mode acquire non-zero Chern numbers. We emphasise that this is a
\emph{prediction}, not an existing result. In the language of this
review the resulting structure would combine M1 (topological closing of
radiation channels) with M3 (Floquet commensurability) in the magnetic
sector, extending the topological-laser construction of Bandres \etal\
\citep{bandres2018topological} to the magnonic domain. The conceptual
unification of these four communities --- photonic BIC, cavity
magnonics, topological photonics, and quantum magnonics --- is the
cleanest current example of the design-rule transfer that motivates
the present review.

\subsection{Plasma wave--particle resonance and current drive (M3 family)}\label{sec:plasmaSR}

If celestial mechanics is the historical home of the M3 locking family,
plasma physics is its most consequential modern application. A magnetised
plasma is a kinetic system supporting a continuous spectrum of waves
whose phase velocities can match either the parallel velocity of the
particles (Landau resonance) or an integer multiple of the cyclotron
frequency (cyclotron resonance) --- in either case the
energy-momentum exchange is a wave--particle resonance whose
effective quality factor scales not as a cavity finesse but as the
inverse collision frequency, which in a hot tokamak plasma can exceed
$10^{6}$. The commensurability condition has the same \emph{form} as the
one that locks Pluto into a 3:2 resonance with Neptune
\citep{wan2001pluto}, and the phase-space island structure transfers
between the two problems. The applications --- non-inductive current
drive, electron- and ion-cyclotron heating, and selective extraction of
alpha-particle energy from a burning plasma --- are the engineering core
of the international fusion-energy programme. We emphasise, consistent
with Sec.~\ref{sec:taxonomy}, that these are M3 locking phenomena, sharp
because collisional decorrelation is slow; they do not satisfy the strict
super-resonance criterion \eqref{eq:SRdef}, and we survey them for the
transfer of phase-locking technique, not as instances of the definition.

\subsubsection{Wave--particle resonance: Landau and cyclotron mechanisms}\label{sec:landau}

The starting point is Landau's 1946 discovery
\citep{landau1946vibrations} that an electrostatic wave in a
collisionless plasma can damp without dissipation: the wave loses
energy to the small fraction of particles whose velocity matches its
phase velocity,
\begin{equation}
\omega - k_\| v_\| = 0,
\label{eq:landau}
\end{equation}
even though the bulk plasma is undisturbed. The damping rate is set
by the slope of the distribution function at $v_\| = \omega/k_\|$,
and it persists in the long-time limit; the apparent paradox between
Landau's reversible Vlasov equation and irreversible damping was
resolved by O'Neil \citep{oneil1965collisionless}, who showed that the
small fraction of resonant particles becomes trapped in the wave
potential, oscillating around the phase-locked point with bounce
frequency
$\omega_B = (e k E_0 / m)^{1/2}$.
Read against Sec.~\ref{sec:M3}, Eq.~\eqref{eq:landau} is a one-channel
M3 condition: the integer commensurability is $1:1$ between the wave
phase and the particle parallel motion.

In the presence of a static magnetic field $\mathbf B_0$ the resonance
condition generalises to
\begin{equation}
\omega - k_\| v_\| = n\,\Omega_c, \qquad n\in\mathbb Z,
\label{eq:cycres}
\end{equation}
where $\Omega_c = eB_0/m$ is the cyclotron frequency
\citep{stix1992waves,brambilla1998kinetic}. The integer $n$ counts the
harmonic of the gyromotion at which the wave is in phase with the
particle: $n=0$ recovers Landau damping, $n=\pm 1$ corresponds to the
fundamental cyclotron resonance, and $|n|>1$ to higher harmonics
accessible at finite Larmor radius. Equation~\eqref{eq:cycres} is the
canonical M3 condition of plasma physics; it spans an entire
two-parameter family $(n, k_\|/k_\perp)$ of integer-locked channels,
each with an effective quality factor
$Q_\mathrm{wp} \sim \omega/\nu_\mathrm{eff}$, where $\nu_\mathrm{eff}$
is the effective decorrelation rate of the resonant particle population
(collisional in cold plasmas, stochastic in turbulent plasmas
\citep{karney1979stochastic}). In a magnetic-confinement fusion plasma
$\nu_\mathrm{eff}\!\sim\!10^{-6}\omega$, placing the wave--particle
resonance among the sharpest resonances known --- sharp, as the
exclusion criteria of Sec.~\ref{sec:taxonomy} insist, because the bare
decorrelation is slow, not because a radiative channel has been
interfered away.

\subsubsection{Lower-hybrid current drive}\label{sec:lhcd}

The lower-hybrid wave is an obliquely propagating electromagnetic wave
whose frequency lies between the ion and electron cyclotron
frequencies,
\(
\omega_\mathrm{LH}^2 \approx \Omega_{ci}\Omega_{ce}
\)
\citep{stix1992waves}. Karney \citep{karney1979stochastic} showed that
when the wave amplitude exceeds the stochasticity threshold, ions are
heated by perpendicular cyclotron resonances arranged in a stochastic
ladder of harmonics $n$; the heating rate exceeds the linear-Landau
estimate by an order of magnitude. Fisch \citep{fisch1987theory}
subsequently identified that the same wave, when launched with a
unidirectional parallel wavenumber spectrum, can be Landau-absorbed
preferentially by suprathermal electrons, thereby driving a
non-inductive plasma current. The FOM is the
current-drive efficiency
\(
\eta_\mathrm{CD} = \langle J/n_e P_\mathrm{rf}\rangle,
\)
which in the Fisch theory is maximised when the wave spectrum is
narrow in $v_\|$ and resonant with electrons of velocity
$v_\|/v_{th,e}\!\sim\!3$. The theory secured Fisch the 1992 APS Award
for Excellence in Plasma Physics Research and, later, the 2005 Maxwell
Prize, and remains the conceptual backbone of every steady-state
tokamak proposal.

Experimentally, lower-hybrid current drive has been pushed to the
megawatt scale in Alcator C-Mod, JT-60U, Tore Supra, and EAST, where
current-drive efficiencies of order $\eta_\mathrm{CD}\!\sim\!3\times
10^{19}\,\mathrm{A\,W^{-1}\,m^{-2}}$ are routinely reported
\citep{bonoli2014review,wesson2011tokamaks}. The full plasma current
on Tore Supra has been sustained non-inductively for several minutes by
lower-hybrid alone --- a locked wave--particle resonance maintained
against collisional decorrelation by continuous RF injection.

\subsubsection{Electron-cyclotron resonance heating and current drive}\label{sec:ecrh}

The electron-cyclotron range exploits the fundamental and
second-harmonic of $\Omega_{ce}$, typically $100$--$200\,$GHz in a
fusion-scale magnetic field of several tesla, to deposit power
locally and selectively in the plasma volume
\citep{prater2004heating,wesson2011tokamaks}. Because the resonance
condition \eqref{eq:cycres} is satisfied on a thin layer where
$B_0(\mathbf r) = m\omega/(en)$, electron-cyclotron resonance
heating (ECRH) is an inherently spatially localised tool: a launched
beam from a gyrotron source intersects the resonance layer and
deposits all of its power within $\sim\!1\,\mathrm{cm}$, which is
unique among RF heating schemes. The same selectivity converts ECRH
into electron-cyclotron current drive (ECCD), with current-drive
efficiencies that approach the Fisch theoretical bound when an
asymmetric wave spectrum is launched.

In modern tokamak operation ECRH/ECCD is used not only to heat and
drive current but to stabilise magnetohydrodynamic instabilities ---
the neoclassical tearing mode in particular is suppressed by depositing
ECCD precisely on the resonant magnetic surface where the mode
nucleates. This is resonant feedback: a wave-driven local resonance is
engineered to extinguish a separate MHD instability, and the design
problem is the same coupled-eigenvalue exercise \eqref{eq:Heff}
familiar from photonics. The ITER plan calls for
$24\,$MW of installed ECRH at $170\,$GHz for precisely these
purposes.

\subsubsection{Ion-cyclotron range of frequencies}\label{sec:icrf}

The ion-cyclotron range, $20$--$80\,$MHz in a fusion-scale plasma,
provides heating through fundamental and harmonic ion resonances
\(
\omega = n\,\Omega_{ci}
\) \citep{stix1992waves,brambilla1998kinetic}. A characteristic
feature is mode conversion: a launched fast magnetosonic wave can
convert at a layer of degenerate dispersion (the ion-ion hybrid
resonance) into a kinetic Alfv\'en wave or an ion-Bernstein wave,
which then transfers energy to the minority-ion species through the
$n=1$ cyclotron resonance. In a deuterium--tritium plasma with a
small hydrogen minority, $80\%$ of the launched ICRF power can be
absorbed by the minority hydrogen on the resonance layer; the
hot-minority distribution then transfers its energy to the
deuterium and tritium thermals through Coulomb collisions
\citep{wesson2011tokamaks}. The FOM is the
hot-minority-tail temperature divided by the bulk temperature, which
in JET ICRF experiments reaches $\sim\!200\,\mathrm{keV}$ from a
$\sim\!10\,$keV bulk --- a twenty-fold resonant tail
enhancement.

\subsubsection{Alpha channeling and the burning plasma}\label{sec:alpha}

A burning fusion plasma generates $3.5\,$MeV alpha particles whose
free-streaming losses degrade confinement and whose collisional
slowing-down heats both electrons and ions indiscriminately. Fisch
and Rax \citep{fisch1992interaction} proposed that an appropriately
phased lower-hybrid wave can extract energy from the alpha population
\emph{selectively}, transferring it to the bulk ions via the wave
rather than via direct collisions, while at the same time driving the
fast alphas radially outward to be removed before they collisionally
heat the electrons. The mechanism is a sequence of locked M3
transitions in two-dimensional gyration--orbit phase space, with
the wave acting as a transducer that diagonalises the
energy--position transport tensor of the alpha distribution.

If \emph{alpha channeling} can be made to work at reactor scale, the
energy economics of a magnetic-fusion power plant improve
substantially, because the alpha energy is
deposited where it is most useful (in the ions, not the electrons)
and the alpha particles are removed before they thermalise. The
construction amounts to a Maxwell-demon-like resonant sorting in
classical phase space \citep{fisch1992interaction}; only
the recent maturation of high-power lower-hybrid sources
\citep{bonoli2014review} has made reactor-scale experimental tests
plausible.

\subsubsection{Integrated tokamak scenarios and multi-channel M3}\label{sec:plasmaintegrated}

Any integrated burning-plasma scenario for ITER or DEMO requires the
\emph{simultaneous} operation of LHCD, ECRH/ECCD, ICRF, and (where
available) alpha channeling. This is a
multi-channel M3 locking problem: the integer
commensurabilities $\omega = n\Omega_c + k_\|v_\|$ must be satisfied
for several different harmonics and species simultaneously. The design
optimisation reduces to a multi-channel commensurability condition on the
wave--particle phase space, with each channel obeying \eqref{eq:cycres}
and the joint performance set by the worst-coupled channel.

We notice that the integer commensurability that locks the
Galilean satellites into the Laplace resonance
\citep{barnes2011laplace,paita2018element} has the same canonical form
\eqref{eq:mmr} as the wave--particle conditions above, with the wave
packets playing the role of tidal forcing and the gyrating ions playing
the role of the satellites. The shared form is a genuine and useful
transfer of dynamical-systems technique across a $10^{12}$-fold
separation in length scale --- though it is a statement about the M3
locking family, not about super-resonance in the strict sense of
\eqref{eq:SRdef}.

\subsection{Emerging frontiers}\label{sec:emerging}

\subsubsection{Time-modulated media beyond photonic time crystals}

The Floquet replicas and topological band structure of time-modulated
photonic media are surveyed in Sec.~\ref{sec:floquettopo} as part of
the consolidated discussion of topological and Floquet super-resonance.
Beyond photonic time crystals proper
\citep{galiffi2022photonics,lyubarov2022amplified,lustig2018topological},
the time-varying-media programme also encompasses synthetic-frequency
gauge fields and frequency-dimension lattices, whose discrete
translational symmetry along the time axis encodes the same M3
integer-commensurability content as the Laplace resonance
\citep{barnes2011laplace} but realised electromagnetically. The
relevant FOM is the gain bandwidth of the time-modulated medium,
which grows with both the modulation depth (the M3 ingredient) and the
spectral overlap of the dispersion replicas; as a gain figure it
belongs, in our taxonomy, to the amplification family.

\subsubsection{Cavity optomechanics and mesoscopic back-action}

Cavity optomechanics \citep{kippenberg2007cavity} is a platform that
combines M2-type coherent enhancement with M1-type channel selection: a
high-$Q$ optical cavity is coupled
to a mechanical mode of frequency $\Omega_m$ through radiation pressure.
The single-photon coupling $g_0$ is parametrically enhanced by the
intracavity photon number $\bar n_{\mathrm{cav}}$ to a field-enhanced
coupling $g=g_0\sqrt{\bar n_{\mathrm{cav}}}$, and the relevant FOM is
the multiphoton cooperativity
$C=4g^2/(\kappa\gamma_m)=4g_0^2\bar n_{\mathrm{cav}}/(\kappa\gamma_m)$,
identical in form to the cavity-QED cooperativity of
Sec.~\ref{sec:quantum}. Two distinct super-resonance ingredients combine
here. The M2 ingredient is the $\sqrt{\bar n_{\mathrm{cav}}}$ Bose
enhancement of the coupling, which is the optomechanical counterpart of
the collective $\sqrt N$ Dicke scaling \citep{dicke1954coherence}; the
M1 ingredient is the resolved-sideband condition $\kappa<\Omega_m$, under
which the optical density of states is engineered so that only the
red (cooling) or blue (amplifying) motional sideband couples to the
cavity, suppressing the unwanted radiation channel by interference much
as in a quasi-BIC. In this regime the field has achieved sideband cooling
of a mechanical mode to its quantum ground state, optomechanically
induced transparency (a back-action analogue of the EIT super-resonance
of Sec.~\ref{sec:plasmonics}), ponderomotive squeezing of light below the
shot-noise limit, back-action-evading measurement, and optomechanical
entanglement; these developments postdate the early perspective of
Ref.~\citep{kippenberg2007cavity} and are reviewed in
Ref.~\citep{aspelmeyer2014cavity}. Because the mechanical
sidebands form a ladder of replicas spaced by $\Omega_m$, a periodically
modulated optomechanical cavity additionally accesses the M3
commensurability family, placing optomechanics --- like cavity
magnonics (Sec.~\ref{sec:magnonics}) --- at the confluence of all three
classes in our taxonomy and making it a natural transducer between
optical, microwave, and mechanical degrees of freedom.

\subsubsection{Gravitational-wave detection and ultralight dark matter}

Gravitational-wave detectors employ high-finesse Fabry--P\'erot
arm cavities (finesse $\mathcal{F}\!\sim\!450$)
\citep{aasi2015advanced,abbott2016observation} with Pound--Drever--Hall
locking \citep{drever1983laser}. Signal recycling and power recycling
compose a coupled-cavity system whose
sensitivity to a gravitational strain $h$ is set by the full transfer
function of the coupled cavities (Sec.~\ref{sec:keyeqs}); quantum
squeezing, now in routine use, further reduces the quantum noise below
the shot-noise level. The detection of ultralight bosonic dark
matter exploits black-hole super-radiance
\citep{brito2020superradiance,press1972rotating} --- an amplification
phenomenon in our taxonomy --- as well as resonant
electromagnetic conversion in neutron-star magnetospheres
\citep{mcdonald2024resonant}, both
amplified over astrophysical times. We develop these applications
in Sections~\ref{sec:gravitational}
and~\ref{sec:darkmatter}.

\section{Applications}\label{sec:applications}

The mechanism of Section~\ref{sec:definition} and the manifestations
catalogued in Section~\ref{sec:manifestations} converge on a number of
applications. We treat first the domains in which interference-suppressed
radiative decay --- or, where noted, one of its M2/M3 relatives --- has
delivered a concrete experimental demonstration: refractometric and
biochemical sensing, sub-diffraction near-field imaging, extreme-field
photonics, gravitational-wave readout, and the integrated mesoscale
(``mesotronic'') platform. We then collect, in a single clearly-labelled
subsection, the further directions that are at present \emph{proposals or
conjectures} rather than demonstrated devices. For each domain we identify
the relevant class, state the figure of merit, and --- for the demonstrated
domains --- the benchmark result. We are deliberately careful not to
present a projected performance as an achieved one.

\subsection{Refractometric, biochemical, and vibrational sensing}\label{sec:sensing}

\begin{figure}[t]
\centering
\includegraphics[width=0.58\linewidth]{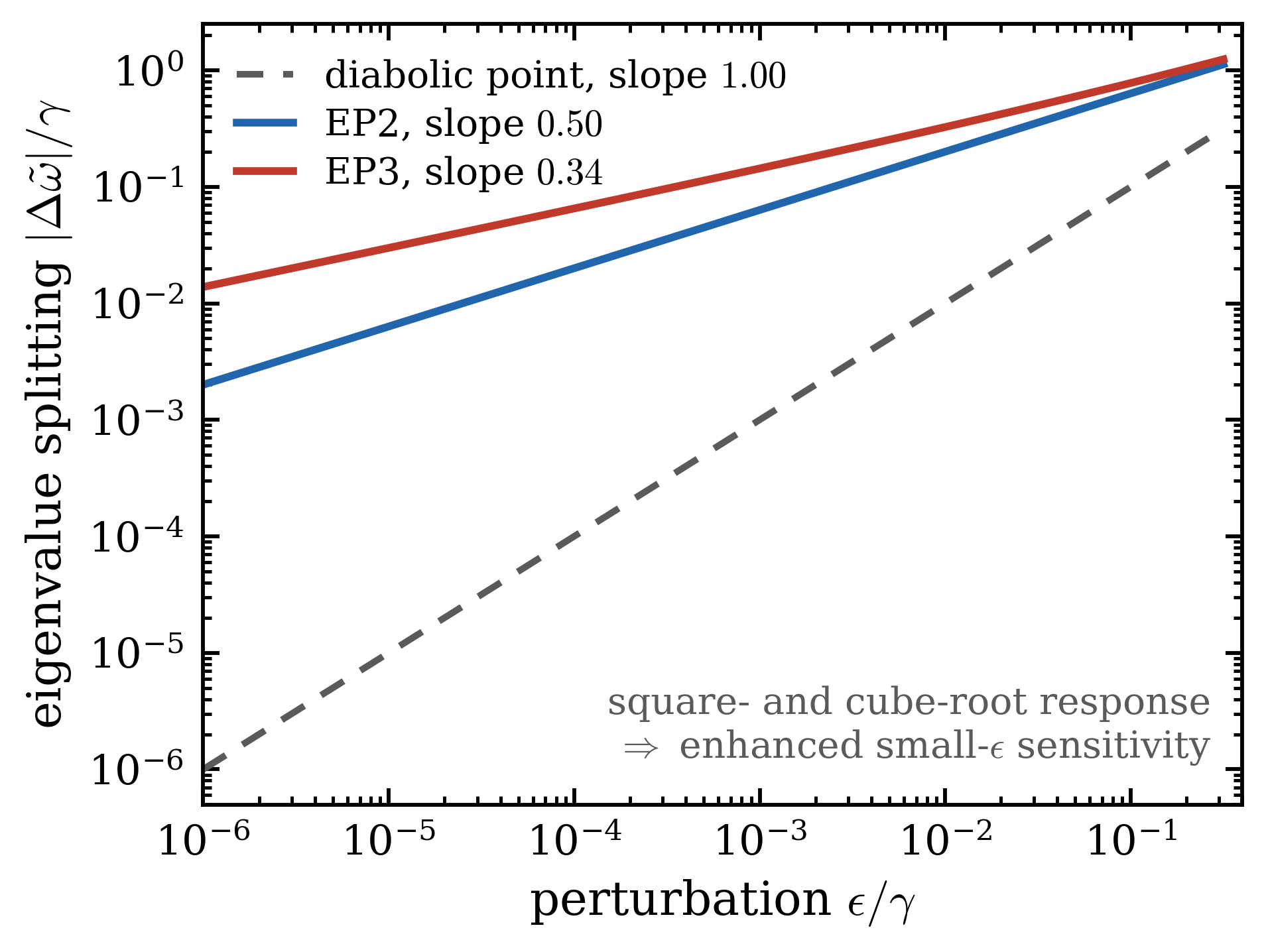}
\caption{Exceptional-point sensing: eigenvalue splitting versus
perturbation strength $\epsilon$, computed for the PT-symmetric dimer
$H_2$ (EP2 at $g=\gamma$), the PT-symmetric trimer $H_3$ (EP3 at
$g=\gamma/\sqrt2$, the configuration of the ternary microring
experiment \citep{hodaei2017enhanced}), and, for reference, an
uncoupled (diabolic) crossing. Log--log slopes fitted over the
small-$\epsilon$ tail are $0.50$, $0.34$, and $1.00$: the square- and
cube-root branch responses of
Refs.~\citep{chen2017exceptional,hodaei2017enhanced}, whose diverging
small-$\epsilon$ slope underlies EP-enhanced refractometric and
nanoparticle sensing (noise permitting; see the caveats discussed in
Sec.~\ref{sec:pt}).}
\label{fig:ep}
\end{figure}

Building on the mechanisms catalogued in
Sections~\ref{sec:mie}--\ref{sec:pt}, the most mature application is
refractometric sensing. The FOM is the
spectral slope $d\lambda_{\mathrm{res}}/dn_{\mathrm{env}}$, which scales
as $Q/\Vmode$ of the underlying mode. Benchmark: WGM cavities reach
single-molecule sensitivity \citep{vollmer2008whispering,
armani2003ultrahigh}; high-order-Fano microspheres and magnetic-WGM
spheres deliver $d\lambda/dn\!\sim\!10^3\,\mathrm{nm/RIU}$
\citep{geints2024magnetic,minin2022extreme,minin2024discovery,
gandhi2020ultrahigh}, ten times the saturation of conventional WGM
sensors. Exceptional-point sensors
\citep{chen2017exceptional,hodaei2017enhanced} achieve the same end with
a different figure of merit --- the square-root branch splitting near
the EP rather than $Q$ itself --- whose slope diverges as the
perturbation vanishes. Figure~\ref{fig:ep} computes the scaling from
the canonical PT-symmetric dimer and trimer: the eigenvalue splitting
responds to a perturbation $\epsilon$ as $\epsilon^{1/2}$ at a
second-order EP and as $\epsilon^{1/3}$ at a third-order EP, against
the linear response of an ordinary (diabolic) crossing --- three orders
of magnitude of splitting enhancement at $\epsilon/\gamma=10^{-6}$ for
the dimer alone. The enhancement is of the \emph{splitting}; whether it
survives as measurement precision depends on the noise budget of the
specific readout (Sec.~\ref{sec:pt})
\citep{langbein2018no,lau2018fundamental}.

 Quasi-BIC refractometric sensors
\citep{koshelev2018asymmetric,azzam2021photonic} reach sensitivities up
to $\sim\!10^3\,\mathrm{nm/RIU}$ with $Q\!>\!10^3$ in CMOS-compatible
silicon metasurfaces. Vibrational fingerprinting (CEIRA, SEIRA) is the same
FOM moved to mid-infrared, with all-dielectric Mie-resonant
arrays \citep{kuznetsov2016optically,koshelev2019meta} now replacing the
historical plasmonic Fano prototype \citep{lukyanchuk2010fano,
lassiter2010fano,khurgin2015deep}. The frontier is multi-analyte parallel
sensing on a single quasi-BIC chip, exploiting Eq.~\eqref{eq:SRdef} to
trade modal volume against bandwidth at the design stage rather than
post-fabrication.

\subsection{Sub-diffraction imaging, super-resolution, and metalenses}\label{sec:imaging}

Resonant encoding of sub-wavelength information (Sec.~\ref{sec:keyeqs})
delivers the central application: a resonant metalens built from a dense
array of coupled subwavelength resonators converts evanescent detail
into temporally coded far-field signals. Benchmark:
Ref.~\citep{li2014resonant} showed, in a numerical feasibility study,
$\lambda/30$-scale imaging at
microwave frequencies by time reversal of the resonant response ---
notably, under the name \emph{super-resonance}, yet one more independent
coinage of the term; the
photonic-hook construction of
Minin and Minin \citep{minin2017diffractive,minin2022janus,
tang2024unusual} positions a $\sim\!\lambda/3$ off-axis focal spot, a
candidate sub-diffraction probe for near-field lithography
\citep{minin2024hook}. Acoustic
super-resolution schemes obey the same logic with $\lambda$ reinterpreted
as the elastic wavelength \citep{maznev2015upholding}. In every case the
gain is scheme-specific --- there is no universal $1/Q$ law for spatial
resolution --- but the narrow M1 modes are the resource on which the
encoding rests.

\subsection{Extreme-field photonics, SERS, and nonlinear conversion}\label{sec:extreme}

The internal-field enhancement \eqref{eq:Eintensity}, computed in
Sec.~\ref{sec:mie} for an idealised lossless sphere, gives
$|E|^2/|E_0|^2\!\sim\!10^6$,
$|H|^2/|H_0|^2\!\sim\!10^7$ for a single $5$--$10\,\mu$m sphere; the
attainable values in real, lossy materials are lower by orders of
magnitude (Sec.~\ref{sec:mie}), but remain large enough to be
technologically interesting. Three application
lanes follow: \textit{(i)}~SERS, with enhancement factors above
$10^{10}$ reported in plasmonic Fano-resonant hot spots
\citep{lukyanchuk2010fano,lassiter2010fano,stockman2011nanoplasmonics},
hybridisation with a dielectric supercavity being the natural extension;
\textit{(ii)}~nonlinear conversion in the sub-wavelength supercavity
volume, where the Koshelev quasi-BIC platform
\citep{koshelev2018asymmetric,koshelev2020subwavelength} delivers
several orders of magnitude of second-harmonic, sum-frequency, and FWM
efficiency over non-resonant references;
\textit{(iii)}~high-harmonic generation in spectrally tailored mesoscale
resonators \citep{minin2025mesotronics,minin2022extreme}. A natural
frontier --- gain-compensated modes
\citep{liberal2014magnetic,gordon2007coated} (an amplification-family
construction) fused with the modal
selectivity of the high-order Fano resonance --- is a research direction,
not a demonstrated result, and its ultimate limits are set by gain
saturation and noise.

\subsection{Gravitational-wave detection and ultralight dark-matter searches}\label{sec:gravitational}\label{sec:darkmatter}

LIGO/Virgo \citep{aasi2015advanced,abbott2016observation} is the most
spectacular deployment of resonant optical storage. The arm
Fabry--P\'erot cavities \citep{drever1983laser}, together with the power-
and signal-recycling cavities, form a coupled-cavity system whose strain
sensitivity is set by its full transfer function (Sec.~\ref{sec:keyeqs});
the same coupled-cavity engineering, on a chip, underlies the Malara
super-resonant CPA \citep{malara2016super}. Quantum squeezing
\citep{tse2019quantum}, now routine, reduces the quantum noise
below the shot-noise level, contributing to a present strain sensitivity
of $h\!\sim\!10^{-23}\,\mathrm{Hz}^{-1/2}$. The astrophysical companion
phenomenon is black-hole super-radiance (Sec.~\ref{sec:celestial}): the
condition
\eqref{eq:superradianceBH} amplifies an ultralight boson of mass
$\sim\!10^{-13}\,$eV over astrophysical times, so that the observed spin
distribution of stellar-mass and supermassive holes constrains
ultralight bosonic dark matter
\citep{press1972rotating,bekenstein1973black,brito2020superradiance}.
Magnetospheric GW-to-EM conversion in neutron stars
\citep{mcdonald2024resonant} provides a complementary resonant-conversion
channel detectable in the radio band. The common thread is resonant
storage and conversion of weak signals --- engineered coupled cavities
on Earth, amplification and resonant conversion in the cosmos --- though,
as the taxonomy of Sec.~\ref{sec:taxonomy} makes clear, only the
terrestrial readout is built from interference-suppressed modes, while
the astrophysical channels are amplification phenomena.

\subsection{Mesotronics and integrated supercavity platforms}\label{sec:mesotronics}

``Mesotronics'' \citep{minin2025mesotronics} --- the integration of the
mesoscale Mie resonators of Sec.~\ref{sec:mie} into chip-scale
devices --- assembles in a single passive element a high computed modal
$Q$ \citep{minin2024discovery}, the (idealised) internal-field
enhancement of \eqref{eq:Eintensity}, sub-wavelength hot-spots, and the
$10^3\,\mathrm{nm/RIU}$ refractometric slope of Sec.~\ref{sec:sensing}.
The material palette spans low-index water and Teflon, intermediate
borosilicate and cenosphere shells, and high-index Si
\citep{minin2024freezing,minin2022janus,minin2022borosilicate,
minin2023cenosphere,yue2020teflon}, reflecting that the high-order Fano
condition is a discrete condition on $(q,m)$ rather than a high-index
requirement. A proposed --- not yet realised --- frontier is the on-chip
cascade of a mesoscale resonator with a quasi-BIC metasurface and a
topological edge waveguide; whether the multiplicative coupled-cavity
composition of Eq.~\eqref{eq:Leff} carries over to such a heterogeneous
stack, and what sensitivity it would reach, are open design questions.

\subsection{Further directions: proposals and conjectures}\label{sec:emergingapps}

The applications above rest on existing experimental demonstrations. We
collect here, separately and explicitly as \emph{proposals or
conjectures}, the further directions in which the framework suggests
opportunities but no device has yet been shown. None of the
performance figures in this subsection should be read as achieved.

\paragraph{Optical trapping and photonic hooks.}\label{sec:trapping}
The internal-field gradient of \eqref{eq:Eintensity} implies trapping
forces well above conventional tweezers; gain-loaded particles
\citep{liberal2014magnetic} (an amplification construction) have been
predicted to enhance these forces further. The photonic-hook geometry  \citep{minin2017diffractive,minin2022janus,tang2024unusual}
yields an off-axis $\sim\!\lambda/3$ trap, and rotating that geometry
\citep{tang2024unusual} couples mechanical angular momentum into the
optical field --- a rotating-scatterer relative of the rotational
amplification of Sec.~\ref{sec:celestial}, albeit one in which no
reflection gain has been demonstrated. These are active research
directions, not established manipulation platforms.

\paragraph{Frequency standards and narrow-linewidth
lasers.}\label{sec:clocks}
The Meiser \etal\ proposal \citep{meiser2009prospects} for a
super-radiant laser with a sub-millihertz linewidth, exploiting the
collective physics of Sec.~\ref{sec:quantum}
\citep{dicke1954coherence,gross1982superradiance}, remains a proposal at
the deepest linewidths; partial demonstrations exist but the projected
ultimate performance has not been reached. Single-photon super-radiance
\citep{jordao2025single} suggests cavity-free clock architectures, again
at the conceptual stage.

\paragraph{Quantum information and on-chip
networks.}\label{sec:quantumnet}
The framework suggests assigning a distinct mechanism to each network
function: Purcell-enhanced quasi-BIC sources
\citep{rybin2017highq,koshelev2018asymmetric,bogdanov2019bound,
munsch2009continuous}, M1 dark-mode magnonic memories
\citep{sukhorukova2022superresonant,zhang2015magnon}, and topologically
protected routing \citep{ozawa2019topological,harari2018topological}.
The integration of all three on one chip is an attractive target that no
group has yet attempted, and we present it as such.

\paragraph{Plasma heating and current drive (M3
family).}\label{sec:fusion}
Wave--particle current drive is, of course, a mature and economically
central technology --- LHCD sustains
$\eta_\mathrm{CD}\!\sim\!3\!\times\!10^{19}\,\mathrm{A\,W^{-1}\,m^{-2}}$
on present tokamaks \citep{fisch1987theory,bonoli2014review,
wesson2011tokamaks} --- but it is an M3 commensurability phenomenon, not
strict super-resonance, and we include it only to note a cross-field
design analogy: ECCD neoclassical-tearing-mode stabilisation
\citep{prater2004heating} poses the same kind of spectral-purity
targeting problem as quasi-BIC metasurface design (Sec.~\ref{sec:bic}).
The Fisch--Rax alpha-channeling scheme \citep{fisch1992interaction} is
flagged as an open problem in Sec.~\ref{sec:outlook}.

\paragraph{Acoustic and THz devices.}\label{sec:acousticTHz}
Sub-MHz acoustic BICs and absorbers \citep{liu2000locally,
cummer2016controlling,ma2016acoustic,hussein2014dynamics} and the
elastic encoding schemes of Sec.~\ref{sec:imaging}
\citep{maznev2015upholding} carry the M1 toolkit into the acoustic
domain. THz quasi-BIC devices in high-index low-loss semiconductors
\citep{rybin2024metaphotonics,koshelev2019meta,kivshar2018meta} are a
natural but still largely unrealised target, for which the
Harris--Hussein broadband elastic construction \citep{harris2026super}
offers a possible phase-coherence design route.

\paragraph{Biomedical directions.}
Photothermal therapy with metal-coated dielectric particles
\citep{liberal2014magnetic,minin2022janus} is a plausible extension of
the SERS and field-enhancement physics above, distinguished from purely
plasmonic approaches by lower intrinsic ohmic loss. We note that more
speculative proposals --- for instance, optomechanical disruption of
aerosolised pathogens by the internal field of a mesoscale droplet ---
require force estimates against realistic, lossy field values
(Sec.~\ref{sec:mie}) and dynamic, rather than steady-state, illumination
before they can be assessed; we therefore do not advance them here as
applications.

\section{Design Rules and Cross-Field Transfer}\label{sec:design}

Sections~\ref{sec:manifestations} and~\ref{sec:applications}
permit us to distil a compact set of design rules. The first is the
super-resonance rule proper; the next two record how the relative
families are engineered, included so that the boundaries stay sharp; the
remainder are practical corollaries.

\textit{Rule 1 (M1 --- super-resonance).}\enspace Identify the partial
amplitudes that contribute to $\gammarad$ and engineer their destructive
interference. This is the field-agnostic recipe for super-resonance in
the strict sense of \eqref{eq:SRdef}. The Friedrich--Wintgen
construction \eqref{eq:fw_imag} is the canonical realisation; the
Tolstoy acoustic array \citep{tolstoy1986superresonant}, the Hsu
photonic BIC \citep{hsu2013observation}, the anapole
\citep{miroshnichenko2015nonradiating}, and the Sukhorukova magnonic
dark mode \citep{sukhorukova2022superresonant} all implement it.

\textit{Rule 2 (M2 --- enhancement, for contrast).}\enspace Choose system
size and material parameters so that a single collective or high-order
amplitude exhausts the spectral weight, or so that $N$ emitters radiate
in phase. This is coherent \emph{enhancement}, not suppression:
Eq.~\eqref{eq:mie_intensity} (high-order Mie),
Dicke super-radiance \citep{dicke1954coherence}, plasmonic Fano
resonance \citep{lukyanchuk2010fano}, and the nuclear giant
dipole resonance \citep{harakeh2001giant} are instances. Its dark-state
complement is, however, M1, and the two are engineered together.

\textit{Rule 3 (M3 --- locking, for contrast).}\enspace Set up exact
integer commensurability between characteristic frequencies and minimise
the libration amplitude around the resonant island. The Laplace
resonance \citep{barnes2011laplace}, the Pluto super-resonance
\citep{wan2001pluto}, lower-hybrid current drive
\citep{fisch1987theory}, and photonic-time-crystal Floquet replicas
\citep{galiffi2022photonics,lyubarov2022amplified} instantiate it. This
is phase-space locking, not radiative suppression.

\textit{Rule 4 (cascade).}\enspace In specific coupled-cavity geometries
operated at mutual resonance, figures of merit compose
multiplicatively rather than additively (Eq.~\eqref{eq:Leff}); the
nested-resonator CPA chip \citep{malara2016super,selim2023enhanced} is
the clearest demonstration. The composition is a property of the
particular coupled-cavity design and must be verified case by case ---
it is not a universal law --- but where it holds, the design effort
spent on a single super-resonance pays off disproportionately.

\begin{figure}[t]
\centering
\includegraphics[width=0.52\linewidth]{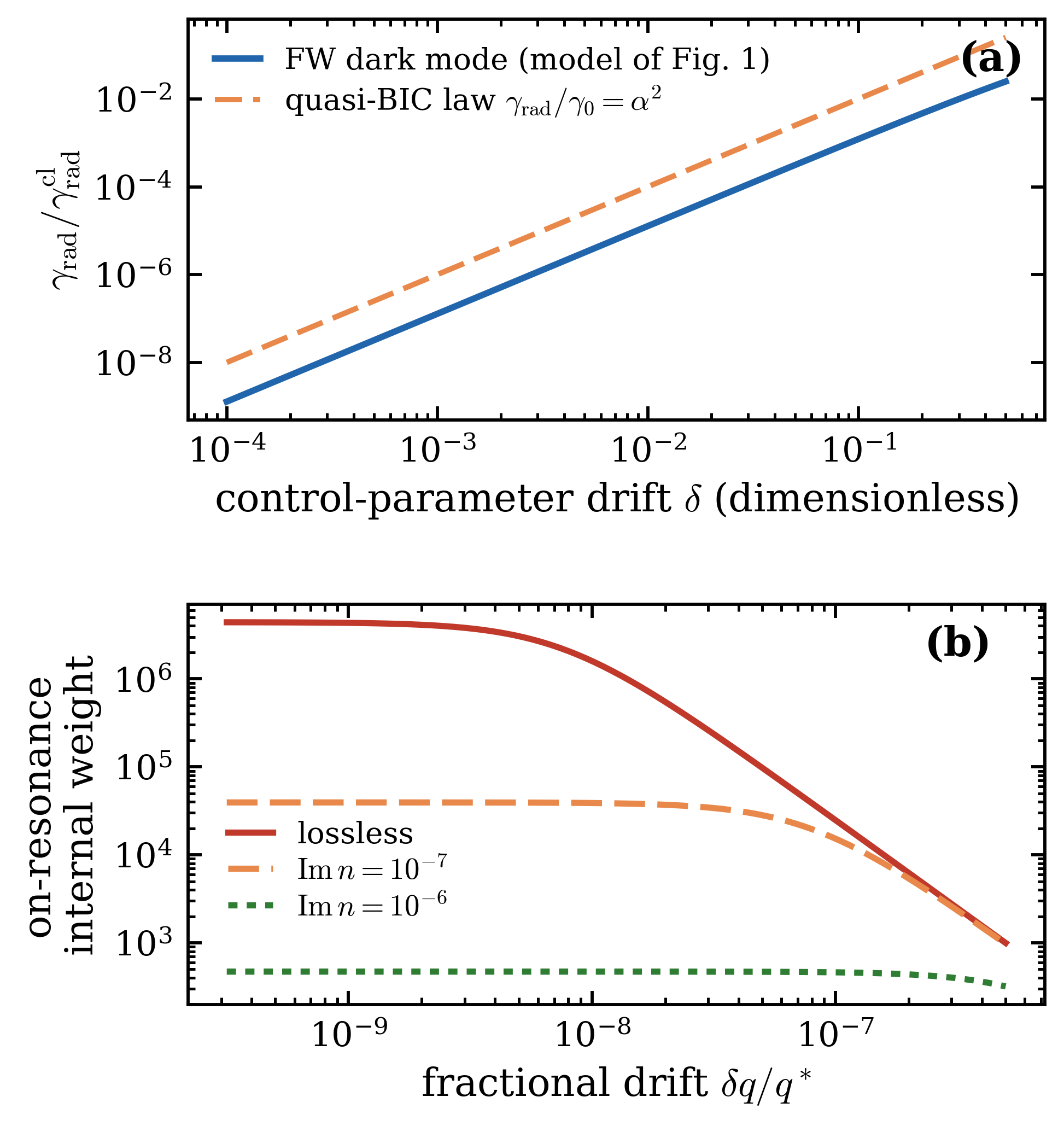}
\caption{The tolerance budget of Rules 5 and 6. (a)~Re-opening of the
radiative channel as the control parameter drifts off the M1 point:
the Friedrich--Wintgen dark width of the model of Fig.~\ref{fig:fw}
versus detuning drift $\delta$, together with the quasi-BIC law
$\gammarad/\gamma_0=\alpha^2$ \citep{koshelev2018asymmetric}. Both are
quadratic: interference zeros are generically second-order in the
control parameter, which is what makes quasi-BIC devices tolerant to
fabrication spread. (b)~Tolerance of the spectral parameter for the
high-order Mie needle of Fig.~\ref{fig:mie}: on-resonance internal
weight versus fractional size-parameter drift $\delta q/q^\ast$, for
three loss levels. The lossless needle loses half its response already
at $\delta q/q^\ast\!\approx\!10^{-8}$ (cf.\ the $10^{-6}$-level
sensitivity bound of Ref.~\citep{chylek1985internal}); absorption
lowers the attainable peak but widens the usable window --- the
loss--tolerance trade-off of Rule~6
\citep{yue2019poynting,yue2020teflon}.}
\label{fig:tol}
\end{figure}

\textit{Rule 5 (tolerance).}\enspace Tolerance-budget the controlling
parameter carefully; an M1 super-resonance is fragile to drift of order
$10^{-6}$ in the size parameter or its analogue
\citep{chylek1985internal,wan2019highorder}, and free-standing BIC
nanocavities demand absolute fabrication accuracies of a few nanometres
at near-infrared frequencies \citep{hoang2022nanocavity}. This is the fundamental
trade-off: the cost of a super-resonance is the precision with which its
parameters must be fixed, and a cascade compounds the requirement.
Active feedback (\eg\ Pound--Drever--Hall
locking \citep{drever1983laser}) is the standard mitigation, and
topological protection (Sec.~\ref{sec:topom1m3}) is the structural
alternative. Figure~\ref{fig:tol} quantifies the budget for the two
distinct senses of ``drift'' that this rule conflates at first reading.
Drift of the \emph{control} parameter away from the interference
condition re-opens the radiative channel only quadratically --- the
generic softness that makes quasi-BICs manufacturable. Drift of the
\emph{spectral} parameter off a fixed needle, by contrast, is punished
on the scale of the linewidth itself: for the Mie resonance of
Fig.~\ref{fig:mie} the half-response range is
$\delta q/q^\ast\!\sim\!10^{-8}$, and Rule~6's deliberate loss buys
tolerance precisely by paying peak response.

\textit{Rule 6 (loss).}\enspace Material loss has a non-monotonic effect:
a small amount of dissipation can enhance the spatial localisation of
the remaining field while damping its peak intensity
\citep{yue2019poynting,yue2020teflon}. Deliberately introducing
$\mathrm{Im}\,n\!\sim\!10^{-5}$ can therefore convert a delocalised
high-$Q$ mode into a sub-wavelength hot-spot device --- a design degree
of freedom, albeit one whose generality across geometries remains to be
mapped.

\textit{Rule 7 (cross-field transfer).}\enspace The interference
mechanism is the same eigenvalue problem in every wave system, yet its
solutions have not been carried across all the communities where they
should operate. Acoustic and elastic analogues of dielectric
supercavities are emerging \citep{harris2026super,lee2025surface};
magnonic counterparts of optical BICs
\citep{sukhorukova2022superresonant} are still in their infancy; THz
quasi-BIC devices are essentially absent from the experimental
literature. Cross-pollination of the M1 toolkit across these fields is,
in our estimation, where the next decade is most likely to yield
surprising returns.

\section{Outlook}\label{sec:outlook}

We close with a list of open problems whose resolution would, in our
estimation, define super-resonance research over the next decade.
They are organised around the M1 mechanism of
Section~\ref{sec:definition}, with the relative families entering where
they genuinely bear on the M1 programme.

\paragraph{Low-loss mesoscale dielectrics for extreme field
concentration.}
The peak field intensities of \eqref{eq:Eintensity} are idealised
lossless-sphere values (Sec.~\ref{sec:mie}); real materials fall short of
them by orders of magnitude. Commonly available optical glasses ---
borosilicate being a representative case --- reach
$\mathrm{Im}\,n\!\sim\!10^{-7}$; materials of substantially higher
purity exist (fibre-grade fused silica, at the dB/km attenuation level,
corresponds to $\mathrm{Im}\,n\!\sim\!10^{-10}$ in the visible) but the
choice of such bulk materials, and of the shapes into which they can be
formed with mesoscale precision, is extremely limited. Absorption,
moreover, is not the only obstacle: a mesoscale resonator operating at
the intensities of \eqref{eq:Eintensity} is acutely sensitive to
fabrication imperfection --- non-uniformity of the density (and hence of
the refractive index) through the resonator volume, together with the
figure accuracy and surface roughness of its boundary, both detune and
broaden the high-order resonance. Quantifying --- and then pushing back
--- this loss- and disorder-limited ceiling on the attainable internal
field is the single largest experimental bottleneck on the M2 axis. The
route likely involves ultra-low-absorption glasses and crystalline
dielectrics, together with direct experimental probes of the internal
hot-spot intensity, which is itself an unsolved measurement problem.

\paragraph{Nonlinear super-resonance.}
The Kerr and free-carrier nonlinearities saturate, and in some regimes
reverse, the super-resonance enhancement
\citep{koshelev2020subwavelength,minin2025mesotronics}. A predictive
theory of nonlinear super-resonance that includes self-consistent
feedback between the field and the material is still lacking. We expect
that the relevant FOM in the nonlinear regime is not $Q$ but
a saturation threshold $|E|^2_{\mathrm{sat}}$, and that the design space
will partition into a low-intensity ``linear M2'' regime, a
``saturating'' regime, and a high-intensity ``dynamic-instability''
regime each with its own optimum.

\paragraph{Quantum limits of super-resonance.}
What is the quantum-mechanical limit of the super-resonance figure of
merit? Squeezing of a super-resonant mode would yield sub-shot-noise
operation; super-resonant single-photon super-radiance
\citep{jordao2025single} suggests that the quantum limits are not
obvious. The intersection of cavity-QED ultrastrong coupling
\citep{frisk2019ultrastrong} with the multi-cavity cascade
\eqref{eq:Leff} is essentially unexplored.

\paragraph{Topology of super-resonance manifolds.}
The set of parameters for which the M1 condition is exact has non-trivial
topology. Whether this topology can be exploited for disorder-robust
super-resonance --- in the spirit of topological lasers
\citep{harari2018topological} --- is an open question, and one where the
phase-space-island structure of the M3 family
\citep{barnes2011laplace} may offer a useful, if formally distinct,
analogy worth making precise.

\paragraph{Cross-field transfer of design rules.}
Many M1 design techniques developed in one community have not been
carried across to others where they should be operative
(Rule 7 of Sec.~\ref{sec:design}). Acoustic and elastic analogues of
dielectric supercavities are emerging
\citep{harris2026super,lee2025surface}; magnonic counterparts of
optical BICs \citep{sukhorukova2022superresonant} are still in their
infancy. This cross-pollination is one of the central
opportunities the present mapping is meant to expose.

\paragraph{Super-resonance and machine learning.}
The high dimensionality of the parameter space and the fragility of the
super-resonance condition (Rule 5 of Sec.~\ref{sec:design}) make
machine-learning-assisted inverse design a natural tool. Physics-informed
neural networks for inverse design of BIC metasurfaces are encouraging
proofs-of-principle; whether they can be made to discover \emph{new}
super-resonance mechanisms --- as opposed to optimising known ones ---
remains an open question.

\paragraph{Industrial scale-up.}
Of all the applications listed in Sec.~\ref{sec:applications}, only WGM
biosensors and gravitational-wave Fabry--P\'erot arms have crossed from
laboratory demonstration to industrial deployment. The path from a
single-particle proof-of-principle (Sec.~\ref{sec:mie}) to a
wafer-scale super-resonant device is itself a research programme that
will require materials, fabrication, and packaging advances commensurate
with those of CMOS photonics three decades ago. The mesotronic platform
\citep{minin2025mesotronics} provides a concrete intermediate target
between the academic demonstration and the manufacturable device.

\paragraph{Super-resonance in the Earth--ionosphere cavity?}
A speculative but concrete target is the global Earth--ionosphere
resonator, which supports both the azimuthal Schumann modes near
$7.83$, $14.3$, and $20.8$~Hz
\citep{schumann1952strahlungslosen,balser1960observations,besser2007synopsis}
and a distinct family of transverse (radial) cavity resonances set by the
waveguide height. Whether an accidental degeneracy between an azimuthal and
a transverse mode could couple to realise a Friedrich--Wintgen dark state
\eqref{eq:fw_imag} --- a geophysical M1 super-resonance --- has, to our
knowledge, not been examined. We flag it as an open question rather than a
prediction, and with an explicit caveat that tempers the prospect: the
cavity is strongly dissipative (observed Schumann quality factors are only
$Q\!\sim\!4$--$6$, dominated by ionospheric ohmic loss rather than
radiation), so any interference-darkened combination would remain limited
by $\gammanr$ and might never reach a genuinely super-resonant $Q$.
Establishing whether the radiative-suppression mechanism survives at all in
such a lossy, inhomogeneous, magnetised-plasma resonator is the substance
of the problem.

\paragraph{Closing remark.}
The prefix ``super-'' has been attached, across many communities, to at
least three physically distinct things: the interference-driven
suppression of a radiative linewidth (M1), the coherent enhancement or
concentration of a coupling (M2), and the integer-commensurability
locking of conservative dynamics (M3), with active amplification as a
fourth relative. The central claim of this review is narrow and, we
believe, defensible: the \emph{first} of these admits a single,
field-agnostic operational definition \eqref{eq:SRdef}, and reading the
literature with that definition in mind discloses a genuine common
structure linking fields that have had little to do with one another ---
a Tolstoy array of acoustic cavities, a quasi-BIC dielectric metasurface,
a coherent-perfect-absorber dark mode, and a magnonic dark mode are all
the same eigenvalue problem. The value of the interdisciplinary mapping
is twofold: it transfers M1 design technique across fields, and ---
equally --- it keeps the genuinely common physics from being diluted by
conflation with its relatives. We trust that the framework will be
sharpened and corrected by the communities whose work we have attempted
to synthesise; the cross-pollination has only begun, with much still left
to explore. In the near future, rapid advances in computing power,
increasingly driven by artificial intelligence, may well uncover
super-resonant regimes and cross-field connections that remain
inaccessible to conventional analysis.

\section*{Acknowledgements}
The research was partially supported by the Tomsk Polytechnic University Development Program.

\section*{Declaration of competing interest}
The authors declare that they have no known competing financial
interests or personal relationships that could have appeared to
influence the work reported in this paper.

\section*{Data availability}
No new experimental data were generated in this review. The Python
scripts that generate all figures are available from the authors upon
reasonable request.

\bibliographystyle{unsrtnat}
\bibliography{references}

\end{document}